\documentclass[aps,prd,notitlepage,nofootinbib,superscriptaddress,groupedaddress,twocolumn]{revtex4-1}
\usepackage[utf8]{inputenc}

\usepackage[table ]{ xcolor}

\usepackage{multirow}

\usepackage{hyperref}
\usepackage{graphicx}
\usepackage{mathtools}
\usepackage{amsmath,amssymb,amsfonts,amsthm,latexsym,stmaryrd,physics,mathrsfs}
\allowdisplaybreaks[4]
\usepackage{marginnote}
\usepackage{color}
\usepackage{bm}
\usepackage{float}
\usepackage{nicefrac}
\usepackage{microtype} 
\usepackage{booktabs}
\usepackage{verbatim}
\usepackage{setspace}
\usepackage[T1]{fontenc}
\usepackage[utf8]{inputenc}

\hypersetup{
	colorlinks=true
,urlcolor=blue
,anchorcolor=blue
,citecolor=blue
,filecolor=blue
,linkcolor=blue
,menucolor=blue
,pagecolor=blue
,linktocpage=true
,pdfproducer=medialab
}

\def\beq{\begin{equation}}
\def\eeq{\end{equation}}
\newcommand{\bea}{\begin{eqnarray}}
\newcommand{\eea}{\end{eqnarray}}
\def\bi{\begin{itemize}}
\def\ei{\end{itemize}}
\def\ba{\begin{array}}
\def\ea{\end{array}}
\def\bfig{\begin{figure}}
\def\efig{\end{figure}}

\def\C{\mathbb{C}}
\def\R{\mathbb{R}}

\def\sgn{\text{sgn}}

\newtheorem{theorem}{Theorem}[section]
\newtheorem{definition}{Definition}[section]

\newtheorem{lemma}[theorem]{Lemma}

\newcommand{\Slc}{\mathrm{SL}(2,\mathbb{C})}
\newcommand{\PSlc}{\mathrm{PSL}(2,\mathbb{C})}

\newcommand{\Su}{\mathrm{SU}(2)}
\newcommand{\PSu}{\mathrm{PSU}(2)}

\def\be{\begin{eqnarray}}
\def\ee{\end{eqnarray}}

\newcommand{\ca}{\mathcal A}

\newcommand{\cb}{\mathcal B}
\newcommand{\cc}{\mathcal C}

\newcommand{\cf}{\mathcal F}

\newcommand{\ch}{\mathcal H}
\newcommand{\ci}{\mathcal I}
\newcommand{\cj}{\mathcal J}
\newcommand{\ck}{\mathcal K}
\newcommand{\cl}{\mathcal L}
\newcommand{\cm}{\mathcal M}
\newcommand{\cn}{\mathcal N}

\newcommand{\calp}{\mathcal P}

\newcommand{\cs}{\mathcal S}
\newcommand{\ct}{\mathcal T}
\newcommand{\cu}{\mathcal U}

\newcommand{\cx}{\mathcal X}
\newcommand{\cy}{\mathcal Y}
\newcommand{\cz}{\mathcal Z}

\newcommand{\scrp}{\mathscr{P}}
\newcommand{\scrq}{\mathscr{Q}}

\newcommand{\fa}{\mathfrak{a}}  \newcommand{\Fa}{\mathfrak{A}}

\newcommand{\fl}{\mathfrak{l}}  
  
\newcommand{\fn}{\mathfrak{n}}  
  
\newcommand{\fp}{\mathfrak{p}}  \newcommand{\Fp}{\mathfrak{P}}
\newcommand{\fq}{\mathfrak{q}}

\newcommand{\fv}{\mathfrak{v}}

\renewcommand{\a}{\alpha}
\renewcommand{\b}{\beta}
\newcommand{\g}{\gamma}
\newcommand{\G}{\Gamma}

\newcommand{\eps}{\varepsilon}

\newcommand{\sig}{\sigma}

\renewcommand{\l}{\lambda}
\renewcommand{\L }{\Lambda}
\renewcommand{\o}{\omega}
\renewcommand{\O}{\Omega}
\renewcommand{\t}{\tau}

\newcommand{\rmd}{\mathrm d}

\newcommand{\lt}{\left}
\newcommand{\rt}{\right}

\newcommand{\lag}{\left\langle}
\newcommand{\rag}{\right\rangle}

\newcommand{\act}{\rhd}

\newcommand{\sn}{\mathscr{N}}

\newcommand{\re}{\mathrm{Re}}
\newcommand{\im}{\mathrm{Im}}

\begin{document}

\title{Four-dimensional Spinfoam Quantum Gravity with Cosmological Constant: Finiteness and Semiclassical Limit}

\author{Muxin Han}
\email{hanm(At)fau.edu}
\affiliation{Department of Physics, Florida Atlantic University, 777 Glades Road, Boca Raton, FL 33431, USA}
\affiliation{Institut f\"ur Quantengravitation, Universit\"at Erlangen-N\"urnberg, Staudtstr. 7/B2, 91058 Erlangen, Germany}

\begin{abstract}

We present an improved formulation of 4-dimensional Lorentzian spinfoam quantum gravity with cosmological constant. The construction of spinfoam amplitudes uses the state-integral model of $\PSlc$ Chern-Simons theory and the implementation of simplicity constraint. The formulation has 2 key features: (1) spinfoam amplitudes are all finite, and (2) With suitable boundary data, the semiclassical asymptotics of the vertex amplitude has two oscillatory terms, with phase plus or minus the 4-dimensional Lorentzian Regge action with cosmological constant for the constant curvature 4-simplex.  

\end{abstract}

\maketitle

\tableofcontents

\section{Introduction}

The spinfoam quantum gravity is the covariant formulation of Loop Quantum Gravity (LQG) in 4 spacetime dimensions \cite{rovelli2014covariant,Perez2012}. 
There are 2 motivations to include the cosmological constant $\L$ in the spinfoam quantum gravity: Firstly, spinfoam models without $\L$ are well-known to have the infrared divergence (see e.g. \cite{Smerlak:2011fna,Riello2013,Dona:2018pxq}), then $\L$ is expected to provide a natural infrared cut-off to make spinfoam amplitudes finite. Secondly, the simplest consistent explanation for the cosmological accelerating expansion is a positive $\L$, so quantum gravity should reproduce $\L$ in the semiclassical regime. Based on these motivations, a satisfactory spinfoam quantum gravity with $\L$ is expected to (1) define finite spinfoam amplitudes, and (2) consistently recover classical gravity with $\L$ in the semiclassical limit. This work covers both positive and negative $\L$.

The semiclassical limit of LQG scales the Planck length $\ell_P\to 0$ while keeping the geometrical area $\fa$ fixed. By the LQG area spectrum $\fa=\g \ell_P^2\sqrt{j(j+1)}$, the semiclassical limit implies the SU(2) spin $j\to\infty$. We do not scale the Barbero-Immirzi parameter $\g$. In presence of $\L$, we require in addition that $\L$ should not scale in the semiclassical limit, then in 4d, the dimensionless quantity $k\propto (|\L|\ell_P^2)^{-1}$ scales as $k\to\infty$ in addition to $j\to\infty$, whereas $j/k\propto |\L| \fa$ is fixed. This suggests that the semiclassical limit of the spinfoam quantum gravity with $\L$ should be a double-scaling limit, i.e. $j,k\to \infty$ while fixing $j/k$. In our following discussion, $k$ becomes the integer Chern-Simons level.

In 3 dimensions, The Turaev-Viro (TV) model \cite{Turaev1992} with quantum group $\Su_\fq$ ($\fq=e^{\pi i/k},\ k\in\mathbb{Z}$) is the spinfoam quantum gravity with $\L$ that satisfy both expectations (1) and (2): It gives finite amplitudes due to the cut-off of spins given by $\Su_\fq$; The vertex amplitude, the 6j symbol of $\Su_\fq$, recovers the Regge action of 3d gravity with $\L>0$ in the semiclassical limit \cite{q6jasymp} \footnote{The semiclassical limit in 3d is the same double-scaling limit since $\fa\propto \ell_P\sqrt{j(j+1)}$ becomes the length and $k^2\propto (\L\ell_P^2)^{-1}$.}.    

\begin{figure}[h]
	\begin{center}
	\includegraphics[width=6cm]{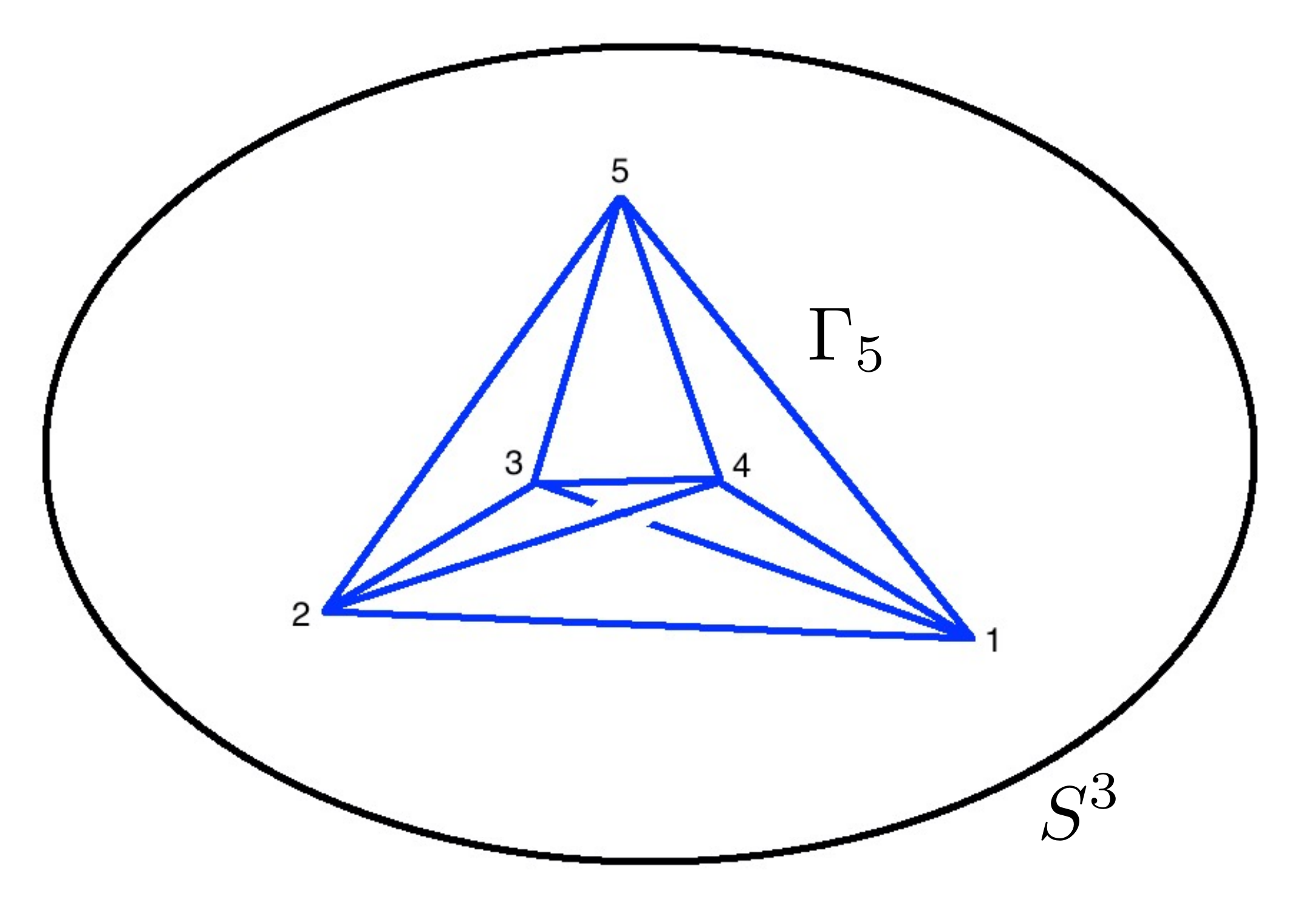}
	\caption{$\G_5$-graph embedded in $S^3$.}
	\label{S3G5}
	\end{center}
\end{figure}

In contrast, a satisfactory 4d spinfoam quantum gravity with $\L$ has not been achieved to satisfy both expectations (1) and (2) in the literature yet. There are 4d spinfoam models based on quantum Lorentz group, as generalizations from the 3d quantum group TV model \cite{NP,QSF,QSF1} (see also e.g. \cite{Dupuis:2013lka,Lewandowski:2008ye} for the LQG kinematics with quantum group). These models produce finite spinfoam amplitudes due to the spin cut-off from the quantum group. But it is difficult to examine the semiclassical limits of these models, due to complexity of their vertex amplitudes in terms of quantum group symbols. More recently, there is a more promising spinfoam model based on the $\Slc$ Chern-Simons (CS) theory instead of quantum group \cite{HHKR}. The vertex amplitude $A^0_v$ of this model is defined to be the CS evaluation of the projective $\Slc$ spin-network function $\Psi_{\G_5}$ based on $\G_5$-graph embedded in $S^3$ (see FIG.\ref{S3G5}):
\be
A^0_v:=\int D\ca D\bar{\ca}\, e^{-iS_{CS}(\ca,\bar{\ca})}\Psi_{\G_5}(\ca,\bar{\ca}),\label{A0v}
\ee
where $S_{CS}$ is the unitary $\Slc$ CS action with the complex level $t=k+\sig$ ($k\in\mathbb{Z}_+$, $\sig \in i\R$) that unifies $\L$ and $\g$ by $k=\mathrm{Re}(t)=\frac{12\pi}{|\L|\ell_P^2\g}$, $\sig=i\mathrm{Im}(t)=ik\g$,
\be
S_{CS}&=&\frac{t}{8\pi}\int_{S^3} \Tr(\ca\wedge \rmd\ca+\frac{2}{3}\ca\wedge \ca\wedge \ca)\nonumber\\
&&+\frac{\bar{t}}{8\pi}\int_{S^3} \Tr(\bar{\ca}\wedge \rmd\bar{\ca}+\frac{2}{3}\bar{\ca}\wedge \bar{\ca}\wedge \bar{\ca}).\label{CSaction}
\ee
$\Psi_{\G_5}$ reduces to the EPRL vertex amplitude \cite{EPRL} when $\ca,\bar{\ca}\to 0$. The derivation of the model \eqref{A0v} from the $\mathrm{BF}_\L$ theory is given in \cite{HHKR} and is reviewed briefly in a moment around \eqref{SHLBF}.

In the semiclassical limit ($j,k\to\infty$, $\sig=ik\g\to i\infty$, keeping $j/k$ fixed), and with suitable boundary condition, $A^0_v$ reproduces the constant curvature 4-simplex geometry and gives the asymptotics as 2 oscillatory terms, with phase plus or minus the Regge action of 4d Lorentzian gravity with $\L$. The sign of $\L$ is not fixed a priori, but rather emerges semiclassically and dynamically from equations of motion and boundary data, as shown in the asymptotic analysis in \cite{HHKR}\footnote{ Firstly, the sign of $\L$ of boundary tetrahedra is determined by the boundary data, then the critical equations from the stationary phase analysis lead the sign of $\L$ to propagate between tetrahedra and 4-simplices. The critical equations has no solution if the boundary tetrahedra fails to have a common sign of $\L$, then the spinfoam amplitude fast suppresses in the semiclassical regime.}. However the drawback of $A^0_v$ is that the formal path integral in \eqref{A0v} is not mathematically well-defined, thus makes the finiteness of the spinfoam amplitude obscure.

In this work, we present an improved formulation of 4d spinfoam quantum gravity with cosmological constant $\L$, which gives \emph{both} finite spinfoam amplitudes \emph{and} the correct semiclassical behavior. We construct a new vertex amplitude $A_v$, which replaces the formal CS path integral in $A^0_v$ by finite sum and finite-dimensional integral, based on the recent state-integral model of complex CS theory \cite{Dimofte2011,levelk,Andersen2014}. The resulting $A_v$ is a bounded function of boundary data. The spinfoam amplitude made by $A_v$ is finite on any triangulation. On the other hand, we are able to apply the stationary phase analysis to the finite-dimensional integral to show that $A_v$ indeed reproduce the constant curvature 4-simplex and the 4d Lorentzian Regge action with $\L$ (positive or negative) in the semiclassical limit. 

The new vertex amplitude $A_v$ is closely related to the partition function $\cz_{S^3\setminus\G_5}$ of the $\PSlc=\Slc/\mathbb{Z}_2$ CS theory on $S^3\setminus\G_5$, which is the complement of an open tubular neighborhood of $\G_5$-graph in $S^3$. $\G_5\subset S^3$ is dual to the triangulation of $S^3$ given by the 4-simplex's boundary. This duality delivers flat connections of the CS theory to decorate the 4-simplex. We adopt the method proposed in \cite{levelk} to explicitly construct $\cz_{S^3\setminus\G_5}$ as a state-integral model with finite sum and finite-dimensional integral (see Section \ref{Complex Chern-Simons theory}). $\cz_{S^3\setminus\G_5}$ quantizes the moduli space $\cl_{S^3\setminus\G_5}$ of $\PSlc$ flat connections on $S^3\setminus\G_5$, and is a wavefunction of flat connection data on the boundary of $S^3\setminus\G_5$. Given a manifold $M$, the moduli space of flat connection with structure group $G$ is the space of $G$-connections modulo gauge transformations with vanishing curvature, equivalent to the character variety of representations of $\pi_1(M)$ in $G$ modulo conjugation \cite{10.2307/2006973}.

The new vertex amplitude $A_v$ contains only finite sums and finite dimensional integrals thus improves the earlier formulation \eqref{A0v}. It is also different from the state-integral model obtained in \cite{hanSUSY} which mainly focuses on the holomorphic block of CS and does not specify the integration cycle \footnote{In addition, the construction here uses different symplectic coordinates from \cite{hanSUSY}.}. $A_v$ has both holomorphic and anti-holomorphic parts of the CS theory. As a key to prove the finiteness, the integration cycle is specified in $A_v$. 

By the construction in \cite{levelk}, the state-integral model converges absolutely if the underlying 3-manifold admits a ``positive angle structure''. Our construction of $\cz_{S^3\setminus\G_5}$ manifests that $S^3\setminus\G_5$ indeed admits a positive angle structure $(\vec{\a},\vec{\b})\in\Fp_{\rm new}$ where $\Fp_{\rm new}$ is a 30-dimensional open convex polytope. The finiteness of $\cz_{S^3\setminus\G_5}$ is a prerequisite for the finiteness of $A_v$ and spinfoam amplitudes on triangulations.

The simplicity constraint needs to be imposed in order to define $A_v$: The derivation of \eqref{A0v} in \cite{HHKR} starts from the Holst-$\mathrm{BF}_\L$ theory on a 4-ball $\cb_4$, which is topologically identical to a 4-simplex
\be
S_{\text{H-BF}_\L}&=&-\frac{1}{2} \int_{\cb_4}\Tr\lt[\left(\star+\frac{1}{\gamma}\right) B \wedge \mathcal{F}(\mathcal{A})\rt]\nonumber\\
&&-\frac{|\Lambda|}{12} \int_{\cb_4}\Tr\left[\left(\star+\frac{1}{\gamma} \right) B \wedge B\rt].\label{SHLBF}
\ee
Considering the formal path integral of $S_{\text{H-BF}_\L}$, integrating out the $so(1,3)$-valued 2-form $B$ gives the action $\frac{3i}{4|\L|}\int_{B_4}\Tr\left[\left(\star+\frac{1}{\gamma} \right)\mathcal{F} \wedge \mathcal{F}\rt]$, which is a total derivative and gives the CS action \eqref{CSaction} on the boundary $S^3\simeq \partial\cb_4$. By the feature of Gaussian integral, integrating out $B$ constraints $|\L| B/3=\cf (\ca)$, which encodes $B$ in the $so(1,3)$ curvature $\cf(\ca)$. On the boundary $S^3$, $\cf(\ca)$ is the $so(1,3)$ curvature of the CS connection $\ca$. Classically $S_{\text{H$\L$BF}}$ reduces to the Holst action of gravity with $\pm|\L|$ when the simplicity constraint $B=\pm e\wedge e$ is imposed, where $e$ is the cotetrad 1-form. At the quantum level, the simplicity constraint must be imposed to the CS partition function in order to obtain the spinfoam vertex amplitude.

By the relation $\cf (\ca)=|\L| B/3$, the simplicity constraint of $B$ can be translated to constraining $\ca$. By the CS symplectic structure, the resulting simplicity constraint can be divide into the first-class and second-class components. The first-class components are imposed \emph{strongly} to $\cz_{S^3\setminus\G_5}$ and restrict certain boundary data to a discrete set $\{2j_{ab}\}_{a<b}$, $a,b=1,\cdots,5$, where $j_{ab}\in \mathbb{N}_0/2$ and $j_{ab}\leq (k-1)/2$. $\{j_{ab}\}_{a<b}$ are analog of SU(2) spins associated to 10 boundary faces of the 4-simplex. Interestingly a consistency condition ``4d area=3d area'' (similar to \cite{generalize}) gives restrictions to the positive angle structure $(\vec{\a},\vec{\b})$. The second-class components of the simplicity constraint have to be imposed \emph{weakly}. We propose coherent states $\Psi_\rho$ peaked at points $\rho$ in the (subspace of) phase space of $\ca$, and apply the simplicity constraint to restrict $\rho$. The restricted $\rho$ is equivalent to the set of 20 spinors $\xi_{ab}\in \C^2$ normalized by the Hermitian inner product, such that for each $a=1,\cdots,5$, $\{{j}_{ab},{\xi}_{ab}\}_{b\neq a}$ are subject to the generalized closure condition of a constant curvature tetrahedra \cite{curvedMink}. In our model, all tetrahedra and triangles are spacelike. We denote the $\rho$ restricted by the simplicity constraint by $\rho_{\vec{j},\vec{\xi}}$. As a result, the vertex amplitude is defined by the inner product
\be
A_v(\vec{j},\vec{\xi})=\langle\bar{\Psi}_{\rho_{\vec{j},\vec{\xi}}}\mid \cz_{S^3\setminus\G_5}\rangle.\label{Avinnerprod}
\ee
where the complex conjugate of $\Psi_\iota$ is conventional. This inner product is a finite-dimensional integral of $L^2$-type. We show that the integral converges absolutely and $A_v$ is a bounded function of $\vec{j},\vec{\xi}$. $A_v$ as an inner product \eqref{Avinnerprod} resembles the idea of $A^0_v$, but now $A_v$ is well-defined.


Given a simplicial complex $\ck$ made by 4-simplices $v$, tetrahedra $e$, and faces $f$, following the general scheme of spinfoam state-sum models, the spinfoam amplitude associated to $\ck$ is defined by
\be
A=\!\!\!\sum_{\left\{j_{f}\right\}}^{(k-1) / 2}\!\!\!\!\!{}' \prod_{f} A_{f}\left(j_{f}\right) \int[\mathrm{d} \xi\rmd \xi'] \prod_{e} A_{e}\left(\vec{j}, \vec{\xi}_{e}, \vec{\xi}_{e}^{\prime}\right) \prod_{v} A_{v}(\vec{j}, \vec{\xi})\nonumber
\ee
where $j_f$ associates to a face $f$ and $\vec{\xi}_e=(\xi_1,\cdots,\xi_4)_e$ associates to a tetrahedron $e$. The CS level $k=\frac{12\pi}{|\L|\ell_P^2\g}\in\mathbb{Z}$ provides the cut-off to the sum over half-integer $0\leq j_f\leq (k-1)/2$. The face and edge amplitudes $A_f,A_e$ are not specified here except for requiring $A_e$ is an Gaussian-like continuous function approaching $\delta(\vec{\xi}_{e}, \vec{\xi}_{e}^{\prime})$ as $j\to\infty$. Given the boundedness of $A_v$, the amplitude $A$ is finite because the sum over $j_f$'s is finite and the integral over $\vec{\xi}$'s is compact. Here $\sum{}'$ indicates that some special spins are excluded in the sum.

When $\ck$ has boundary, the boundary data of $A$ are $j_f,\vec{\xi}_e$ for boundary faces $f$ and boundary tetrahedra $e$. These data are deformations of the data of coherent intertwiners in spin-network states. We conjecture that the boundary states of $A$ are $\fq$-deformed spin-network states of quantum group $\Su_{\fq}$ with $\fq$ root of unity.

After accomplishing the finiteness of spinfoam amplitude with $\L$, we demonstrate the correct semiclassical behavior fo the new vertex amplitude $A_v$ in Section \ref{Semiclassical analysis}. $A_v$ in \eqref{Avinnerprod} as a finite-dimensional integral can be expressed in the form $\int e^{k \ci}$ where $\ci$ depends on $j$'s only by $j/k$. Therefore we use the stationary phase analysis to study the semiclassical behavior of $A_v$ as $j,k\to\infty$ keeping $j/k$ fixed: The dominant contribution of $A_v$ comes from critical points, i.e. solutions of the critical equation $\delta \ci=0$. Given any boundary data $\{j_{ab},\xi_{ab}\}$ corresponding to the geometrical boundary of a nondegenerate convex constant curvature 4-simplex, there are exactly 2 critical points, which are 2 flat connections $\Fa,\widetilde{\Fa}\in\cl_{S^3\setminus\G_5}$ having geometrical interpretations as the constant curvature 4-simplex. $\Fa,\widetilde{\Fa}$ give the same 4-simplex geometry but opposite 4d orientations. $\Fa,\widetilde{\Fa}$ are analogous to the 2 critical points related by parity in the EPRL vertex amplitude \cite{semiclassical}. As a result, the asymptotic behavior of $A_v$ is given up to an overall phase by
\be
A_v&=&\lt(\sn_{+}e^{iS_{\rm Regge}+C}+\sn_{-}e^{-iS_{\rm Regge}-C}\rt)\label{asympformula}\\
&&\times \lt[1+O\lt({1}/{k}\rt)\rt],\nonumber
\ee
where $\sn_{\pm}$ are non-oscillatory and relate to the Hessian matrix of $\ci$. In the exponents, 
\be
S_{\rm Regge}&=&\frac{\L k\g}{12\pi}\lt(\sum_{a<b}\fa_{ab}\Theta_{ab}-\L |V_4|\rt).
\ee
is the 4d Lorentzian Regge action with $\L$ of the constant curvature 4-simplex reconstructed by $\Fa$ or $\widetilde{\Fa}$. The gravitational coupling is effectively given by $\ell^2_P=\frac{12\pi}{|\L| k\g}$. $C$ is an undetermined geometry-independent integration constant. 
This semiclassical result of $A_v$ is similar to the one related to $A_v^0$ in \cite{HHKR,HHKRshort,3dblockHHKR}.

Lastly, it is known that the formalism of state-integral models that we use to construct $\cz_{S^3\setminus\G_5}$ excludes the contributions from abelian flat connections \cite{Dimofte2011,levelk,33revisit}. This does not cause trouble for us since abelian flat connections only relate to degenerate tetrahedron geometries, which we exclude in the model.

\begin{figure}[t]
	\begin{center}
	\includegraphics[width=5cm]{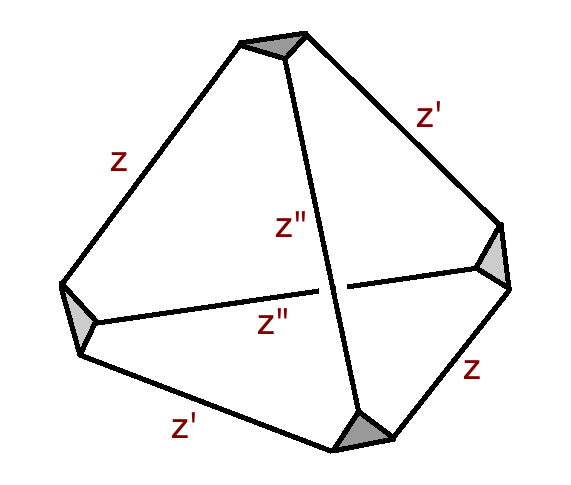}
	\caption{An ideal tetrahedron.}
	\label{tetrahedron}
	\end{center}
\end{figure}

The paper is organized as follows. In Section \ref{Complex Chern-Simons theory} we construct the state-integral model of $\cz_{S^3\setminus\G_5}$, including the discussion of ideal triangulation of $S^3\setminus\G_5$, a brief review of $\PSlc$ CS theory on an ideal tetrahedron, defining convenient phase space coordinates, constructing octahedron partition function then the partition function $\cz_{S^3\setminus\G_5}$, and the discussion of coherent states. In Section \ref{Spinfoam amplitude with cosmological constant}, we impose simplicity constraint and construct $A_v$, then we construct the spinfoam amplitude $A$ on simplicial complex and prove the finiteness, we also discuss the relation between boundary data of $A$ and LQG spin-networks, and various choices that we make in the definition of $A$. In section \ref{Semiclassical analysis}, we derive the asymptotic behavior of $A_v$ in the semiclassical limit.

\section{Complex Chern-Simons theory on $S^3\setminus\G_5$}\label{Complex Chern-Simons theory}

The purpose of this section is to construct the complex CS theory on the 3-manifold $S^3\setminus\G_5$. In Section \ref{Ideal triangulation of},  we firstly review the ideal triangulation of $S^3\setminus\G_5$ (see also \cite{hanSUSY}). As the building block, the CS theory on the ideal tetrahedron is reviewed in Section \ref{Complex Chern-Simons theory on ideal tetrahedron}. Then as an intermediate step, we construct the CS partition function on the idea octahedron in Section \ref{Octahedron partition function}, since the ideal triangulation of $S^3\setminus\G_5$ is made by 5 ideal octahedra. Section \ref{Phase space coordinates of} define the phase space coordinates of the CS theory on $S^3\setminus\G_5$ and the symplectic transformation from the phase space coordinates of the CS theory on the octahedra. The symplectic transformation defines the Weil-like transformations which relate the octahedron partition functions to the CS partition function on $S^3\setminus\G_5$, as discussed in Section \ref{S3-G5 partition function}. In Section \ref{Coherent state}, we discuss the coherent state of the CS theory, as will be useful for the spinfoam model.

\subsection{Ideal triangulation of $S^3\setminus\G_5$}\label{Ideal triangulation of}

\begin{widetext}

	\begin{figure}[H]
		\begin{center}
		\includegraphics[width=16cm]{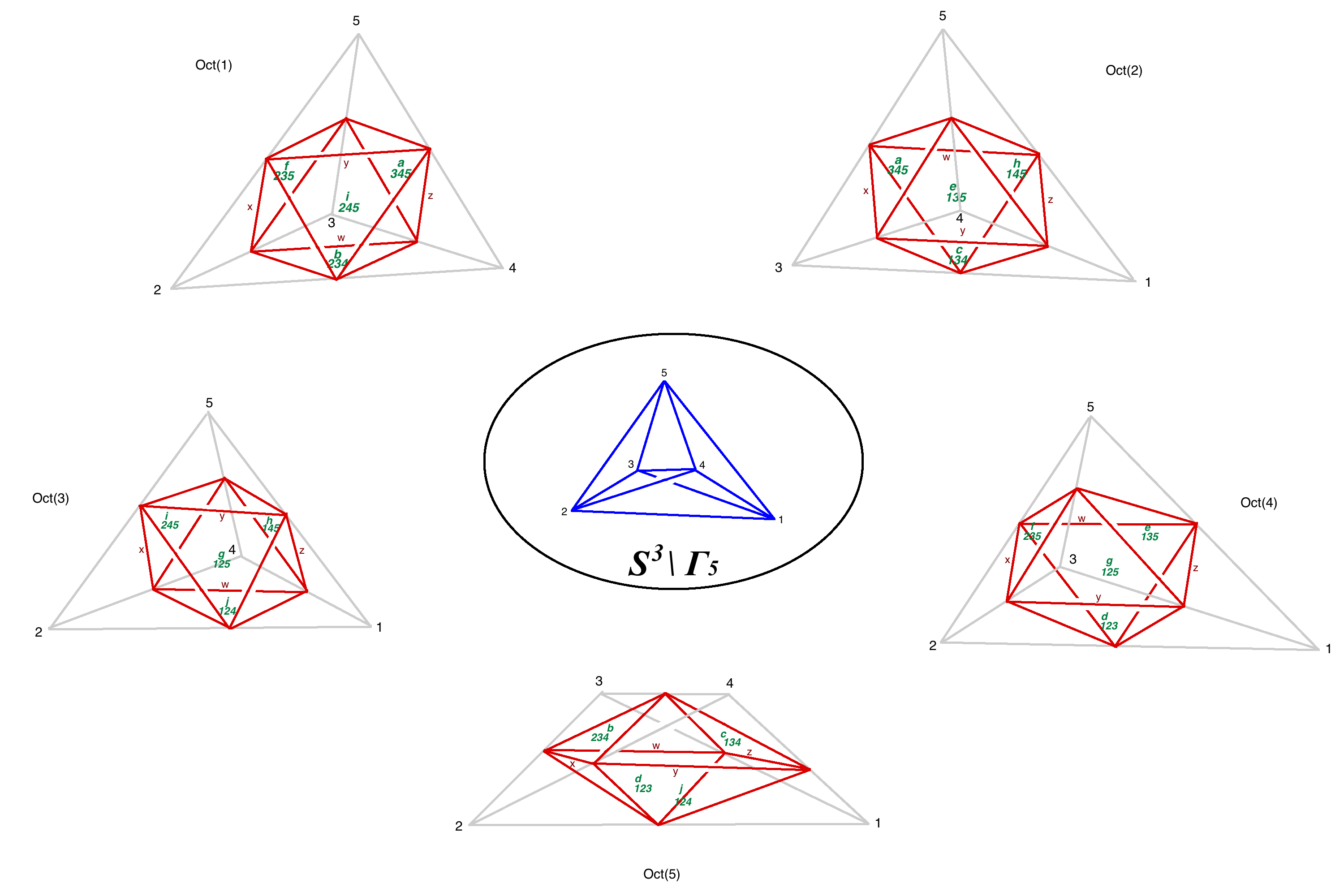}
		\caption{The decomposition of $S^3\setminus\G_5$ with 5 ideal octahedra (red), each of which can be decomposed into 4 ideal tetrahedra. The truncations of octahedron vertices are not drown in the figure. The faces with green label $a,b,c,d,e,f,g,h,i,j$ are the faces where a pair of octahedra are glued. Two ideal octahedra are glued through a pair of faces having the same label. In each ideal octahedron, we have chosen the edges with red label $x,y,z,w$ to form the equator of the octahedron. This ideal triangulation firstly appears in \cite{hanSUSY}.}
		\label{5oct}
		\end{center}
		\end{figure}
		
\end{widetext}

The 3-manifold $M_3=S^3\setminus\G_5$ is the complement in $S^3$ of an open tubular neighborhood of $\G_5$-graph (see FIG.\ref{5oct}). $M_3$ can be triangulated by a set of (topological) ideal tetrahedra. An ideal tetrahedron $\Delta$ is a tetrahedron whose vertices are located at infinities. It is convenient to truncate the vertices to define the ideal tetrahedron as the ``truncated tetrahedron'' as in FIG.\ref{tetrahedron}. There are 2 types of boundary components for the ideal tetrahedron: (a) the original boundary of the tetrahedron, and (b) the boundaries created by truncating tetrahedron vertices. Following e.g. \cite{Dimofte2011,DGG11,DGV}, the type-(a) boundary is called \emph{geodesic boundary}, and the type-(b) boundary is called \emph{cusp boundary}.

$M_3$ also has 2 types of boundary components: (A) the boundaries created by removing the open ball containing vertices of the graph, and (B) the boundaries created by removing tubular neighborhoods of edges. Here each type-(A) boundary component is a $4$-holed sphere. Each type-(B) boundary component is an annulus which begins and ends at a pair of holes of two type-(A) boundaries. The type-(A) boundary is called the \emph{geodesic boundary} of $M_3$, and the type-(B) boundary is called the \emph{cusp boundary}. An ideal triangulation decomposes $M_3$ into a set of ideal tetrahedra, such that the geodesic boundary of $M_3$ is triangulated by geodesic boundaries of the ideal tetrahedra, while the cusp boundary of $M_3$ is triangulated by cusp boundaries of the ideal tetrahedra. This ideal triangulation of $S^3\setminus\G_5$ is \emph{not} the triangulation of $S^3$ dual to $\G_5$ (the latter is given by the boundary of the 4-simplex). It is important to distinguish this two triangulations.

\begin{figure}[t]
	\begin{center}
	\includegraphics[width=8cm]{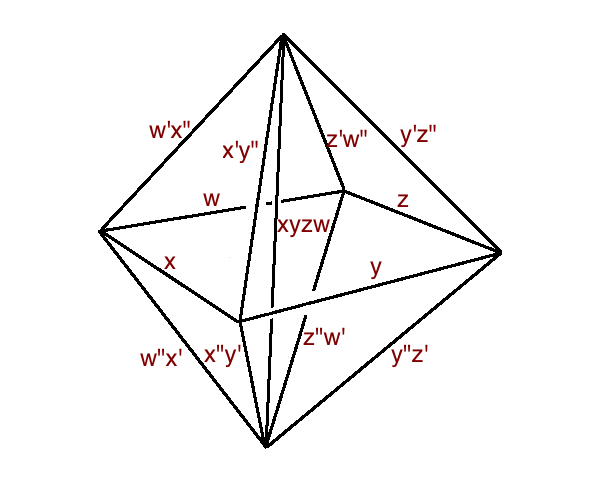}
	\caption{Chosen the equator edges with labels $x,y,z,w$, an ideal octahedron can be subdivided into 4 ideal tetrahedra by drawing a vertical line connecting the remaining 2 vertices which doesn't belong to the equator. Vertices are truncated, although truncations are not shown in the figure. }
	\label{oct_coordinate}
	\end{center}
	\end{figure}

Here the geodesic boundary of $S^3\setminus\G_5$ consists of five 4-holed spheres $\{\cs_a\}_{a=1}^5$, while the cusp boundary consists of 10 annuli $\{\ell_{ab}\}_{a<b}$. The $\G_5$-graph in FIG.\ref{5oct} motivates to subdivides $S^3\setminus\G_5$ into 5 tetrahedron-like regions (5 grey tetrahedra in FIG.\ref{5oct}, whose vertices coincide with the vertices of the graph). Every tetrahedron-like region should actually be understood as an ideal octahedron (with vertices truncated). The octahedron faces triangulate the 4-holed spheres, and the octahedron cusp boundaries (created by truncating vertices) triangulate the annuli. The way of gluing 5 ideal octahedra to form $S^3\setminus\G_5$ is shown in FIG.\ref{5oct}. Each ideal octahedron can be subdivided into 4 idea tetrahedra as shown in FIG.\ref{oct_coordinate}. A specific way of subdividing the octahedron is specified by a choice of octahedron equator. As a result, $S^3\setminus\G_5$ is triangulated by 20 ideal tetrahedra.

Given $M_3$ with both geodesic and cusp boundaries, a \emph{framed} $\PSlc$ flat connection on $M_3$ is an $\PSlc$ flat connection $A$ on $M_3$ with a choice of flat section $s$ (called the \emph{framing flag}) in an associated $\mathbb{CP}^1$ bundle over every cusp boundary (see e.g. \cite{DGV,GMN09,FG03}). The flat section $s$ can be viewed as a $\C^2$ vector field on a cusp boundary, defined up a complex rescaling and satisfying the flatness equation $(\rmd-A)s=0$ ($\rmd$ is the exterior derivative). Consequently the vector $s(\fp)$ at a point $\fp$ of the cusp boundary is an eigenvector of the holonomy of $A$ around the cusp based at $\fp$. The eigenvector fixes the Weyl symmetry. Similarly, a framed flat connection on $\partial M_3$ is a flat connection $\Fa$ on $\partial M_3$ with the same choice of {framing flag} on every cusp boundary. In addition, if a cusp boundary component of certain 3-manifold is a small disc, such as the boundaries created by truncating of tetrahedron vertices, the holonomy of any framed flat connection $\Fa$ around the disc is unipotent. The moduli space of framed $\PSlc$ flat connections on $\partial (S^3\setminus\G_5)$ is denoted by $\calp_{\partial (S^3\setminus\G_5)}$ which is a symplectic manifold with the Atiyah-Bott symplectic form. The moduli space of framed $\PSlc$ flat connections on $S^3\setminus\G_5$ is denoted by $\cl_{S^3\setminus\G_5}$ which is a Lagrangian submanifold in $\calp_{\partial (S^3\setminus\G_5)}$.  


\subsection{Complex Chern-Simons theory on ideal tetrahedron}\label{Complex Chern-Simons theory on ideal tetrahedron}

Given the ideal triangulation, the building block of the CS theory on $S^3\setminus\G_5$ is the theory on an ideal tetrahedron $\Delta$. In this subsection, we review main results of the CS theory on $\Delta$, and refer to e.g. \cite{Dimofte2011,DGV,levelk} for details. The boundary $\partial\Delta$ of the ideal tetrahedron is a sphere with 4 cusp discs. We denote by $\calp_{\partial\Delta}$ the phase space of $\PSlc$ CS theory on $\Delta$. $\mathcal{P}_{\partial\Delta}$ is the moduli space of $\PSlc$ flat connections on a 4-holed sphere, where the holonomy around each hole is unipotent.

\begin{figure}[h]
	\begin{center}
	\includegraphics[width=5cm]{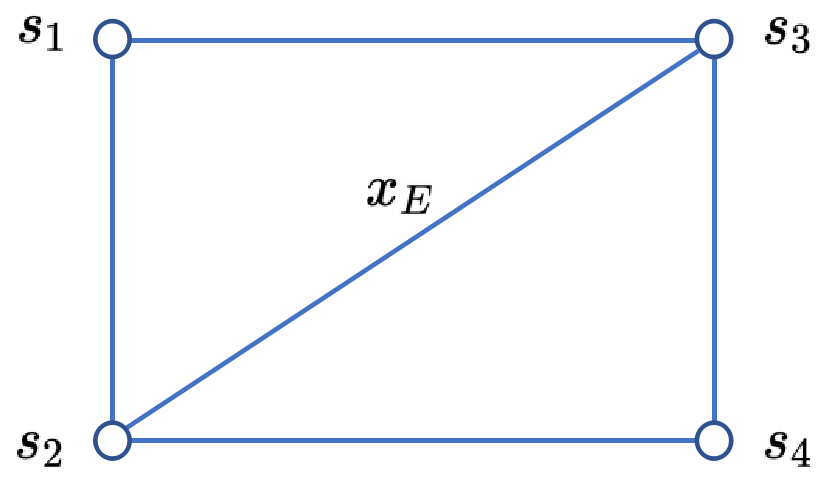}
	\caption{The quadrilateral in the 2d ideal triangulation for defining $x_E$}
	\label{FG}
	\end{center}
\end{figure}

The moduli space of $\PSlc$ flat connections on a $n$-holed sphere can be described as the following: A 2-sphere in which $n$ discs are removed is a $n$-holed sphere. We make a 2d ideal triangulation of the $n$-holed sphere such that edges in the triangulation end at the boundary of the holes. For example, the boundary of the ideal tetrahedron is an ideal triangulation of the 4-holed sphere. The 2d ideal triangulation has $3(n-2)$ edges. Each edge $E$ associates to a coordinate $x_E$ of the moduli space. Given a framed flat connection, $x_E$ is a cross-ratio of 4 framing flags $s_1,s_2,s_3,s_4$ associated to the vertices of the quadrilateral containing $E$ as the diagonal (see FIG.\ref{FG}), 
\be
x_E=\frac{\lag s_1\wedge s_2\rag\lag s_3\wedge s_4\rag}{\lag s_1\wedge s_3\rag\lag s_2\wedge s_4\rag}
\ee
where $\lag s_i\wedge s_j\rag$ is an $\Slc$ invariant volume on $\C^2$, and is computed by parallel transporting $s_1,\cdots,s_4$ to a common point inside the quadrilateral by the flat connection. The set of $\{x_E\}_E$ are the Fock-Goncharov (FG) edge coordinates of the moduli space of $\PSlc$ flat connections on the $n$-holed sphere. The correspondence between $\{x_E\}_E$'s and framed $\PSlc$ flat connections on $\cs_a$ is 1-to-1 \cite{FG03}. By the ``snake rule'' \cite{DGV,GMN09}, $\PSlc$ holonomies on the $n$-holed sphere can be expressed as $2\times 2$ matrices whose entries are functions of $\{x_E\}$. In particular, the eigenvalue $\l$ of the counterclockwise holonomy (of the flat connection) around a single hole relates to $x_E$ by
\be
\prod_{E\ \text{around hole}}(-x_E)=\l^2.
\ee
It is convenient to lift it to a logarithmic relation
\be
\sum_{E\ \text{around hole}}(\chi_E-i\pi)=2L,\label{XEL}
\ee
where $x_E=e^{\chi_E},\quad \l=e^L$. The moduli space has a natural Poisson structure with
\be
\{\chi_E,\chi_{E'}\}=\epsilon_{E,E'},
\ee 
where $\epsilon_{E,E'}\in {0,\pm1,\pm2}$ counts the number of oriented triangles shared by $E,E'$, $\epsilon_{E,E'}=+1$ if $E'$ occurs to the left of $E$ in a triangle. Note that the moduli space of $\PSlc$ flat connections on any $n$-holed sphere is not a symplectic manifold unless $\l$ of all holes are fixed.

Applying to the boundary of the ideal tetrahedron, we denote the FG coordinates at edges around a given hole (cusp disc) by $z,z',z''$ (see FIG.\ref{tetrahedron}). The trivial holonomy around each hole gives that 
\be
zz'z''=-1
\ee
The similar conditions for all 4 cusps identify the FG coordinates at opposite edges. As a result, we find
\be 
\mathcal{P}_{\partial \Delta}=\left\{z, z^{\prime}, z^{\prime \prime} \in \mathbb{C}^{*} \mid z z^{\prime} z^{\prime \prime}=-1\right\} \simeq\left(\mathbb{C}^{*}\right)^{2}.
\ee
$\mathcal{P}_{\partial \Delta}$ is a symplectic manifold since the holonomy eigenvalues at all holes are fixed. The Atiyah-Bott symplectic form is $\Omega=\frac{\rmd z^{\prime\prime}}{z^{\prime\prime}}\wedge \frac{\rmd z}{z} $. We also define the logarithmic phase space coordinates $Z=\log(z),Z^{\prime}=\log(z'),Z^{\prime \prime}=\log(z'')$ with canonical lifts that satisfy
\be
&&Z+Z^{\prime}+Z^{\prime \prime}=i \pi, \label{ZZZipi}\\
&&\left\{Z, Z^{\prime\prime}\right\}_\O=\left\{Z^{\prime\prime}, Z^{\prime }\right\}_\O=\left\{Z^{\prime }, Z\right\}_\O=1.
\ee

The $\PSlc$ CS theory at levels $k\in \mathbb{Z},\sig\in i\R$ endows the following symplectic form $\o_{k,\sig}$ on $\calp_{\partial\Delta}$:
\be
\omega_{k, \sigma}&:=&\frac{1}{4 \pi}(t \Omega+\bar{t} \bar{\Omega}), \quad t:=k+\sigma, \quad \bar{t}:=k-\sigma
\ee
$k,\sig$ relates to the cosmological constant $\L$ by 
\be
k=\frac{12\pi}{|\L|\ell_P^2\g},\quad \sig&=&ik\g\label{kandLambda}
\ee
where $\g$ is The Barbero-Immirzi parameter \cite{HHKR}. We use the following parametrization to change from $\g$ to $b$ \cite{levelk} 
\be
i \g&=&\frac{1-b^2}{1+b^2},\quad b^2=\frac{1-i \gamma }{1+i \gamma },\\
\frac{4 \pi i}{t}&=&\frac{2 \pi i}{k}\left(1+b^{2}\right), \quad \frac{4 \pi i}{\bar{t}}=\frac{2 \pi i}{k}\left(1+b^{-2}\right) .
\ee
with complex $b$ satisfying 
\be
\mathrm{Re}(b)>0, \quad\mathrm{Im}(b)\neq 0, \quad |b|=1.
\ee	
We reparametrize $z,z''$ and define $\widetilde{z},\widetilde{z}''$ by 
\be
z&=&\exp \lt[\frac{2 \pi i}{k}(-i b \mu-m)\rt], \\
\widetilde{z}&=&\exp \lt[\frac{2 \pi i}{k}\left(-i b^{-1} \mu+m\right)\rt], \label{zbarz}\\
z''&=&\exp \lt[\frac{2 \pi i}{k}(-i b \nu-n)\rt], \\
\widetilde{z}''&=&\exp \lt[ \frac{2 \pi i}{k}\left(-i b^{-1} \nu+n\right)\rt],\label{zbarzpp}
\ee
where $(m,n)$ are real and periodic $(m\sim m+k,n\sim n+k)$. When $(\mu,\nu)$ are real, $\widetilde{z},\widetilde{z}''$ are complex conjugates of $z,z''$. But in the following, $(\mu,\nu)$ will be analytic continued away from the real axis. 
$\omega_{k, \sigma}$ written in terms of $\mu,\nu, m, n$ gives
\be
\omega_{k, \sigma}=\frac{2 \pi}{k}(\rmd \nu \wedge \rmd \mu-\rmd n \wedge \rmd m).
\ee

The quantization of $(\calp_{\partial\Delta},\o_{k,\sig})$ promotes $\mu,\nu,m,n$ to operators $\bm{\mu}, \mathbf{m} , \bm{\nu}, \mathbf{n}$ satisfying the commutation relations
\be
[\boldsymbol{\mu},\boldsymbol{\nu}]=[\mathbf{n},\mathbf{m}]=-\frac{k}{2 \pi i},  \quad[\boldsymbol{\nu}, \mathbf{m}]=[\boldsymbol{\mu}, \mathbf{n}]=0. \label{CCR}
\ee
The variables $m,n$ are both canonical conjugate and periodic, so the spectra of $\mathbf{m}, \mathbf{n}$ are discrete and bounded: $m,n\in \mathbb{Z}/k\mathbb{Z}$. A representation of \eqref{CCR} is the kinematical Hilbert space
\be
\ch^{(k,\sig)}_{kin}=L^2(\R)\otimes \C^k
\ee
For any wave function $f(\mu|m)\in \ch_{kin}^{(k,\sig)}$ where $\mu\in\R$ and $m\in\mathbb{Z}/k\mathbb{Z}$, the actions of $\bm{\mu}, \mathbf{m} , \bm{\nu}, \mathbf{n}$ are given by 
\be
&&\boldsymbol{\mu} f(\mu| m)=\mu f(\mu| m), \quad e^{-\frac{2 \pi i}{k} \mathbf{m}} f(\mu | m)=e^{-\frac{2 \pi i}{k} m} f(\mu| m), \nonumber\\
&&\boldsymbol{\nu} f(\mu| m)=\frac{k}{2 \pi i} \partial_{\mu} f(\mu| m), \quad e^{-\frac{2 \pi i}{k} \mathbf{n}} f(\mu| m)=f(\mu| m+1).\nonumber\\
\label{mumnunoperator}
\ee

We also define the operators corresponding to $z,z'',\widetilde{z},\widetilde{z}''$	
\be
\bm{z} &=&\exp\lt[ \frac{2 \pi i}{k}(-i b \boldsymbol{\mu}-\mathbf{m})\rt], \\
\widetilde{\bm{z}}&=&\exp\lt[ \frac{2 \pi i}{k}\left(-i b^{-1} \boldsymbol{\mu}+\mathbf{m}\right) \rt],\\
\bm{z}'' &=&\exp \lt[\frac{2 \pi i}{k}(-i b \boldsymbol{\nu}-\mathbf{n})\rt], \\
\widetilde{\bm{z}}''&=&\exp \lt[\frac{2 \pi i}{k}\left(-i b^{-1} \boldsymbol{\nu}+\mathbf{n}\right)\rt],
\ee
They satisfy $q$- and $\widetilde{q}$-Weyl algebras
\be
&&\bm{z}\bm{z}''=q\bm{z}''\bm{z}, \quad \widetilde{\bm{z}}\widetilde{\bm{z}}''=\widetilde{q}\widetilde{\bm{z}}''\widetilde{\bm{z}},\nonumber\\
&&\bm{z}\widetilde{\bm{z}}''=\widetilde{\bm{z}}''\bm{z},\quad \widetilde{\bm{z}}\bm{z}''=\bm{z}''\widetilde{\bm{z}},\nonumber\\
&&q=\exp \lt(\frac{4 \pi i}{t}\rt)=\exp\lt[\frac{2 \pi i}{k}\left(1+b^{2}\right)\rt], \\ 
&&\widetilde{q}=\exp \lt(\frac{4 \pi i}{\bar{t}}\rt)=\exp \lt[\frac{2 \pi i}{k}\left(1+b^{-2}\right)\rt].
\ee

The above discussion focuses on flat connections on the boundary $\partial\Delta$. Only a subset of the flat connections on the boundary can be extend into the bulk. The moduli space of $\PSlc$ flat connection on the ideal tetrahedron $\Delta$, denoted by $\cl_\Delta$, is a holomorphic Lagrangian submanifold in $\calp_{\partial\Delta}$. $\cl_\Delta$ can be expressed as the holomorphic algebraic curve in terms of $z,z''$ (see e.g. \cite{Dimofte2011,DGV}):
\be
\mathcal{L}_{\Delta} =\left\{z^{-1}+z^{\prime \prime}-1=0\right\} \subset \mathcal{P}_{\partial \Delta},
\ee
and similarly for the anti-holomorphic variables $\widetilde{z},\widetilde{z}''$. In the quantum theory, we promote the algebraic curve to the quantum constraints imposed on wave functions
\be
\lt(\bm{z}^{-1}+\bm{z}^{\prime \prime}-1\rt)\Psi_\Delta(\mu|m)=\lt(\widetilde{\bm{z}}^{-1}+\widetilde{\bm{z}}^{\prime \prime}-1\rt)\Psi_\Delta(\mu|m)=0.\nonumber
\ee
The solution is the quantum dilogarithm function (see e.g. \cite{levelk,Imamura:2013qxa,Faddeev:1995nb,Kashaev1996hyperbolic})
\be\label{qdilog}
\Psi_\Delta(\mu\mid m)=\begin{dcases}
\prod_{j=0}^{\infty} \frac{1-q^{j+1} z^{-1}}{1-\widetilde{q}^{-j} \widetilde{z}^{-1}} & |q|<1 ,\\
\prod_{j=0}^{\infty} \frac{1-\widetilde{q}^{j+1} \widetilde{z}^{-1}}{1-q^{-j} z^{-1}} & |q|>1 .
\end{dcases}
\ee
$\Psi_\Delta(\mu|m)$ is the CS partition function on the ideal tetrahedron $\Delta$. $\Psi_\Delta(\mu|m)$ defines a meromorphic function of $\mu\in\C$ for each $m\in\mathbb{Z}/k\mathbb{Z}$, and is analytic in $b$ in each regime $\mathrm{Im}(b)>0$ and $\mathrm{Im}(b)<0$ (correspondingly $|q|<1$ and $|q|>1$). The poles and zeros of $\Psi_\Delta(\mu|m)$ are
\be
&&\mu_{\rm pole/zero} =i b u+i b^{-1} v ,\quad\text{with}\quad u,v \in \mathbb{Z}, \nonumber\\ 
&&u-v=-m+ k\mathbb{Z} \quad \begin{cases}
\text { zeroes: } & u,v \geq 1 ,\\
\text { poles: } & u,v \leq 0.
\end{cases}\label{polesmu}
\ee
Poles of $\Psi_{\Delta}$ are in the lower-half plane 
\be
\mathrm{Im}(\mu_{\rm pole})=\mathrm{Re}(b)(u+v)\leq 0.
\ee
$\Psi_{\Delta}(\mu|m)$ is holomorphic in $\mu$ when $\mathrm{Im}(\mu)> 0$.

The asymptotic behavior of $\Psi_\Delta(\mu|m)$ as $\mathrm{Re}(\mu)\to \infty$ with fixed $\mathrm{Im}(\mu)$ is
\be
\Psi_\Delta(\mu|m)&=&\begin{cases}
O(1) & \operatorname{Re}(\mu) \rightarrow+\infty \\
\exp \left[\frac{i \pi}{k}\left(\mu-\frac{i}{2} Q\right)^{2}+O(1)\right] & \operatorname{Re}(\mu) \rightarrow-\infty
\end{cases},\nonumber\\
Q&=&b+{b}^{-1}>0.
\ee
The asymptotic behavior indicates that $\Psi_\Delta(\mu|m)$ does not belong to the Hilbert space $\ch_{kin}^{(k,\sig)}$ but is a tempered distribution. $\Psi_\Delta(\mu|m)$ is analytic in the upper-half plane $\mathrm{Im}(\mu)>0$. We have the following useful observation from the asymptotic behavior: Let $\alpha>0$, then
\be
&&\left|e^{-\frac{2 \pi}{k} \beta \mu} \Psi_\Delta(\mu+i \alpha| m)\right| \nonumber\\
&\sim& \begin{cases}
	\exp\lt[ -\frac{2 \pi}{k} \beta \mu\rt] & \mu \rightarrow \infty \\
	\exp \lt[-\frac{2 \pi}{k} \mu(\alpha+\beta-{Q} / 2)\rt] & \mu \rightarrow-\infty
	\end{cases} .
\ee
Therefore $e^{-\frac{2 \pi}{k} \beta \mu} \Psi_\Delta(\mu+i \alpha| m)$ is a Schwartz function of $\mu$ if $\a,\b$ is inside the open triangle $\Fp(\Delta)$:
\be
\Fp(\Delta)=\{(\a,\b)\in\R^2|\a,\b>0,\ \a+\b<Q/2\}.
\ee
The Fourier transform $\int\rmd \mu\,e^{\frac{2 \pi i}{k} \nu \mu} \Psi_\Delta(\mu| m)$ is convergent if the integration contour is shifted away from the real axis while $\a=\mathrm{Im}(\mu)$, $\beta=\mathrm{Im}(\nu)$ belong to $\Fp(\Delta)$. $\a,\b$ can be understood as angles associated with coordinates $z,z''$ in the context of hyperbolic geometry. $(\a,\b)\in\Fp(\Delta)$ is called a ``positive angle structure'' of $\Delta$ \cite{levelk,Andersen2014}.

\subsection{Octahedron partition function}\label{Octahedron partition function}

Four ideal tetrahedra are glued to form an ideal octahedron as shown in FIG.\ref{oct_coordinate}. The phase space $\calp_{\partial\mathrm{oct}}$ is a symplectic reduction from 4 copies of $\calp_{\partial\Delta}$: The FG edge coordinates $\{x_E\}$ of $\calp_{\partial\mathrm{oct}}$ a product of the tetrahedron edge coordinates. In general for any edge on the boundary or in the bulk, it associates \cite{DGV}
\be
x_E&=&\prod(z,z',z''\ \text{incident at}\ E)\quad\text{or}\nonumber\\
\chi_E&=&\sum(Z,Z',Z''\ \text{incident at}\ E),\label{edgesum}
\ee 
as product or sum over all the tetrahedron edge coordinates incident at the edge $E$. For boundary edges, $x_E$ are the FG coordinates of $\calp_{\partial\mathrm{oct}}$. The lift of $\chi_E=\log(x_E)$ is determined by the lifts of $Z,Z',Z''$ of ideal tetrahedra. For the bulk edge, $x_E$ or $\chi_E$ is rather a constraint which is denoted by $c_E=\exp(C_E)$, satisfying
\be
c_E=1\ \ \ \text{or}\ \ \ C_E=2\pi i, 
\ee
because the flat connection holonomy around a bulk edge is trivial. We denotes the edge coordinates in 4 copies of $\calp_{\partial\Delta}$ by $X,Y,Z,W$ and their double primes. All the edge coordinates of $\calp_{\partial\mathrm{oct}}$ are expressed in FIG.\ref{oct_coordinate}, where we have a single constraint at the bulk edge
\be
C=X+Y+Z+W=2\pi i\label{CXYZW}
\ee
We make a symplectic transformation in $\calp_{\partial\Delta}\times\calp_{\partial\Delta}\times \calp_{\partial\Delta}\times\calp_{\partial\Delta}$ from the tetrahedron coordinates $(X,X'')$,$(Y,Y'')$,$(Z,Z'')$,$(W,W'')$ to a set of new symplectic coordinates $(X,P_X),(Y,P_Y),(Z,P_Z),(C,\G)$, where
\be
P_X&=&X''-W'',\ \ P_Y=Y''-W'',\nonumber\\ 
P_Z&=&Z''-W'',\ \ \G=W''\label{PXYZW}
\ee
and each pair are canonical conjugate variables, Poisson commutative with other pairs. The octahedron phase space $\calp_{\partial\mathrm{oct}}$ is a symplectic reduction by imposing the constraint $C=2\pi i$ and removing the ``gauge orbit'' variable $\G$. A set of symplectic coordinates of $\calp_{\partial\mathrm{oct}}$ are given by $\vec{\phi}=(X,Y,Z)$, $\vec{\pi}=(P_X,P_Y,P_Z)$. The Atiyah-Bott symplectic form $\O$ implies 
\be
\{\phi_i,\pi_j\}_\O=\delta_{ij},\quad \{\phi_i,\phi_j\}_\O=\{\pi_i,\pi_j\}_\O=0.\label{octsymp}
\ee

The CS partition function on the ideal octahedron, $Z_{\rm oct}$, is a product of 4 tetrahedron partition function 
followed by the restriction on the quantum deformed constraint surface $e^C=q$, $e^{\widetilde{C}}=\widetilde{q}$ \footnote{The quantum deformation is necessary to make the partition function invariant under 3d Pachner move (see e.g. \cite{Dimofte2011}).}:  
\be
&&Z_{\rm oct}(\mu_{X},\mu_{Y},\mu_{Z}|m_{X},m_{Y},m_{Z})\nonumber\\
&=&\Psi_\Delta \left(\mu_{X}|m_{X}\right) \Psi_\Delta \left(\mu_{Y}|m_{Y}\right) \Psi_\Delta \left(\mu_{Z}|m_{Z}\right)\nonumber\\
&& \Psi_\Delta \left(i Q-\mu_{X}-\mu_{Y}-\mu_{Z}|-m_{X}-m_{Y}-m_{Z}\right)\nonumber
\ee
The octahedron partition function gives the following asymptotics behavior
\be
&&\lt|e^{-\frac{2\pi}{k}\sum_{i}\b_{i}\mu_{i}}Z_{\rm oct}\lt(\{\mu_{i}+i\a_{i}\}\mid\{m_{i}\}\rt)\rt|\nonumber\\
&\sim&\begin{cases}
e^{ -\frac{2 \pi}{k}\mu_{X}\lt(\a_{X}+\b_{X}+\a_{Y}+\a_{Z}-{Q} / 2\rt)} \quad & \mu_{X}\to\infty\\
e^{-\frac{2\pi}{k}\mu_{X}\lt(\a_{X}+\b_{X}-Q/2\rt)} \quad & \mu_{X}\to-\infty
\end{cases}\nonumber
\ee
where $i=X,Y,Z$. The similar behaviors are satisfied for $\mu_{Y}\to\pm\infty$ or $\mu_{Z}\to\pm\infty$. Therefore $e^{-\frac{2\pi}{k}\sum_{i}\b_{i}\mu_{i}}Z_{\rm oct}\lt(\{\mu_{i}+i\a_{i}\}\mid\{m_{i}\}\rt)$ is a Schwartz function of $\mu_X,\mu_Y,\mu_Z$, if $(\a_{X},\b_{X},\a_{Y},\b_{Y},\a_{Z},\b_{Z})\in\R^6$ is contained by the open polytope $\Fp(\mathrm{oct})$ defined by the following inequalities
\be
&&\a_{X},\a_{Y},\a_{Z}>0,\quad \a_X+\a_Y+\a_Z<Q,\nonumber\\
&&\a_{X}+\b_{X}<\frac{Q}{2},\quad \a_{Y}+\b_{Y}<\frac{Q}{2},\quad \a_{Z}+\b_{Z}<\frac{Q}{2}, \nonumber\\
&& \a_{X}+\a_{Y}+\a_Z+\b_X>\frac{Q}{2},\quad \a_{X}+\a_{Y}+\a_Z+\b_Y>\frac{Q}{2},\nonumber\\
&&\a_{X}+\a_{Y}+\a_Z+\b_Z>\frac{Q}{2}.\label{octinequality}
\ee 
To see $\Fp(\mathrm{oct})$ is not empty, Appendix \ref{A plot for the polytope} shows a plot FIG.\ref{polytope} of the intersection between $\Fp(\mathrm{oct})$ and the plane of $\a_X=\a_Y=\a_Z$, $\b_X=\b_Y=\b_Z$. $(\vec{\a},\vec{\b})\in \Fp({\rm oct})$ is a positive angle structure of the ideal octahedron.

Following \cite{levelk}, we consider any $2N$-dimensional phase space $(\calp,\o)$ with Darboux coordinates $(\mu_i,m_i)$ and $(\nu_i,m_i)$ such that $\o=\frac{2 \pi}{k}\sum_{i=1}^n(d \nu_i \wedge d \mu_i-d n_i \wedge d m_i)$. The phase space associates with an ``angle space'' ($\calp_{\rm angle},\o_{\rm angle}$) whose universal cover is $T^*\R^N\simeq \R^{2N}$, the Darboux coordinates of $\calp_{\rm angle}$ are 
\be
\a_i=\mathrm{Im}(\mu_i),\quad \b_i=\mathrm{Im}(\nu_i)
\ee 
so that $\o_{\rm angle}=\sum_{i=1}^N d\b_i\wedge d\a_i$. Given a $2N$-dimensional open convex symplectic polytope $\Fp\in\calp_{\rm angle}$, we define $\pi(\Fp)$ to be the projection of $\Fp$ to the base of $T^*\R^N$, with coordinates $\vec{\a}$, then define 
\be
\operatorname{strip}(\Fp):=\left\{\vec{\mu} \in \mathbb{C}^{N} \mid \operatorname{Im}(\vec{\mu}) \in \pi(\Fp)\right\}.
\ee
We define the functional space
\be
\begin{aligned}
&\mathcal{F}_{\Fp}:=\Big\{\text { holomorphic functions } f: \operatorname{strip}(\Fp) \rightarrow \mathbb{C} \text { s.t. }\\
&\forall(\vec{\alpha}, \vec{\beta}) \in \Fp, \text { the function } e^{-\frac{2 \pi}{k} \vec{\mu} \cdot \vec{\beta}} f(\vec{\mu}+i \vec{\alpha}) \in \mathcal{S}\left(\mathbb{R}^{N}\right)\nonumber\\
&\text { is Schwartz class }\Big\} .
\end{aligned}
\ee
We have the convergence for any $f\in \mathcal{F}_{\Fp}$
\be
\int \rmd^N \mu\, e^{\frac{2\pi i}{k}\vec{\mu}\cdot\vec{\nu}} f(\vec{\mu})<\infty
\ee
when the integration contour is shifted away from the real axis while $\vec{\a}=\mathrm{Im}(\vec{\mu})$, $\vec{\beta}=\mathrm{Im}(\vec{\nu})$ belong to $\Fp$. 
$f\in \mathcal{F}_{\Fp}$ implies the Fourier transform of $f$ also belongs to $\mathcal{F}_{\Fp}$.

To accommodate partition functions of complex Chern-Simons theory at level $k$, we define 
\be
\cf_{\Fp}^{(k)}=\cf_{\Fp}\otimes_\C (V_k)^{\otimes N},\quad V_k\simeq \C^k.
\ee
As examples, the tetrahedron partition function $\Psi_\Delta$ belongs to $\cf^{(k)}_{\Fp(\Delta)}$ with $N=1$, and the octahedron partition function $Z_{\rm oct}$ belongs to $\cf_{\Fp({\rm oct})}^{(k)}$ with $N=3$.

\subsection{Phase space coordinates of $\calp_{\partial (S^3\setminus\G_5)}$}\label{Phase space coordinates of}

The geodesic boundary of $S^3\setminus\G_5$ consists of five 4-holed spheres, denoted by $\cs_{a=1,\cdots,5}$. In FIG.\ref{5oct}, each $\cs_a$ are made by the triangles from the geodesic boundaries of the octahedra. We compute all FG edge coordinates $\chi^{(a)}_{mn}$ ($a$ labels the 4-holed sphere and $mn$ labels the edge $E$) of flat connections on $\cs_{a=1,\cdots,5}$ by Eq.\eqref{edgesum}, and list them in Table \ref{edges} in Appendix \ref{Darboux coordinates of P}.

The phase space $\calp_{\partial(S^3\setminus\G_5)}$ is the moduli space of framed $\PSlc$ flat connections on the 2d boundary $\partial(S^3\setminus\G_5)$. We choose the Darboux coordinates of $\calp_{\partial(S^3\setminus\G_5)}$ as follows: First of all, the complex Fenchel-Nielsen (FN) length variable $\l_{ab}^2=e^{2L_{ab}}$ are squared eigenvalues of $\PSlc$ holonomies meridian to the 10 annuli $\ell_{ab}$ connecting 4-holed spheres $\cs_a$ and $\cs_b$. They relate edge coordinates $\chi^{(a)}_{mn}$ by \eqref{XEL}. 
Ten $2L_{ab}$ are linear combinations of of $(X_a,P_{X_a}),(Y_a,P_{Y_a}),(Z_a,P_{Z_a})$ from 5 Oct($a$) with integer coefficients. Their expressions are given in Appendix \ref{Darboux coordinates of P}. The resulting $L_{ab}$ are mutually Poisson commutative and commuting with all edge coordinates $\chi^{(a)}_{mn}$. 

All $L_{ab}$ commutes with 4-holed sphere edge coordinates $\chi^{(a)}_{mn}$. $\calp_{\partial(S^3\setminus\G_5)}$ is complex 30-dimensional. Among the Darboux coordinates, the position variables include ten $2L_{ab}$ and 5 variables $\cx_a$ $(a=1,\cdots,5)$, one for each 4-holed sphere. We choose $\cx_a$ to be one of $\chi^{(a)}_{mn}$:
\be
\cx_1&=&\chi^{(1)}_{25},\quad
\cx_2=\chi^{(2)}_{15},\quad
\cx_3=\chi^{(3)}_{15},\nonumber\\
\cx_4&=&\chi^{(4)}_{15},\quad
\cx_5=\chi^{(5)}_{14}.
\ee



We denote the conjugate momentum variables by $\ct_{ab}$ and $\cy_a$, and denote
\be
\mathscr{Q}_I=(2L_{ab},\cx_a),\quad \mathscr{P}_I=(\ct_{ab},\cy_a),\quad I=1,\cdots,15,\nonumber
\ee
where $I$ labels the boundary components $(\ell_{ab},\cs_a)$. The momentum variables $\ct_{ab}$ conjugate to $2L_{ab}$ are called the twist variables. On each $\cs_a$, the momentum variable $\cy_a$ conjugate to $\cx_a$ turns out to be also FG edge coordinates up to sign and $2\pi i$.
\be
\cy_1&=&\chi^{(1)}_{23}. \quad
\cy_2=\chi^{(2)}_{14},\quad
\cy_3=\chi^{(3)}_{45}-2\pi i,\nonumber\\
\cy_4&=&-\chi^{(4)}_{35}+2\pi i,\quad
\cy_5=\chi^{(5)}_{34}-2\pi i.
\ee
Explicit expressions of $2L_{ab},\ct_{ab},\cx_{a},\cy_a$ in terms of $(X_a,P_{X_a}),(Y_a,P_{Y_a}),(Z_a,P_{Z_a})$ are given in Appendix \ref{Darboux coordinates of P}.

There exists a linear symplectic transformation from $\vec{\Phi} \equiv (X_{a}, Y_{a}, Z_{a} )_{a=1}^{5}$ and $\vec{\Pi} \equiv (P_{X_{a}}, P_{Y_{a}}, P_{Z_{a}} )_{a=1}^{5}$ to $\vec{\mathscr Q},\vec{\mathscr P}$
\be
\left(\begin{array}{c}
\vec{\mathscr Q} \\
\vec{\mathscr P}
\end{array}\right)=\left(\begin{array}{ll}
\mathbf{A} & \mathbf{B} \\
-(\mathbf{B}^{T})^{-1} & \mathbf{0}
\end{array}\right)\left(\begin{array}{l}
\vec{\Phi} \\
\vec{\Pi}
\end{array}\right)+i \pi\left(\begin{array}{l}
\vec{t} \\
\vec{0}
\end{array}\right),\label{ABCDt}
\ee
such that all entries in $\mathbf{A}, \mathbf{B},\vec{t}$ are integers. $\vec{t}$ is a 15-dimensional vector. $\mathbf{A}, \mathbf{B}$ are $15\times 15$ blocks satisfying that $\mathbf{A} \mathbf{B}^T$ is a symmetric matrice. Matrices $\mathbf{A}, \mathbf{B}, \vec{t}$ are given explicitly in Appendix \ref{Symplectic transformation}
   
Following from \eqref{octsymp}, the Atiyah-Bott symplectic form $\O$ on $\calp_{\partial(S^3\setminus\G_5)}$ is expressed as
\be
\O&=&\sum_{I=1}^{15}\rmd \mathscr{P}_I\wedge \rmd \mathscr{Q}_I\nonumber\\
&=&2\sum_{a<b}\rmd \ct_{ab}\wedge \rmd L_{ab}+\sum_{a=1}^{5}\rmd \cy_a\wedge \rmd \cx_a.\label{Odarboux3fold}
\ee
The coordinates $\vec{\scrq},\vec{\scrp}$ are used below for constructing the CS partition function of $S^3\setminus \G_5$. We sometimes use the notations $\scrq_{ab}=2L_{ab},\ \scrq_a=\cx_a,\ \scrp_{ab}=\ct_{ab},\ \scrp_a=\cy_a$ in our following discussion.

It is remarkable that there is no additional constraint for gluing octahedra to form $S^3\setminus\G_5$, since gluing octahedra does not produce additional bulk edge. Therefore $\calp_{\partial(S^3\setminus\G_5)}\simeq\times_{a=1}^5\calp_{\partial {\rm oct}(a)}$. It is simply a symplectic transformation from the octahedra Darboux coordinates $\vec{\Phi},\vec{\Pi}$ to $\mathscr{P}_I, \mathscr{Q}_I$ of $\calp_{\partial(S^3\setminus\G_5)}$. The moduli space of framed flat connections on each octahedron is a Lagrangian submanifold $\cl_{\mathrm{oct}(a)}\subset\calp_{\partial {\rm oct}(a)}$. Then $\times_{a=1}^5\cl_{\mathrm{oct}(a)}\simeq \cl_{S^3\setminus \G_5}$ is a Lagrangian submanifold in $\times_{a=1}^5\calp_{\partial {\rm oct}(a)}\simeq \calp_{\partial(S^3\setminus\G_5)}$. Given any five framed flat connections on five octahedra respectively, they define a flat connection on $S^3\setminus\G_5$.

\subsection{$S^3\setminus \G_5$ partition function}\label{S3-G5 partition function}


The symplectic matrix in \eqref{ABCDt} can be decomposed into generators
\begin{widetext}
\be
\left(\begin{array}{ll}
\mathbf{A} & \mathbf{B} \\
-(\mathbf{B}^{T})^{-1} & \mathbf{0}
\end{array}\right)=\left(\begin{array}{cc}
\mathbf{0} & -\mathbf{I} \\
\mathbf{I} & \mathbf{0}
\end{array}\right)\left(\begin{array}{cc}
\mathbf{I} & \mathbf{0} \\
\mathbf{A} \mathbf{B}^{T} & \mathbf{I}
\end{array}\right)\left(\begin{array}{cc}
-\left(\mathbf{B}^{-1}\right)^{T} & \mathbf{0} \\
\mathbf{0} & -\mathbf{B}
\end{array}\right) .\label{DecompSympl}
\ee
\end{widetext}

We start with a product of 5 octahedron partition functions, each of which associates to an octahedron in the decompostion of $S^3\setminus \G_5$
\be
Z_\times(\vec{\mu}\mid \vec{m})&=&\prod_{a=1}^5 Z_{\rm oct}(\mu_{X_a},\mu_{Y_a},\mu_{Z_a}|m_{X_a},m_{Y_a},m_{Z_a})\nonumber\\
&\in& \cf_{\Fp(\mathrm{oct})^{\times 5}}^{(k)}.
\ee
The generators of the symplectic transformation is represented as Weil-like action on $Z_\times$ according to the order in \eqref{DecompSympl} \cite{levelk,Dimofte2011}.

\emph{1. U-type transformation:}
\be
U&=&\left(\begin{array}{cc}
-\left(\mathbf{B}^{-1}\right)^{T} & \mathbf{0} \\
\mathbf{0} & -\mathbf{B}
\end{array}\right),
\ee
\be
Z_1(\vec{\mu}\mid \vec{m})&=& (U\act Z_{\times})(\vec{\mu}\mid \vec{m})\nonumber\\
&=&\sqrt{\det( -\mathbf{B})}\,Z_{\times}\lt(-\mathbf{B}^T\vec{\mu}\mid-\mathbf{B}^T\vec{m}\rt),\label{Utrans}
\ee
where $\sqrt{\det( -B)}=4 i $. That all entries of $\mathbf{B}$ are integers guarantees that $Z_1$ is well-defined for $\vec{m}\in\mathbb{Z}/k\mathbb{Z}$. In addition, $Z_\times \in \cf^{(k)}_{\Fp(\mathrm{oct})^{\times 5}}$ indicates that the following function is Schwartz class when $(\vec{\a},\vec{\b})\in\Fp(\mathrm{oct})^{\times 5} $, 
\be
&&e^{-\frac{2\pi}{k}(-\mathbf{B}^T\vec{\mu})\cdot \vec{\b}}Z_{\times}\lt(-\mathbf{B}^T\vec{\mu}+i\vec{\a}\mid \vec{m}\rt)\nonumber\\
&=&e^{-\frac{2\pi}{k}\vec{\mu}\cdot (-\mathbf{B}\vec{\b})}Z_{\times}\lt(-\mathbf{B}^T(\vec{\mu}-i(\mathbf{B}^{-1})^T\vec{\a})\mid \vec{m}\rt).
\ee
where $\mu_i\in\R$. Therefore $Z_1$ belongs to $\cf_{\Fp_1}^{(k)}$ where $\Fp_1=U\circ \Fp(\mathrm{oct})^{\times 5} $ with $U$ acting on the angle space $\calp_{\rm angle}$ as symplectic transformation. \\

\emph{2. T-type transformation:}
\be
&& T=\left(\begin{array}{cc}
\mathbf{I} & \mathbf{0} \\
\mathbf{A} \mathbf{B}^{T} & \mathbf{I}
\end{array}\right),
\ee
\be
&&Z_2(\vec{\mu}\mid \vec{m})=(T\act Z_1)(\vec{\mu}\mid \vec{m})\label{Ttrans}\\
&=&(-1)^{\vec{m} \cdot \mathbf{A} \mathbf{B}^{T} \cdot \vec{m}} e^{\frac{i \pi}{k}(-\vec{\mu} \cdot \mathbf{A} \mathbf{B}^{T} \cdot \vec{\mu}+\vec{m} \cdot \mathbf{A} \mathbf{B}^{T} \cdot \vec{m})}Z_1(\vec{\mu}\mid \vec{m}).\nonumber
\ee
All entries of $\mathbf{A} \mathbf{B}^{T}$ are integers so that $Z_2$ is well-defined for $\vec{m}\in (\mathbb{Z}/k\mathbb{Z})^{15}$. $Z_1\in \cf_{\Fp_1}^{(k)}$ implies that the following function is Schwartz class when $(\vec{\a},\vec{\b})\in\Fp_1 $,
\be
&&e^{-\frac{2\pi}{k}\vec{\mu}\cdot \vec{\b}}Z_1(\vec{\mu}+i\vec{\a}\mid \vec{m})\nonumber\\
&=&\text{phase}\, \cdot\, e^{-\frac{2\pi}{k}\vec{\mu}\cdot( \vec{\b}+ \mathbf{A} \mathbf{B}^{T}\cdot \vec{\a})}Z_2(\vec{\mu}+i\vec{\a}\mid \vec{m}).
\ee
Therefore $Z_2$ belongs to $\cf_{\Fp_2}^{(k)}$ where $\Fp_2=T\circ \Fp_1 $.\\

\emph{3. S-type transformation:}
\be
S&=&\left(\begin{array}{cc}
\mathbf{0} & -\mathbf{I} \\
\mathbf{I} & \mathbf{0}
\end{array}\right),
\ee
\be
&&Z_3(\vec{\mu}\mid \vec{m})=(S\act Z_2)(\vec{\mu}\mid \vec{m})\label{Strans}\\
&=&\frac{1}{k^{15}} \sum_{\vec{n} \in (\mathbb{Z} / k \mathbb{Z})^{15}} \int_\cc \rmd^{15} \nu\, e^{\frac{2 \pi i}{k}\left(-\vec{\mu}\cdot\vec{\nu}+\vec{m}\cdot\vec{ n}\right)} Z_2(\vec{\nu}\mid \vec{n}).\nonumber
\ee
If we set $\a_i=\mathrm{Im}(\mu_i)$ and $\b_i=\mathrm{Im}(\nu_i)$ ($i=1,\cdots,15$),
\be
e^{\frac{2 \pi i}{k}\left(-\vec{\mu}\cdot\vec{\nu}\rt)} Z_2(\vec{\nu}\mid \vec{n})
&=& \lt[e^{\frac{2 \pi }{k}\vec{\a}\cdot\mathrm{Re}(\vec{\nu})}Z_2(\mathrm{Re}(\vec{\nu})+i\vec{\b}\mid \vec{n})\rt]\nonumber\\
&\times& e^{\frac{2\pi i}{k}\lt[-\mathrm{Re}(\vec{\mu})\cdot\mathrm{Re}(\vec{\nu})+\vec{\a}\cdot\vec{\b}\rt]+\frac{2\pi}{k}\mathrm{Re}(\vec{\mu})\cdot\vec{\b}}\nonumber
\ee
is a Schwartz function in $\mathrm{Re}(\vec{\nu})$, when $(\vec{\b},-\vec{\a})\in \Fp_2$ (the function in the square bracket is a Schwartz function, $e^{\frac{2\pi i}{k}\lt[-\mathrm{Re}(\vec{\mu})\cdot\mathrm{Re}(\vec{\nu})\rt]}$ is a phase), or equivalently
\be
(\vec{\a},\vec{\b})\in \Fp_3=S\circ \Fp_2=S\circ T\circ U\circ\Fp(\mathrm{oct})^{\times 5}.\label{STUFpoct}
\ee
Given any $(\vec{\a},\vec{\b})\in \Fp_3$, let $\mathrm{Im}(\mu_i)=\a_i$ and the integration contour $\cc$ defined such that $\mathrm{Im}(\nu_i)=\b_i$, then $Z_3(\vec{\mu}\mid \vec{m})$ converges absolutely, and $Z_3\in \cf_{\Fp_3}^{(k)}$. As far as the contour $\cc$ satisfies the condition $\mathrm{Im}(\nu_i)=\b_i$, $(\vec{\a},\vec{\b})\in \Fp_3$, $Z_3(\vec{\mu}\mid \vec{m})$ is independent of choices of $\cc$, i.e. choices of $\b_i$, due to the analyticity of $Z_2$ and the fast decay of the integrand at the infinity.\\



\emph{4. Affine shift} \footnote{The affine shifted classical coordinate $X+i\pi t$ $(t\in\mathbb{Z})$ has the quantum deformation $X+(i\pi+\frac{\hbar}{2}) t$ when entering the partition function \cite{Dimofte2011}. In terms of the exponential variables, the affine shift is given by $(-q^{\frac{1}{2}})^te^X=(-q^{\frac{1}{2}})^tx$. Here we define $q^{\frac{1}{2}}=e^{\frac{\hbar}{2}}$ where $\hbar=\frac{2\pi i}{k}(1+b^2)$. If we parametrize $e^X=\exp[\frac{2 \pi  i}{k} \left(-i b \mu -m\right)]$, the affine shift $X\to X+(i\pi+\frac{\hbar}{2}) t$ corresponds to $\mu \to \mu +\frac{1}{2} i \left(b+{b}^{-1}\right) t$, $m\to m$, and adding an overall $(-1)^t$ to $e^X$.}:
\be
\sig_{\vec{t}}:&& \left(\begin{array}{l}
\vec{X} \\
\vec{P}
\end{array}\right)\mapsto \left(\begin{array}{l}
\vec{X} \\
\vec{P}
\end{array}\right)+i \pi\left(\begin{array}{l}
\vec{t} \\
\vec{0}
\end{array}\right),
\ee
\be
\cz_{S^3\setminus \G_5}(\vec{\mu}\mid \vec{m})&=&(\sig_{\vec{t}}\act Z_3)(\vec{\mu}\mid \vec{m})\nonumber\\
&=&Z_3\lt(\vec{\mu}-\frac{iQ}{2}\vec{t}\mid \vec{m}\rt).\label{affinetrans}
\ee
We have $\cz_{S^3\setminus \G_5}\in \cf_{\Fp_{\rm new}}^{(k)}$, where 
\be
\Fp_{\rm new}&&\,=\sig'_{\vec{t}}\circ \Fp_3=\sig'_{\vec{t}}\circ S\circ T\circ U\circ\Fp(\mathrm{oct})^{\times 5},\\
\sig'_{\vec{t}}:&&\ \left(\begin{array}{l}
	\vec{\a} \\
	\vec{\b}
	\end{array}\right)\mapsto\left(\begin{array}{l}
		\vec{\a}' \\
		\vec{\b}'
		\end{array}\right):= \left(\begin{array}{l}
	\vec{\a}+\frac{Q}{2}\vec{t} \\
	\vec{\b}
	\end{array}\right).\nonumber
\ee

The resulting $\cz_{S^3\setminus \G_5}(\vec{\mu}\mid\vec{m})$ is the CS partition function on $S^3\setminus \G_5$.
$\Fp_{\rm new}$ is obviously non-empty since $\Fp(\mathrm{oct})$ is non-empty. Every $(\vec{\a},\vec{\b})\in \Fp_{\rm new}$ is a positive angle structure of $S^3\setminus\G_5$, and leads to the absolute convergence of $\cz_{S^3\setminus \G_5}(\vec{\mu}\mid\vec{m})$.

$\vec{\mu},\vec{m}$ relate to $\{{\scrq}_I, {\widetilde \scrq}_I\}_{I=1,\cdots,15}$ by
\be
&&\mu_{I}=\frac{k\left(\widetilde{\mathscr{Q}}_{I}^{\prime}+\mathscr{Q}_{I}^{\prime}\right)}{2 \pi\left(b+b^{-1}\right)}, \quad m_{I}=\frac{i k\left(\mathscr{Q}_{I}^{\prime}-b^{2} \widetilde{\mathscr{Q}}_{I}^{\prime}\right)}{2 \pi\left(b^{2}+1\right)},\label{muImI}\\
&&\scrq_I'=\scrq_I-i\pi t_I,\qquad \widetilde{\scrq}_I'=\widetilde{\scrq}_I-i\pi t_I
\ee
or in terms of exponentials
\be
(-1)^{t_I}e^{\scrq_I}&=&\exp \left[\frac{2 \pi i}{k}(-i b \mu_I-m_I)\right], \\
(-1)^{t_I}e^{\widetilde{\scrq}_I}&=&\exp \left[\frac{2 \pi i}{k}\left(-i b^{-1} \mu_I+m_I\right)\right].
\ee
Consider the shifts $\scrq_I\to \scrq_I+2\pi i p_I$, $\widetilde{\scrq}_I\to \widetilde{\scrq}_I-2\pi i \widetilde{p}_I$ ($p_I,\widetilde{p}_I\in\mathbb{Z}$) which leave $e^{\scrq_I},e^{\widetilde{\scrq}_I}$ invariant, Fixing $\mathrm{Im}(\mu_I)=\a_I$ implies $\widetilde{p}_I={p}_I$, then the shifts reduce to the gauge freedom $m_I\to m_I+kp_I$ in $\mathbb{Z}/k\mathbb{Z}$.

\subsection{Coherent states}\label{Coherent state}

Given the 4-holed sphere $\cs_a$, we transform the corresponding phase space coordinates from $\cx_a,\cy_a,\widetilde{\cx}_a,\widetilde{\cy}_a$ to $\mu_a,\nu_a,m_a,n_a$ by
\be
\cx_a-i\pi t_a&=&\frac{2\pi i}{k}(-i b\mu_{a}-m_{a}),\label{cxa}\\
\widetilde{\cx}_a-i\pi t_a&=&\frac{2\pi i}{k}(-i b^{-1}\mu_{a}+m_{a}),\\
\cy_a&=&\frac{2 \pi i}{k}(-i b \nu_{a}-n_{a}), \\
\widetilde{\cy}_a&=&\frac{2 \pi i}{k}\left(-i b^{-1} \nu_{a}+n_{a}\right),\label{cya}
\ee
where $\mu_{a}$ is the component in $\vec{\mu}\in\mathrm{strip}(\Fp_{\rm new})$. These coordinates parametrize $\PSlc$ flat connections on $\cs_a$ with fixed $e^{2L_{ab}}$ at holes. The moduli space of $\PSlc$ flat connections on $\cs_a$ is locally $\C^6$, but fixing $e^{2L_{ab}}$ reduces the space to locally $\C^2$. Let's fix $\mathrm{Im}(\mu_{a})=\alpha_{a}$ and focus on degrees of freedom $\mathrm{Re}(\mu_{a}),m_{a}$. In the following discussions of this section, we use $\mu_a\in\R$ to represent $\mathrm{Re}(\mu_a)$. We define the Hilbert space
\be
\ch_{\cs_a}=L^2(\R)\otimes_\C V_k,\quad V_k\simeq \C^k.
\ee
containing functions of $\mu_{a}\in\R,m_{a}\in \mathbb{Z}/k\mathbb{Z}$. 
Operators $\bm{\mu}_a,\bm{\nu}_a,\mathbf{m}_a,\mathbf{n}_a$ on $\ch_{\cs_a}$ are defined in the same way as in \eqref{mumnunoperator}. We suppress the $a$ index in following discussions of this section.

We firstly focus on $L^2(\R)$ and define the ``annihilation operator'' and coherent state $\psi_z(\mu)$ labelled by $z\in\C$. $\psi_z(\mu)$ satisfies
\be
\frac{1}{\sqrt{2}}\lt(\sqrt{\frac{2\pi}{k}}\bm{\mu}+i\sqrt{\frac{2\pi}{k}}\bm{\nu}\rt)\psi^0_z(\mu)
=\sqrt{\frac{k}{2\pi}}z\psi^0_z(\mu).\nonumber
\ee
The solution is
\be
\psi^0_z(\mu)
=\lt(\frac{2}{k}\rt)^{1/4}e^{-\frac{\pi}{k}\left(\mu-\frac{k}{\pi \sqrt{2}} \operatorname{Re}\left(z\right)\right)^{2}} e^{i \sqrt{2}  \mu \operatorname{Im}\left(z\right)},\label{coheretntpsiz0}
\ee
where $\psi^0_z(\mu)$ is normalized by the standard $L^2$-norm. 
The coherent state label $z$ relates to the classical phase space coordinates $\mu_0,\nu_0$ be
\be
z=\frac{1}{\sqrt{2}}\frac{2\pi}{k}(\mu_0 +i\nu_0).\label{zmunu}
\ee
We can multiply to $\psi^0_z$ a prefactor that relates to the polytope $\Fp_{\rm new}$, namely, for each $\cs_a$ we define
\be
\psi_{z_a}(\mu_a)=e^{-\sqrt{2}\b_a\mathrm{Re}(z_a)}\psi^0_{z_a}(\mu_a),\label{coheretntpsiz}
\ee
where $\b_a$ is the component in $(\vec{\a},\vec{\b})\in \Fp_{\rm new}$. The prefactor does not affect the semiclassical behavior of $\psi_z$, but relates to the finiteness of the amplitude. Note that $\{\b_a\}_{a=1}^5$ cannot be all zero, because e.g. $\b_1=\a_{Z_2}+\a_{Z_3}>0$ by \eqref{octinequality}. It is still a viable choice to work with the normalized coherent state $\psi^0_{z_a}$, then certain requirements should be implemented to the spinfoam edge amplitude, we come back to this point in Section \ref{Ambiguities}.

We denote the coherent state in $V_k$ by $\xi_{(x,y)}(m)$ where $(x,y)\in[0,2\pi)\times[0,2\pi)$ and $m\in\mathbb{Z}/k\mathbb{Z}$ \cite{gazeau2009coherent},
\be 
\xi_{(x,y)}(m)&=&\lt({\frac{2}{k}}\rt)^{\frac{1}{4}} e^{-\frac{i k x y}{4 \pi }} \label{xixym}\\
&\times&\sum_{n\in\mathbb{Z}}e^{-\frac{k }{4 \pi }\left(\frac{2 \pi  m}{k}-2 \pi  n-x\right)^2} e^{-\frac{i k  }{2 \pi }y\left(\frac{2 \pi  m}{k}-2 \pi  n-x\right)}.\nonumber
\ee
$(x,y)$ relates to the classical phase space coordinates $m_0,n_0$ by
\be
y=\frac{2\pi}{k}n_0,\quad x=\frac{2\pi}{k}m_0,\quad \text{mod}\ 2\pi\label{xymn}
\ee
$\xi_{(x,y)}(m)$ satisfy the over-completeness relation in $V_k$
\be
\frac{k}{4\pi^2}\int_{\mathbb{T}^2}\rmd x\rmd y\,\xi_{(x,y)}(m)\bar{\xi}_{(x,y)}(m')=\delta_{m,m'}.
\ee

We define coherent states in $\ch_{\cs_a}$ by tensor products
\be
\psi_{z_a}\otimes\xi_{(p_a,q_a)} \in \ch_{\cs_a}
\ee
$z_a,\bar{z}_a,x_a,y_a$ coordinatize the part of the phase space associate to $\cs_a$, they form a coordinate system on the moduli space of $\PSlc$ flat connections on $\cs_a$ (with fixed $e^{2L_{ab}}$). We have the following relation 
\be
\bar{\psi}_{z_a}\otimes\bar{\xi}_{(x_a,y_a)}=\psi_{\bar{z}_a}\otimes\xi_{(x_a,-y_a)}.\label{complexcc000}
\ee


We product the coherent states over five $\cs_a$, 
\be
\Psi_\rho\lt(\{{\mu}_a\}\mid\{{m}_a\}\rt)
&=&
\prod_{a=1}^5\psi_{z_a}(\mu_a)\xi_{(x_a,y_a)}(m_a)\in \otimes_a\ch_{\cs_a}\nonumber\\
\rho&=&\{z_a,x_a,y_a\}_{a=1}^5.
\ee
where $\mu_a\in\R$. The partition function $\cz_{S^3\setminus\G_5}$ is a function of $\vec{\mu},\vec{m}$ including $\mu_a,m_a$. We consider the (partial) $L^2$ inner product between $\cz_{S^3\setminus\G_5}$ and $\bar{\Psi}_\rho$ (this may be understood as $\cz_{S^3\setminus\G_5}$ acting on $\bar{\Psi}_\rho$ since $\cz_{S^3\setminus\G_5}$ is a tempered distribution),
\begin{widetext}
\be
\cz_{S^3\setminus\G_5}(\iota)=\langle \bar{\Psi}_\rho\mid \cz_{S^3\setminus\G_5}\rangle_{\otimes_a\ch_{\cs_a}}=\sum_{\{{m}_a\}\in(\mathbb{Z}/k\mathbb{Z})^{5}}\int_{\R^{5}}\prod_{a=1}^5\rmd \mu_a\ \cz_{S^3\setminus\G_5}(\vec{\mu}+i\vec{\a}\mid \vec{m})\,\Psi_\rho\lt(\{{\mu}_a\}\mid\{\vec{m}_a\}\rt),\label{Zsigma1}
\ee
where $\vec{\mu}+i\vec{\a}\in\mathrm{strip}(\Fp_{\rm new})$. 
$\cz_{S^3\setminus\G_5}(\iota)$ is a function of
\be
\iota&=&\lt(\{\mu_{ab}+i \a_{ab},m_{ab}\}_{a<b},\{z_a,x_a,y_a\}_{a=1}^5,\{\a_a,\b_a\}_{a=1}^5\rt),\quad \mu_{ab}\in\R,\ m_{ab}\in\mathbb{Z}/k\mathbb{Z},\ z_a\in\C,\ (x_a,y_a)\in\mathbb{T}^2\label{sigmalabel}
\ee
which includes the position variables of annuli and both the position and momentum variables of 4-holed spheres. $\iota$ determines a unique $\PSlc$ flat connection on each $\cs_a$: Given an $\iota$ and by \eqref{zmunu} and \eqref{xymn}, $z_a,x_a,y_a$ determine phase space coordinates that relate to FG coordinates by \eqref{cxa} - \eqref{cya}). The resulting FG coordinates and $e^{2L_{ab}}$ given by $\mu_{ab},m_{ab}$ of the same $\iota$ determine a unique $\PSlc$ flat connections on $\cs_a$.

\begin{theorem}\label{finitevertex}
Fixing the annulus data $\{\mu_{ab},m_{ab}\}_{a<b}$, $|\cz_{S^3\setminus\G_5}(\iota)|$ is bounded for all $\{z_a,x_a,y_a\}_{a=1}^5$.
\end{theorem}

\textbf{Proof:} 
In $\cz_{S^3\setminus\G_5}(\iota)$, the sum over $\vec{m}'$ is finite, and for any $m$,
\be
\xi_{(x,y)}(m)=\frac{\sqrt[4]{2} e^{-\frac{k y (y+i x)}{4 \pi }} \vartheta _3\left(\frac{1}{2} \left(-\frac{2 \pi  m}{k}+x-i y\right),e^{-\frac{\pi }{k}}\right)}{k^{3/4}}\nonumber
\ee
is smooth in $(x,y)\in[0,2\pi)\times[0,2\pi)\simeq \mathbb{T}^2$, thus $|\xi_{(x,y)}(m)|$ is bounded on $\mathbb{T}^2$ for any $m$. Therefore the boundedness of $\cz_{S^3\setminus\G_5}(\iota)$ is implied by the boundedness of the following integral for all $\vec{m}$
\be
&&\lt|\int_{\R^{5}}\prod_{a=1}^5\rmd \mu_a'\ \cz_{S^3\setminus\G_5}(\vec{\mu}'+i\vec{\a}\mid \vec{m}')\prod_{a=1}^5\psi_{z_a}(\vec{\mu}'_a)\rt|\nonumber\\
&=&\lt|e^{-\sqrt{2}\sum_a\b_a\mathrm{Re}(z_a)}\int_{\R^{5}}\prod_{a=1}^5\rmd \mu_a'\ \cz_{S^3\setminus\G_5}(\vec{\mu}'+i\vec{\a}\mid \vec{m}')\prod_{a=1}^5\psi^0_{z_a}(\vec{\mu}'_a)\rt|\nonumber\\
&\leq&\lt(\frac{1}{k}\rt)^{\frac{5}{4}}e^{-\sqrt{2}\sum_a\b_a\mathrm{Re}(z_a)}\int_{\R^{5}}\prod_{a}\rmd \mu'_a\lt| \cz_{S^3\setminus\G_5}(\vec{\mu}'+i\vec{\a}\mid \vec{m}')\,e^{-\frac{2\pi}{k}\sum_a\b_a\mu_a'}\rt|\prod_{a=1}^5\lt|\psi_{z_a}^0(\vec{\mu}'_a)e^{\frac{2\pi}{k}\b_a\mu_a'}\rt|\nonumber\\
&\leq& C\lt(\frac{1}{k}\rt)^{\frac{5}{4}}e^{-\sqrt{2}\sum_a\b_a\mathrm{Re}(z_a)}\int_{\R^{5}}\prod_{a}\rmd \mu_a'\,e^{-\frac{\pi}{k}\sum_{a}\lt(\mu'_a-\frac{k}{\pi\sqrt{2}}\mathrm{Re}(z_a)\rt)^2}e^{\frac{2\pi}{k}\sum_a\b_a\mu_a'}\nonumber\\
&=&Ck^{\frac{5}{4}}\,e^{\sum_a\frac{\pi  \beta_a^2 }{k}}\label{upperboundcz}
\ee
\end{widetext}
In the third step we use $\mathcal{Z}_{S^{3} \backslash \Gamma_{5}} \in \mathcal{F}_{\mathfrak{P}_{\text {new }}}^{(k)}$, thus as a function of $\mu'_a$ ($a=1,\cdots,5$), $\forall\, (\vec{\a},\vec{\b})\in \Fp_{\rm new}$
\be
e^{-\frac{2\pi}{k} \sum_a\mu'_a\b_a}\cz_{S^3\setminus\G_5}(\vec{\mu}'+i\vec{\a}\mid \vec{m})\in \cs(\R^{5}),
\ee
$C$ is the upper bound of $|e^{-\frac{2\pi}{k} \sum_a\mu'_a\b_a}\cz_{S^3\setminus\G_5}(\vec{\mu}'+i\vec{\a}\mid \vec{m})|$.

$\Box$

\section{Spinfoam amplitude with cosmological constant}\label{Spinfoam amplitude with cosmological constant}

The purpose of this section is to impose the simplicity constraint to $\cz_{S^3\setminus\G_5}(\iota)$ in order to relate the CS partition function to the spinfoam vertex amplitude in 4d. The simplicity constraint turns out to reduce the $\PSlc$ flat connection to $\PSu$ on five $\cs_a$'s. Based on the resulting vertex amplitude, we define the spinfoam amplitude with $\L$ on any simplicial complex and prove its finiteness, as well as discuss several related perspectives.

\subsection{Simplicity constraint and vertex amplitude}

In the simplical context with $\L=0$, the simplicity constraint (in the EPRL/FK model) imposes that for any spacelike tetrahedron $e$, there exists a timelike unit vector $N^I$ in 4d Minkowski space such that $B^{IJ}_fN_J=0$ where $B^{IJ}_f$ ($f=1,\cdots,4$) are bivectors associated to 4 faces $f$. The simplicity constraint and closure condition endow every $e$ a convex geometrical tetrahedron in flat space. Indeed, $B_f^{IJ}$ satisfying the constraint are equivalent to 3d vectors $\fa_f{\fn}^I_f=\frac{1}{2}\epsilon^{IJKL}N_J B_{KL}$ ($\fn^I\fn_I=1$) in the plane normal to $N^I$. Then the BF closure condition $\sum_{f=1}^4B^{IJ}=0$ implies $\sum_{f=1}^4\fa_f{\fn}^I_f=0$, which endows $e$ a convex geometrical tetrahedron (whose face areas and normals are $\fa_f$ and ${\fn}_f^I$) by Minkowski's theorem \cite{Minkowski}. At the quantum level, the first-class part of the simplicity constraint, i.e. the diagonal simplicity constraint $\epsilon_{IJKL}B_f^{IJ}B_f^{KL}=0$ are imposed strongly to the states, whereas the second-class part of the simplicity constraint are imposed weakly \cite{EPRL,FK,generalize}.

In presence of nonvanshing $\L$, $\G_5\subset S^3$ is the dual graph of the tiangulation of $S^3$ given by the 4-simplex's boundary. Each node of $\G_5$, or each $\cs_a\subset\partial(S^3\setminus\G_5)$, is dual to a boundary tetrahdron $e_a$ of the 4-simplex. Each link of $\G_5$, or each annulus $\ell_{ab}\subset\partial(S^3\setminus\G_5)$, is dual to a boundary triangle $f_{ab}$ of the 4-simplex. All tetrahedra and triangles are spacelike similar to the EPRL/FK model. Given any $e_a$, the generalization of closure condition is the defining equation of $\PSlc$ flat connections on the 4-holed sphere $\cs_a$: $O_4O_3O_2O_1=1$ where $O_{f=1,\cdots,4}\in\PSlc$ are holonomies around 4 holes based at a common point $\fp_a\in\cs_a$. By non-abelian stokes theorem, we identify $O_f=e^{|\L| B_f/3}\in \mathrm{SO}(1,3)^+$ due to the relation $\cf(\ca)=|\L| B/3$ from integrating out $B$ in \eqref{SHLBF}. Here $\cf(\ca)$, as the curvature of CS connection $\ca$ on $S^3$, is proportional to the delta function supported on $\G_5$ (equivalent to that $\ca$ is flat on $S^3\setminus\G_5$). Namely $\cf(\ca)=\frac{|\L|}{3}B_f\delta^2(x)\rmd x^1\wedge \rmd x^2 $ on the face $f$ coordinated by $(x^1,x^2)$ tranverse to an edge of $\G_5$ at $\vec{x}=0$. $O_4O_3O_2O_1=1$ with $O_f=e^{|\L| B_f/3}$ reduces to the linear closure condition $\sum_{f=1}^4B_f=0$ as $\L\to 0$. Moreover, the simplicity constraint $B^{IJ}_fN_J=0$ for all $f=1,\cdots,4$ restrict $O_{f=1,\cdots,4}$ to a common PSU(2) subgroup sablizing the timelike vector $N^I$. The result in \cite{curvedMink} shows that restricting all $O_f$ to the subgroup $\PSu$ endows $e$ a convex geometrical tetrahedron with constant curvature. The effect of restricting $O_f$ to $\PSu$ is analogous to the simplicity constraint reviewed above. This motivates us to define this restriction to be the simplicity constraint in presence of nonvanishing $\L$ \cite{Han:2017geu}:

\begin{definition}\label{simplicity}

Semiclassically in presence of nonvanishing cosmological constant, the simplicity constraint restricts the moduli spaces of $\PSlc$ flat connections on 4-holed spheres to the ones that can be gauge-transformed to $\PSu\simeq \mathrm{SO}(3)$ flat connections. 

\end{definition}

\subsubsection{First-class constraints:}

Our proposal is to quantize and impose the simplicity constraint to $\cz_{S^3\setminus\G_5}(\iota)$. Firstly, flat connections on all $\cs_{a}$ are $\PSu$ implies $e^{2L_{ab}}\in \mathrm{U(1)}$, or equivalently $\mu_{ab}=0$ for all annulus $\ell_{ab}$. However due to the presence of $\a_{ab}=\mathrm{Im}(\mu_{ab})$, at the quantum level we may have to decide whether we impose
\be
\Re(\bm{\mu}_{ab})\cz_{S^3\setminus\G_5}(\iota)=0,\quad\text{or}\quad\bm{\mu}_{ab}\cz_{S^3\setminus\G_5}(\iota)=0.  \label{choicemu}
\ee
In either case, these 10 constraint are first-class since $\{\bm{\mu}_{ab}\}_{a<b}$ are commutative, thus they can be imposed strongly to $\cz_{S^3\setminus\G_5}(\iota)$. $\{\bm{\mu}_{ab}\}_{a<b}$ are mulitplication operators acting on $\cz_{S^3\setminus\G_5}(\iota)$. The former choice restricts 
\be
\mathrm{Re}(\mu_{ab})=0,\quad \forall\ \ell_{ab}\label{Remu=0}
\ee
in $\iota$. The latter choice restricts both $\mathrm{Re}(\mu_{ab})$ and the positive angle structure
\be
\mathrm{Re}(\mu_{ab})=0,\quad \text{and}\quad \a_{ab}=0,\quad \forall\ \ell_{ab}.\label{strongerch}
\ee
thus is much stronger than the former choice. However the semiclassical limit of the theory is insensitive to the choices: Consider the former (weaker) choice, $e^{2L_{ab}}$ determined by $\iota$ is given by
\be
e^{2L_{ab}}&=&(-1)^{t_{ab}}\exp \left[\frac{2 \pi i}{k}\left( b \a_{ab}-m_{ab}\right)\right]\nonumber\\
&=&\exp \left[\frac{2 \pi i}{k}\left( b \a_{ab}-\lt(m_{ab}+t_{ab}\frac{k}{2}\rt)\right)\right]\nonumber\\
&=&\exp \left[\frac{2 \pi i}{k}\left( b \a_{ab}+\lt(2j_{ab}+\frac{\epsilon_{ab}}{2}\rt)\right)\right]\label{e2Labjab}
\ee
where $\a_{ab}=\mathrm{Im}(\mu_{ab})$. In the last step, since $-(m_{ab}+t_{ab}\frac{k}{2})\in \mathbb{Z}/k\mathbb{Z}$ (or $\mathbb{Z}/k\mathbb{Z}+1/2$) if $k$ is even (or odd), we have introduce the half-integer ``spin'' $j_{ab}$ such that $-\lt(m_{ab}+t_{ab}\frac{k}{2}\rt)=2j_{ab}+\frac{\epsilon_{ab}}{2}$ mod $k\mathbb{Z}$ where
\be
 \epsilon_{ab}&=&\begin{cases}
	\frac{1-(-1)^{t_{ab}}}{2} & k\ \text{odd}\\
	0 & k\ \text{even}
\end{cases}\\
j_{ab}&=& 0,\frac{1}{2},\cdots, \frac{k-1}{2}\label{jrange}.
\ee
The double-scaling limit $j_{ab},k\to \infty$ with $j_{ab}/k$ fixed is the semiclassical limit for the spinfoam amplitude with cosmological constant (see Section \ref{Semiclassical analysis} for discussion). In this limit, $e^{2L_{ab}}$ is insensitive to $\a_{ab},\epsilon_{ab}$ since they do not scale with $k$
\be
e^{2L_{ab}}\to\exp \left[\frac{4 \pi i}{k}j_{ab}\right]\in \mathrm{U(1)}.\label{conjclass}
\ee
Both choices in \eqref{choicemu} lead to the same semiclassical result. At least semiclassically, each holonomy around holes on $\cs_a$ can be individually conjugated to PSU(2), while $j_{ab}/k$ determines the conjugacy class of the holonomy.

The stronger choice \eqref{strongerch} is indeed viable. We can have $(\vec{\a},\vec{\b})\in \Fp_{\rm new}$ with ten $\a_{ab}=0$, because for instance all ten $\a_{ab}=0$ can be given by $\a_{X_a}=\a_{Y_a}=\a_{Z_a}=Q/4$ and $\b_{X_a}=\b_{Y_a}=\b_{Z_a}=0$ ($a=1,\cdots,5$), which satisfy \eqref{octinequality}. The simplicity constraint results in restrictively $e^{2L_{ab}}\in \mathrm{U(1)}$ when $\a_{ab}=0$, whereas $e^{2L_{ab}}\not\in \mathrm{U(1)}$ for other $\a_{ab}\neq 0$. $\a_{ab}=0$ is a preferred choice because $e^{2L_{ab}}\in\mathrm{U(1)}$ implies that after imposing the simplicity constraint, the area from the 4d bivector $B_f$ coincides with the face area of 3d tetrahedron at the quantum level: Recall the discussion above Definition \ref{simplicity}. We diagonalize an $O_f\in\PSlc$ by a gauge transformation 
\be
O_f&=&\pm\mathrm{diag}(e^{L_{ab}},e^{-L_{ab}})=\pm e^{\mathrm{Re}(L_{ab})\bm{\sig}^3+i\mathrm{Im}(L_{ab})\bm{\sig}^3}\nonumber\\
&\leftrightarrow& e^{2\mathrm{Re}(L_{ab})\mathbf{K}^3-2\mathrm{Im}(L_{ab})\mathbf{L}^3}=e^{\frac{|\L|}{3}B_f}\in \mathrm{SO(1,3)}^+\nonumber
\ee 
where $\mathrm{Im}(L_{ab})\in[0,\pi)$ and $\mathbf{K}^3,\mathbf{L}^3$ are $ so(1,3)$ generators. We obtain $\frac{|\L|}{3}B_f=2\mathrm{Re}(L_{ab})\mathbf{K}^3-2\mathrm{Im}(L_{ab})\mathbf{L}^3$ for the preferred lift of $B_f$. Then $L_{ab}$ relates to the area from the 4d bivector, $|B_f|=|\frac{1}{2}\mathrm{Tr}(B_f^{2})|^{1/2}$, by $\frac{|\L|}{3}|B_f|=2|\mathrm{Re}(L_{ab})^2-\mathrm{Im}(L_{ab})^2|^{1/2}$. Restricting $\a_{ab}=0$ and the simplicity constraint $\mathrm{Re}(\mu_{ab})=0$ result in that 
\be
\!\!\!\!\!\!\!\!\! \frac{|\L|}{3}|B_f|=2\mathrm{Im}(L_{ab})=\frac{4\pi }{k}(j_{ab}+\epsilon_{ab}/4)\equiv\frac{|\L|}{3}\fa_{ab},\label{Bfaab}
\ee 
where $\fa_{ab}$ is the face area of 3d tetrahedron (this is implied by the generalized closure condition, see \cite{curvedMink} or the discussion below). Both $\cz_{S^3\setminus\G_5}$, $\Psi_\iota$ are functions of $L_{ab}$, thus both the 4d and 3d area operators, $\frac{|\L|}{3}|B_f|=2|\mathrm{Re}(L_{ab})^2-\mathrm{Im}(L_{ab})^2|^{1/2}$ and $\frac{|\L|}{3}\fa_{ab}=2\mathrm{Im}(L_{ab})$, act as multiplications. The above shows that these two operators coincide when $\a_{ab}=0$. A similar consisency constraint ``$\text{4d area} = \text{3d area}$'' has also been imposed to the EPRL model \cite{generalize}.

However to keep discussions general, we still use the weaker version \eqref{Remu=0} and keep $\a_{ab}$ general in the following discussion. But we prefer $\a_{ab}= 0$ by the above argument.


\subsubsection{Second-class constraints}

The first-class part of the simplicity constraint and $j_{ab}$ fix $e^{2L_{ab}}$ on 10 annuli. Classically, fixing $e^{2L_{ab}}$ reduces the moduli space of $\PSlc$ flat connections on $\cs_a$ to 2 complex dimensions whose Darboux coordinates $\vartheta,\varphi\in \C$ are studied in \cite{Nekrasov:2011bc}, with $\{\vartheta,\varphi\}=1$ (they are the complexification of $\theta,\phi$ in Section \ref{Integration over PSU(2) flat conn}). Constraining flat connections to $\PSu$ restricts $\im(\vartheta)=\im(\varphi)=0$. The restriction gives second-class constraints due to the noncommutativity of $\vartheta,\varphi$. By the lessons from the EPRL/FK model, the constraints has to be imposed weakly at the quantum level. Our strategy is to impose the constraints to the label $(z_a,x_a,y_a)$ where the coherent state $\Psi_\rho$ is peaked. $(z_a,x_a,y_a)$ is a point in the moduli space of $\PSlc$ flat connections on $\cs_a$ with fixed $e^{2L_{ab}}$'s. We restrict $(z_a,x_a,y_a)$ to the subspace of flat connections that can be gauge transformed to PSU(2). 

Classically, our simplicity constraint is an analog of the linear simplicity constraint in the EPRL/FK model, as discussed at the beginning of this subsection. At the quantum level, although all spinfoam models impose the second-class simplicity constraint weakly, here the constraint is imposed to the coherent state labels, similar to the FK model \cite{FK}, but different from the EPRL model where the constraint is imposed by a master constraint operator.

Although the following discussion does not assume large $j_{ab}$, in the following discussion before Eq.\eqref{xiframing}, we ignore $\a_{ab}$ so that $e^{2L_{ab}}\in \mathrm{U(1)}$ is assumed, since only the semiclassical simplicity constraint are concerned here. After Eq.\eqref{xiframing} we take into account generally $\a_{ab}\neq 0$ and $e^{2L_{ab}}\not\in \mathrm{U(1)}$ at the quantum level. 

On the 4-holed sphere $\cs_a$, flat connections that can be gauge-transformed to PSU(2) are described by four $\PSlc$ 
holonomies $O_1,O_2,O_3,O_4$ that can be simultaneously conjugated to $\PSu$. $O_1,O_2,O_3,O_4$ are based at a common point $\fp$, and each of them travels around a hole of $\cs_a$. As holonomies of flat connection, they satisfy the generalized closure condition
\be
O_4O_3O_2O_1=1.\label{HHHH}
\ee
This equation is invariant under $\PSlc$ gauge transformation. We apply the gauge transformation to make all $O_i\in\PSu$ and treat \eqref{HHHH} as an equation of $\PSu$ holonomies. The conjugacy class of each $O_i$ has been fixed by \eqref{conjclass}, which specifies the squared eigenvalue of $O_i$. There exists a lift from $O_i$ to $H_i\in\Su$ such that 
\be
&&H_i= M(\xi_i)\left(\begin{array}{cc}
	  \pm e^{\frac{2\pi i}{k}j_i} & 0 \\
	0 & \pm e^{-\frac{2\pi i}{k}j_i}
	\end{array}\right)M(\xi_i)^{-1},\label{Hi}\\
&&M(\xi)=\left(\begin{array}{cc}
		\xi^1 & -\bar{\xi}^2 \\
	  \xi^2 & \bar{\xi}^1
	  \end{array}\right),
\ee
satisfying 
\be
H_4H_3H_2H_1=1\label{liftHHHH}
\ee
In each $H_i$, we neglect $\epsilon_{ab}$ when discussing the parametrization of PSU(2) flat connections
\be
j_i=j_{ab},\nonumber
\ee 
for $\ell_{ab}$ ends at the hole labelled by $i$, and similarly for $t_i$. $\xi_i=(\xi_i^1,\xi_i^2)^T$ is defined up to a complex scaling by the above formula of $H_i$. If we fix that $\det(M(\xi_i))=1$, 
\be
&&\vec{n}_i=\xi_i^\dagger\vec{\bm{\sig}}\xi_i,\quad i=1,\cdots,4,\\
&&\text{where}\ \vec{\bm{\sig}}=(\bm{\sig}^1,\bm{\sig}^2,\bm{\sig}^3)\ \text{are Pauli matrices}\nonumber
\ee
give 4 unit 3-vector in $\R^3$. The geometrical interpretation of \eqref{HHHH} relates the holonomies to a geometrical 3d tetrhedron with constant curvature (see \cite{curvedMink,HHKR} or Theorem \ref{tetrareconstr}), in which $\frac{4\pi}{k}j_i=\frac{|\L|}{3}\fa_i$ is the face area and $\vec{n}_i$ are face normals parallel transported to a common vertex of the tetrahedron\footnote{$\frac{4\pi}{k}j_i=\frac{|\L|}{3}\fa_i$ mismatches \eqref{Bfaab} if $\epsilon_{ab}\neq 0$, but it is not a problem since here we discuss coherent state labels, whereas \eqref{Bfaab} is about operator eigenvalues.}. $\{\vec{n}_i\}_{i=1}^4$ relates to the outward pointing normals $\{\fn_i\}_{i=1}^4$ of the tetrahedron by $\fn_i=\sgn(\L)\vec{n}_i$. Eq.\eqref{liftHHHH} with $H_i=e^{\L \vec{v}_i\cdot \vec{\bm{\sig}}}$ reduces to the flat closure condition $\sum_i\vec{v}_i=0$ for small $\L$. 

To clarify our convention, consider $\ell_{ab}$ connecting the $i$th hole of $\cs_a$ to the $j$th hole of $\cs_b$. We choose the framing flag $s_{\ell_{ab}}$ of $\ell_{ab}$ such that on $\cs_a$, the eigenvector of the holonomy $O_i\equiv O_{ab}$, $\xi_i\equiv \xi_{ab}$, coincides with $s_{\ell_{ab}}$ parallel transported to the common base point $\fp_a\in\cs_a$ of $\{O_i\}_{i=1}^4$. If our convention is \eqref{HHHH} on both $\cs_a$ and $\cs_b$, the parallel transport of $O_i\equiv O_{ab}$ of $\cs_a$ gives $O_j^{-1}\equiv O_{ba}$ of $\cs_b$, i.e. $G_{ab}^{-1}O_{ab}G_{ab}=O_{ba}$ with a holonomy $G_{ab}$ along $\ell_{ab}$. $s_{\ell_{ab}}$ evaluated at a point $\fp_b\in\cs_b$ gives $\xi_{ba}$ as the eigenvector of $O_{ba}$ with upper eigenvalue $\pm e^{2\pi i j_i/k}$. But $\xi_{ba}$ does not equal to $\xi_j=(\xi_j^1,\xi_j^2)^T$ on $\cs_b$ but equals to $(-\bar{\xi}_j^2,\bar{\xi}_j^1)^T$ in the convention of \eqref{Hi} \footnote{The inverse of $H_i$ in \eqref{Hi} can be written as $H_i^{-1}= \pm M'(\xi_i)\mathrm{diag}(e^{\frac{2\pi i}{k}j_i}, e^{-\frac{2\pi i}{k}j_i})M'(\xi_i)^{-1}$ where $M'(\xi)$ is given by \eqref{Mxiprim}.}.

In case that the minus sign present in \eqref{Hi}, we write $-e^{\frac{2\pi i}{k}j}=e^{-\frac{2\pi i}{k}j'}$ where $j'=k/2-j$, then Eq.\eqref{Hi} can be rewritten as
\be
&&H_i= M'(\xi_i)\left(\begin{array}{cc}
	 e^{\frac{2\pi i}{k}j_i'} & 0 \\
  0 & e^{-\frac{2\pi i}{k}j_i'}
  \end{array}\right)M'(\xi_i)^{-1},\label{Hiprim}\\
&&M'(\xi)=\left(\begin{array}{cc}
	-\bar{\xi}^2 &-\xi^1  \\
	\bar{\xi}^1 & -\xi^2 
	\end{array}\right)\label{Mxiprim}
\ee
In case of plus sign in \eqref{Hi}, we set $j'=j$. Flipping $+\to -$ in \eqref{Hi} correspond to $j\to k/2-j$ and $M(\xi)\to M'(\xi)$.

\begin{lemma}\label{trianineq}

The lifts $H_{i=1,\cdots,4}\in\Su$ satisfy $H_4H_3H_2H_1=1$ exist if and only if $j'_{i=1,\cdots,4}$ satisfy the triangle inequality, i.e. there exists $J$ such that
\be
&&|j'_1-j'_2|\leq J\leq \mathrm{min}\lt(j'_1+j'_2,k-j'_1-j'_2\rt),\label{trigineq1}\\
&&|j'_3-j'_4|\leq J\leq \mathrm{min}\lt(j'_3+j'_4, k-j'_3-j'_4\rt).\label{trigineq2}
\ee

\end{lemma}

The proof of this Lemma is given in Appendix \ref{Proof of Lemma}. \eqref{trigineq1} and \eqref{trigineq2} agree with the
spin-coupling rule of $\Su_\fq$ with $\fq=e^{\pi i/(k+2)}$.

\begin{lemma}\label{Oeqntrig}

$O_4O_3O_2O_1=1$ has solution $O_i\in\PSu$ if $j_{i}$ given by \eqref{conjclass} equals either $j'_i$ or $k/2-j'_i$ where $\{j'_i\}$ satisfy the triangle inequality \eqref{trigineq1} and \eqref{trigineq2}. 

\end{lemma}

\textbf{Proof:} Given a solution $H_i\in\Su$ to $H_4H_3H_2H_1=1$, Both $\pm H_i$ projects to $O_i\in \PSu$ solving $O_4O_3O_2O_1=1$. If $H_i$ is given by \eqref{Hiprim} with $j'=k/2-j$, 
\be
-H_i= M(\xi_i)\left(\begin{array}{cc}
	e^{\frac{2\pi i}{k}(k/2-j_i')} & 0 \\
 0 & e^{-\frac{2\pi i}{k}(k/2-j_i')}
 \end{array}\right)M(\xi_i)^{-1}.\nonumber
\ee
Since both $\pm H_i$ are allowed for the $\PSu$ equation, $j_{i}$ is given by the squared eigenvalue \eqref{conjclass} of either $H_i$ or $-H_i$, thus can be either $j'_i$ or $k/2-j'_i$. 

$\Box$

We restrict $j_{ab}$ to satisfy the condition in Lemma \ref{Oeqntrig} so that $O_4O_3O_2O_1=1$ has solution at every $\cs_a$. The triangle inequality in Lemma \ref{trianineq} is the analog of the triangle inequality for SU(2) intertwiners in spinfoam models without cosmological constant.

\begin{figure}[h]
	\begin{center}
	\includegraphics[width=6cm]{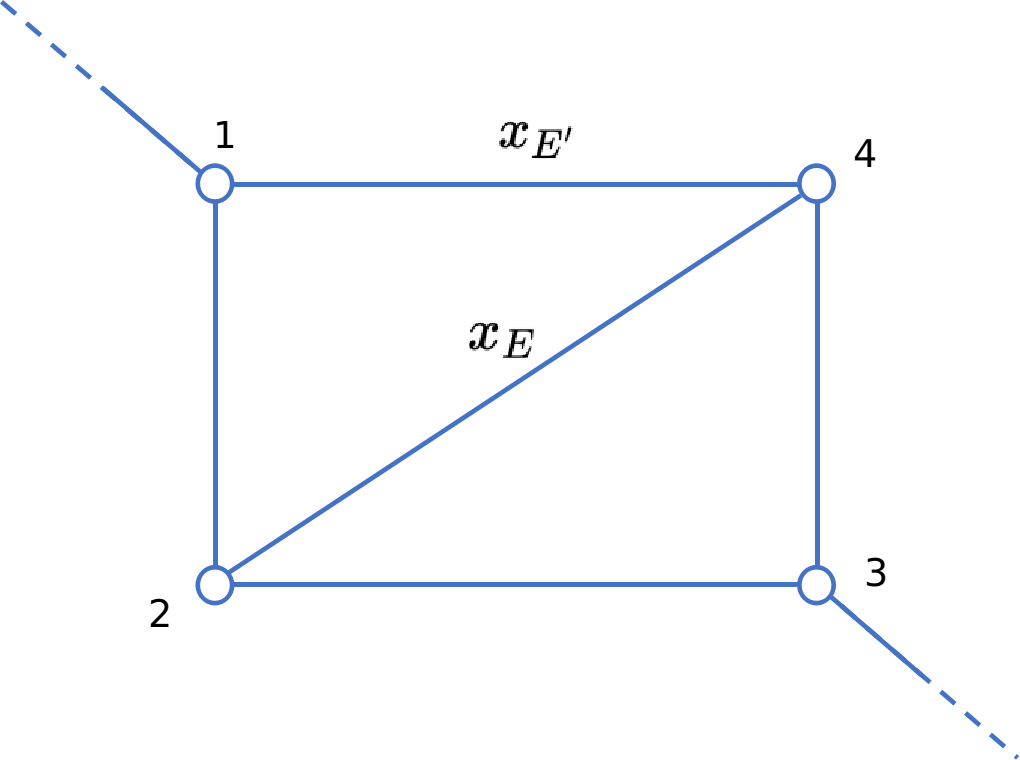}
	\caption{An ideal triangulation of 4-holed sphere.}
	\label{FGinxi}
	\end{center}
	\end{figure}

The eignvector of the holonomy $O_i$, $\xi_i'=(\xi_i^1,\xi_i^2)^T$ or $(-\bar{\xi}_j^2,\bar{\xi}_j^1)^T$, is the framing flag $s_\ell$ (of $\ell$ connecting the hole $i$) parallel transported to the base point $\fp$ of $O_i$, i.e.
\be
\xi_{i}'=s_\ell(\fp),\quad \fp\in\cs_a,\label{xiframing}
\ee
the FG coordinates on $\cs_a$ can be expressed in terms of $\xi'_i$: Without loss of generality, we assume that $\fp$ is inside the quadrilateral shown in FIG.\ref{FGinxi}, and each $O_i$ travels around the hole $i$ counterclockwise. We have 
\be
x_E(\vec{j},\vec{\xi})&=&\frac{\langle \xi'_1\wedge\xi'_2\rangle\langle\xi'_{4}\wedge\xi'_3\rangle}{\langle \xi'_1\wedge\xi'_{4}\rangle\langle\xi'_{2}\wedge\xi'_3\rangle},\nonumber\\
x_{E'}(\vec{j},\vec{\xi})&=&\frac{\langle O_4\xi'_3\wedge\xi_1'\rangle\langle\xi'_{4}\wedge\xi'_2\rangle}{\langle O_4\xi'_3\wedge\xi'_{4}\rangle\langle\xi'_{1}\wedge\xi'_2\rangle},\label{FGintxi}
\ee
Here $O_4$ is given by 
\be
&&O_4= M(\xi_4')\left(\begin{array}{cc}
	  \pm e^{L_{ab}} & 0 \\
	0 & \pm e^{-L_{ab}}
	\end{array}\right)M(\xi_4')^{-1}.
\ee
where $\pm e^{L_{ab}}=\pm\exp \left[\frac{ \pi i}{k}\left(b \alpha_{a b}+\left(2 j_{a b}+\frac{\epsilon_{a b}}{2}\right)\right)\right]$ for $\ell_{ab}$ attached to the 4th hole. 
$x_{E'}$ is independent of $\pm$ sign. Both $x_E(\vec{j},\vec{\xi}),x_{E'}(\vec{j},\vec{\xi})$ are invariant under the $\PSlc$ gauge transformation of \eqref{HHHH}: $O_i\to hO_i h^{-1}$, $\xi_i'\to h\xi_i'$.  

The correspondence between $\{x_E\}_E$'s and framed $\PSlc$ flat connections on $\cs_a$ is 1-to-1 \cite{FG03}, so $x_E,x_{E'}$ given by \eqref{FGintxi} and four $e^{2L_{ab}}$ at the holes uniquely determine a $\PSlc$ flat connection labelled by $\vec{j},\vec{\xi}$. This connection reduces to $\PSu$ when $\a_{ab}=0$. We choose $E,E'$ to be such that $x_{E},x_{E'}$ equals $e^{\cx_a},e^{\cy_a}$ in $(e^{\scrq_I},e^{\scrp_I})$. We lift $x_E,x_{E'}$ to logarithmic coordinates $\chi_E=\log(x_E),\chi_{E'}=\log(x_{E'})$ (the lift is uniquely given by \eqref{edgesum} and the lifts of ideal-tetrahedra coordinates), and obtain $\cx_a,\cy_a$ as functions of $\vec{j},\vec{\xi}$. By \eqref{cxa} - \eqref{cya}, we have $\mu_a,\nu_a,m_a,n_a\in\R$ as functions of $\vec{j},\vec{\xi}$. Furthermore, by \eqref{zmunu} and \eqref{xymn}, we obtain uniquely the functions $z_a(\vec{j},\vec{\xi})$, $ x_a(\vec{j},\vec{\xi})$, and $ y_a(\vec{j},\vec{\xi})$.


Recall \eqref{sigmalabel}, the implementation of the simplicity constraint restricts the label $\iota$ to the subspace
\be
\iota_{\vec{j},\vec{\xi}}&=&\lt(\lt\{0,m_{ab}\rt\}_{a<b},\ \lt\{\rho^{(a)}_{\vec{j},\vec{\xi}}\rt\}_{a=1}^5\rt),\nonumber\\
\rho^{(a)}_{\vec{j},\vec{\xi}}&=&\lt(z_a(\vec{j},\vec{\xi}),x_a(\vec{j},\vec{\xi}),y_a(\vec{j},\vec{\xi})\rt),\nonumber
\ee
where $\vec{j}=\{j_{ab}+\epsilon_{ab}/4\}_{a<b}$ and $\vec{\xi}=\{\xi_{ab}\}_{a,b=1,\cdots,5}$. $m_{ab}$ relates to $j_{ab}$ by \eqref{e2Labjab}. Here $\vec{j}$ have to satisfy the condition in Lemma \ref{Oeqntrig} so that the solution $O_{i=1,\cdots,4}\in\PSu$ to Eq.\eqref{HHHH} exists. $\vec{\xi}$ are eigenvectors of the solution $O_{i=1,\cdots,4}$. 

Therefore the simplicity constraint restrict the partition function $\cz_{S^3\setminus\G_5}(\iota)$ in \eqref{Zsigma1} to 
\be
\cz_{S^3\setminus\G_5}\lt(\iota_{\vec{j},\vec{\xi}}\rt)\equiv A_v(\vec{j},\vec{\xi}),
\ee
which is defined to be the spinfoam vertex amplitude with cosmological constant.

Note that only 2 FG coordinates $x_E,x_{E'}$ out of 6 are used in $z_a,x_a,y_a$. Only these 2 coordinates are restricted to be \eqref{FGintxi}. Other four FG coordinates $x_{E''}\neq x_E,x_{E'}$ may not be simultaneously expressed in terms of $\vec{j},\vec{\xi}'$ as \eqref{FGintxi} when $\a_{ab}\neq 0$, since otherwise $\l^2=\prod_{E\ \text{around hole}}x_E$ would belong to U(1), whereas generally $e^{2 L_{ab}}\not\in \mathrm{U(1)}$ for $\a_{ab}\neq 0$. 
However other four $x_{E''}\neq x_E,x_{E'}$ are absent in the coherent label. $\rho^{(a)}_{\vec{j},\vec{\xi}}$ is generally an $\PSlc$ flat connection, but reduces to $\PSu$ when $\a_{ab}=0$ or in the semiclassical limit.

\subsection{SU(2) flat connections on $\cs_a$ and 4-gon}\label{Integration over PSU(2) flat conn}

A simple counting degrees of freedom shows that $\vec{\xi}$'s solving $O_4O_3O_2O_1=1$ modulo PSU(2) gauge transformations generically span real 2-dimensional space. This 2-dimensional space is denoted by $\cm_{\vec{j}}$. 
$x_E,x_{E'}$ in \eqref{FGintxi} are densely defined functions on $\cm_{\vec{j}}$.

A description of $\cm_{\vec{j}}$ \cite{Nekrasov:2011bc} generalizes the Kapovich-Millson phase space description \cite{polygon,polygon1}: We lift to the cover space $\widetilde{\cm}_{\vec{j}}$ the moduli space of SU(2) flat connection with fixed $\vec{j}$. $\widetilde{\cm}_{\vec{j}}$ is the moduli space of solutions to $H_4H_3H_2H_1=1$ with 
\be
H_i= M(\xi_i)\left(\begin{array}{cc}
	e^{\frac{2\pi i}{k}j_i} & 0 \\
  0 &  e^{-\frac{2\pi i}{k}j_i}
  \end{array}\right)M(\xi_i)^{-1}.\nonumber
\ee
where $j_i=j_{ab}$ of annuli $\ell_{ab}$ connecting to the holes.



Given the 4-dimensional complex vector space $V=\mathrm{Mat}_{2\times 2}(\C)\simeq\C^4$ of complex $2\times 2$ matrices, we endow $V$ with the complex metric $\langle X, Y\rangle=-\frac{1}{2}\lt[\operatorname{Tr}(X Y)-\operatorname{Tr} X \operatorname{Tr} Y\rt]$. If we write $X=x^0 I+\sum_{a=1}^3x^a\bm{\sigma}_a$ and $Y=y^0 I+\sum_{a=1}^3y^a\bm{\sigma}_a$, $\langle X,Y\rangle$ is the complexified Minkowski metric on $\C^4$: $\langle X, Y\rangle=x^0y^0-\sum_{a=1}^3x^ay^a$. 
$\Su$ is the unit 3-sphere in $V_{\R}\simeq\R^4\subset V$ defined by 
\be
&&H=h^0+i\sum_{a=1}^3 h^a\bm{\sigma}_a,\quad h_0,h_a\in\R,\nonumber\\
&&\lag H,H\rag=(h^0)^2+\sum_{a=1}^3(h^a)^2=1.\nonumber
\ee
When restricting $h^0+i\sum_{a=1}^3 h^a\bm{\sigma}_a$ with $h_0,h_a\in\R$, $\lag\cdot,\cdot\rag$ becomes the Euclidean metric on $\R^4$ and induces the spherical metric of $S^3$ on SU(2).

\begin{figure}[h]
	\begin{center}
	\includegraphics[width=6cm]{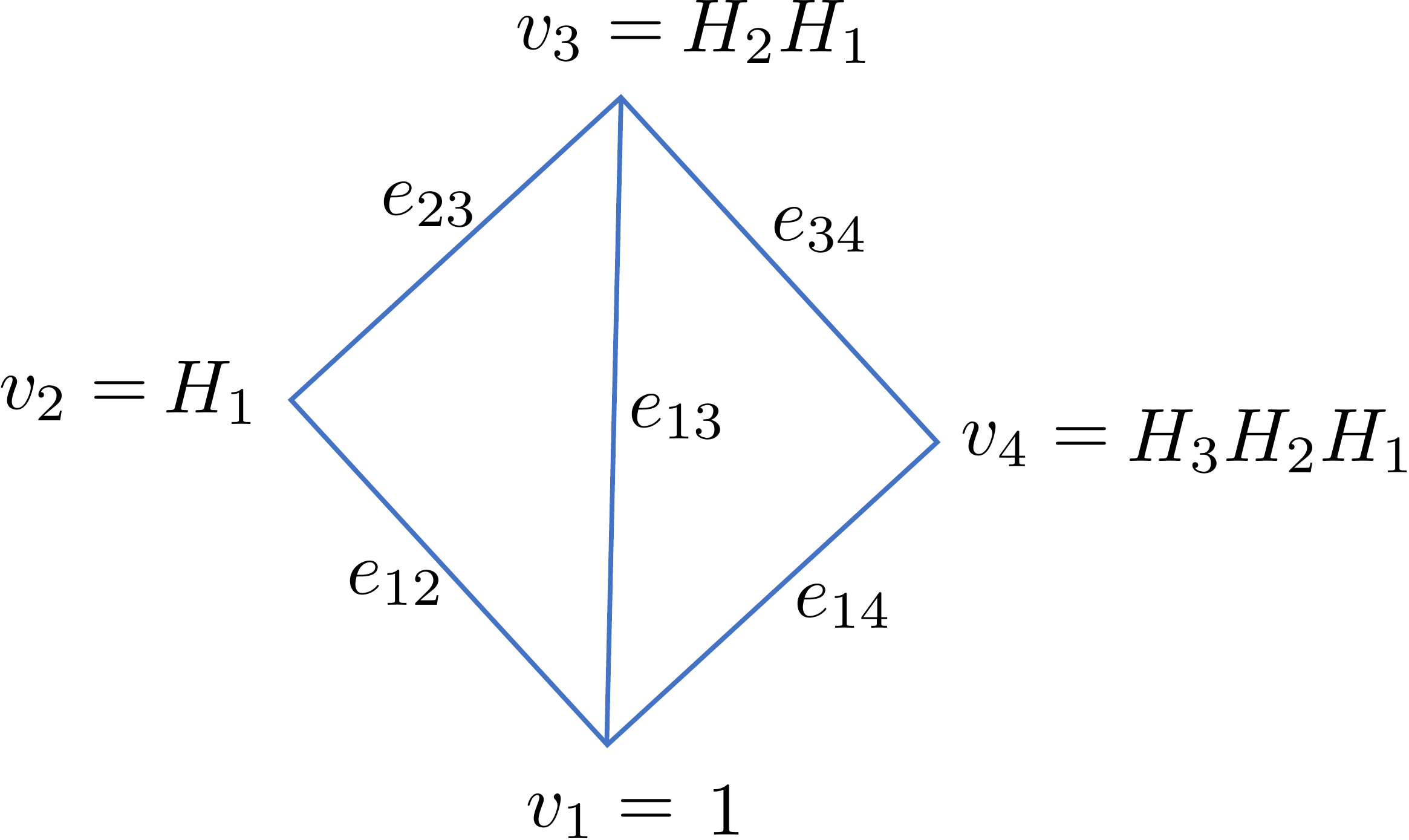}
	\caption{The 4-gon in SU(2) determined by $H_4H_3H_2H_1=1$.}
	\label{4gon}
	\end{center}
	\end{figure}

Given $H_{1,\cdots,4}\in\Su$ satisfying $H_4H_3H_2H_1=1$, The set of $H_i$ determines 4 points $v_1,\cdots,v_4$ in $\Su$ where
\be
v_1=1,\ v_2= H_1,\ v_3=H_2H_1,\ v_4=H_3H_2H_1.\nonumber
\ee
We firstly assume the generic situation that $v_1,\cdots,v_4$ are linearly independent in $\R^4$. Any pair $(v_i,v_j)$ viewed as 2 vectors in $\R^4$ determines a 2-plane $E_{ij}=\mathrm{Span}_\R(v_i,v_j)\subset \R^4$. The intersection between $E_{ij}$ and $\Su$ is the geodesic $e_{ij}$ connecting $v_i,v_j$ ($\Su$ is the unit 3-sphere in $\R^4$)
\be
e_{ij}&=&E_{ij}\cap \Su=\{t_1v_i+t_2v_j\mid \nonumber\\
&&t_1^2+t_2^2+2t_1t_2\lag v_i,v_j\rag=1,\ t_1,t_2\geq0\}.\nonumber
\ee
The vertices $v_i$ and edges $e_{12},e_{23},e_{34},e_{14}$ made a 4-gon in $\Su$. The geodesic distance $\theta_{ij}$ between $v_i$ and $v_j$ is given by
\be
\cos(\theta_{ij})=\lag v_i,v_j\rag\equiv c_{ij},\quad \theta_{ij}\in(0,\pi).\nonumber
\ee
The lengths of $e_{12},e_{23},e_{34},e_{14}$ are $a_i=\theta_{i,i+1}$ such that
\be
\cos(a_i)=\Tr(H_i)/2.\nonumber
\ee 
We draw the diagonal geodesic connecting $v_{1},v_3$. $\theta_{13}$ is the length of the diagonal.

The face $f_{ijk}$ with the vertices $v_i,v_j,v_k$ is the intersection of $F_{ijk}=\mathrm{Span}_\R(v_i,v_j,v_k)$ and $\Su$
\be
f_{i j k}&=&F_{i j k} \cap \Su=\left\{t_{1} v_{i}+t_{2} v_{j}+t_{3} v_{k} \mid t_{1}, t_{2}, t_{3} \geq 0,\right.\nonumber\\
&&\left.t_{1}^{2}+t_{2}^{2}+t_{3}^{2}+2 t_{1} t_{2} c_{i j}+2 t_{1} t_{3} c_{i k}+2 t_{2} t_{3} c_{j k}=1\right\}\nonumber
\ee
The unit normal $n_{ijk}$ of $F_{ijk}$ is defined by $\lag f,n\rag=0$, $\forall\,f\in F_{ijk}$, and $\lag n,n\rag=1$. A choice of orientation of $F_{ijk}$ corresponds to the sign of $n$. We define the bending angle $\phi_{ij}\in (0,\pi)$ by
\be
\cos(\phi_{ij})=\lag n_{ikl},n_{jkl}\rag.
\ee

$\theta=\theta_{13},\phi=\phi_{24}$ are symplectic coordinates of $\widetilde{\cm}_{\vec{j}}$ \cite{Nekrasov:2011bc}. Up to isometries of $S^3$, $(\theta,\phi)$ determines a unique 4-gon in $S^3\simeq \Su$ whose geodesic edge lengths relate to the conjugacy classes of $H_i$. Indeed, geodesic edge lengths $a_i,\theta\in(0,\pi)$ uniquely determine two triangles sharing the diagonal $e_{13}$, up to isometries of $S^3$. We break the translational symmetry by fixing $v_1=1$. The remaining symmetry is the rotation leaving $v_1=(1,0,0,0)\in\R^4$ invariant. We use the freedom of the rotation to fix the position of $v_2,v_3$ of the triangle $(v_1,v_2,v_3)$. Fixing the position of the triangle $(v_1,v_2,v_3)$ breaks the continuous rotational symmetry. $v_1,v_2,v_3$ determine the hyperplane $F_{123}\subset\R^4$. The freedom of $v_4$ is equivallent to choosing the hyperplane $F_{134}$, which is determined by the bending angle $\phi$ up to a parity symmetry with respect to $F_{123}$. This parity symmetry can be fixed by choosing the orientaion of the bending flow in addition, i.e. fixing the orientation of $n_{123}\wedge n_{134}$ (see Appendix \ref{thetatheta}). As a result, $v_1,\cdots,v_{4}\in \Su$ are uniquely determined by $(\theta,\phi)$ once we fix $v_1=1$ and the rotation symmetry. $v_2=H_1,v_3=H_2H_1,v_4=H_3H_2H_1$ determines $H_{1,\cdots,4}$ with $H_4=(H_3H_2H_1)^{-1}$. 
By \eqref{Hi} and the given $\{j_i\}_{i=1}^4$, we obtain all $\xi_i$ as the eigenvector of $H_i$ whose squared eigenvalue is $e^{4\pi ij_i/k}$. We normalize $\xi_i$'s by $\det(M(\xi_i))=1$ up to individual phases. As a result, all $\xi_i$'s are functions of $j_i$ and $\theta,\phi$. Appendix \ref{thetatheta} provides an algorithm to determine $\xi_i$'s from $\theta,\phi$ in practise.


For any function $f$ on ${\cm}_{\vec{j}}$, $f$ can be lifted to a function on $\widetilde{\cm}_{\vec{j}}$ and is invariant under $H_i\to -H_i$. we define the following integral on ${\cm}_{\vec{j}}$
\be
\int_{{\cm}_{\vec{j}}}\rmd \xi\, f=\int\rmd\theta\wedge\rmd\phi\, f 
\ee
This integral on the right-hand side is over compact domain, thus is finite provided that $|f|$ is bounded. The degenerate 4-gons with $\theta,\phi=0$ is included as boundaries of the integral. This integral is needed for gluing vertex amplitudes to construct spinfoam amplitudes on complexes. 


It may happen that for certain $\vec{j}$, $\widetilde{\cm}_{\vec{j}}$ only contain degenerate 4-gon (i.e. becoming a $n$-gon with $n<4$) where a vector $v_i$ is a linear combination of another 2 vectors $v_j,v_k$ in $\R^4$. In this case the dimension of $\widetilde{\cm}_{\vec{j}}$ is less than 2, thus the above integral is ill-defined.
The degenerate 4-gon leads to at least two $H_i$'s belonging to a U(1) subgroup in SU(2). It sometimes gives a pair of collinear $\xi_i'$'s that result in ill-defined $x_E,x_{E'}$ on entire $\widetilde{\cm}_{\vec{j}}$ (see \eqref{FGintxi}). 
We set the contribution from $\vec{j}$ such that $\dim(\widetilde{\cm}_{\vec{j}})<2$ to vanish in the spinfoam amplitude. In particular, it set the contribution of $j_{i}=0$ to vanish.

\subsection{Finite spinfoam amplitude on simplicial complex }
 
Given a simplicial complex $\ck$ made by a finite number of 4-simplices, we associate each 4-simplex with a vertex amplitude as a function on $\times_{a=1}^5 {\cm}_{\vec{j}_a}$ when fixing $\vec{j}$
\be
A_v(\vec{j},\vec{\xi})=\mathcal{Z}_{S^{3} \setminus \Gamma_{5}}(\iota_{\vec{j}, \vec{\xi}})
\ee 
where $\iota_{\vec{j}, \vec{\xi}}=({j}_{ab},\rho^{(a)}_{\vec{j},\vec{\xi}})$. When gluing a pair of 4-simplices by identifying a pair of tetrahedra, we identify 4 spins $j_{f}$ (of tetrahedron face areas) for the pair of tetrahedra, we associate $\rho_{\vec{j},\vec{\xi}}=(z(\vec{j}, \vec{\xi}), x(\vec{j}, \vec{\xi}), y(\vec{j}, \vec{\xi}))$ (of the tetrahedron shape) to one tetrahedron and associate 
\be
J\rho_{\vec{j},\vec{\xi}}=\left(\overline{z(\vec{j}, \vec{\xi})},\, x(\vec{j}, \vec{\xi}),\, -y(\vec{j}, \vec{\xi})\right)
\ee
to the other tetrahedron (recall \eqref{complexcc000}). We may define the gluing of the pair of vertex amplitudes by
\be
\int_{{\cm}_{\vec{j}}}\rmd \xi\,\mathcal{Z}_{S^{3} \setminus \Gamma_{5}}(\vec{j},\rho_{\vec{j},\vec{\xi}})\,\mathcal{Z}_{S^{3} \setminus \Gamma_{5}}(\vec{j},J\rho_{\vec{j},\vec{\xi}}),\label{intZZ}
\ee
where we only focus on variables associated to the pair of tetrahedra identified by gluing. $\int_{{\cm}_{\vec{j}}}\rmd \xi$ is an analog of integrating SU(2) coherent intertwiners in the EPRL model. The gluing defined by \eqref{intZZ} identify at the quantum level $\vec{\xi}$ between the pair of tetrahedra. Generally speaking it may only be necessary to identify $\vec{\xi}$ semiclassically, i.e. gluing 4-simplices by identifying 2 tetrahedra with shape-matching only semiclassically. Thus we define the more general gluing by  
\be
\int_{{\cm}_{\vec{j}}}\rmd \xi\rmd\xi'\,\mathcal{Z}_{S^{3} \setminus \Gamma_{5}}(\vec{j},\rho_{\vec{j},\vec{\xi}})\, A_e(\vec{j},\vec{\xi},\vec{\xi}')\, \mathcal{Z}_{S^{3} \setminus \Gamma_{5}}(\vec{j},J\rho'_{\vec{j},\vec{\xi}'}),\nonumber\\\label{intZZgen}
\ee
where $A_e$ is called the edge amplitude. $A_e$ is a function of $\vec{j},\vec{\xi},\vec{\xi}'$ relating to the tetrahedron $e$ ($A_e$ may depend on $k,\gamma$ which is implicit in the formula). The precise form $A_e$ is not determined in this work, but we require that $A_e$ is Gaussian-like continuous function peaked at $\vec{\xi}=\vec{\xi}'$ and suppressed elsewhere. $A_e$ approaches to $\delta(\vec{\xi},\vec{\xi}') $ when $j\to\infty$. Choices of integration measures of $\vec{\xi},\vec{\xi}'$ are included in choices $A_e$.

Given any simplicial complex $\ck$, we associate a ``spin'' $j_f=0,\frac{1}{2},\cdots,\frac{k-1}{2}$ to each (internal or boundary) face $f$, and associate to each (internal or boundary) tetrahedron $e$ a PSU(2) flat connection labelled by ${\vec{j},\vec{\xi}}$ on the 4-holed sphere. These data enter vertex amplitudes $A_v=\mathcal{Z}_{S^{3} \setminus \Gamma_{5}}(\iota_{\vec{j}, \vec{\xi}})$, edge amplitudes $A_e(\vec{j},\vec{\xi},\vec{\xi}')$, and face amplitudes $A_f(j_f)$. We construct the full spinfoam amplitude $A$ on $\ck$ by integrating over $\rho_{\vec{j},\vec{\xi}}$ of all internal tetrahedra $e$ and summing over $j_f$ of all internal faces.
\be
A=\!\!\!\sum_{\{j_f\}}^{(k-1)/2}\!\!\!\!\!{}'\prod_fA_f(j_f)\int[\rmd \xi\rmd\xi']\prod_{e} A_e(\vec{j},\vec{\xi}_e,\vec{\xi}'_e)\prod_vA_v(\vec{j},\vec{\xi}).\nonumber
\label{fullamplitudeA}\\
\ee
We put subscript $e$ to manifest that $A_e$ only depend on variables relating to $e$. $\int[\rmd \xi]$ is a product of integrals \eqref{intZZgen} over all internal tetrahedra $e$. $A_f(j_f)$ is an undetermined face amplitude. $\prod_v$ products over all 4-simplices. $\sum{}'_{\{j_f\}}$ sums $j_f$ at all internal faces in $\ck$. The sum of each $j_f$ is \emph{finite} by \eqref{jrange}. The cosmological constant relating to $k$ provides a cut-off to the sum over spins. $\sum{}'$ indicates that we exclude $j_f$'s that do not satisfy the triangle inequality or lead to $\widetilde{\cm}_{\vec{j}}$ of dimension less than 2.

\begin{theorem}

The amplitude $A$ is finite for any choice of simplicial complex.

\end{theorem}
 
\textbf{Proof:} $|A_v|$ is bounded because of Theorem \ref{finitevertex}. $|A_e|$ is bounded since it is continuous on the compact space of $\vec{\xi}_e,\vec{\xi}'_e$. The integral in $A$ integrates a function whose absolute value is bounded on a compact domain, thus is absolutely convergent. Then the finite sum over $j_f$ implies the finiteness of $A$.

$\Box$




\subsection{Boundary data}\label{Boundary data}

The boundary data of the spinfoam amplitude $A$ relates to the kinematical states of LQG up to a deformation. The boundary of the 4d simplicial complex $\ck$ is a 3d simplicial complex $\partial\ck$. The dual complex $\partial\ck^*=\G$ is an (oriented) graph with links $\fl\subset \G$ dual to faces $f\subset \partial\ck$ and nodes $\fv\in\G$ dual to tetrahedra $e\subset\partial \ck$. The boundary data of $A$ color every link by a spin $j_\fl=0,\frac{1}{2},\cdots,\frac{k-1}{2}$, and color every node $\fv$ an element $\rho_\fv=\cm_{\vec j}$. 
There is an 1-to-1 correspondence between $\rho_\fv$ and a convex constant curvature tetrahedron (up to degenerate tetrahedra) whose face areas are determined by $j_\fl$ of $\fl$ adjacent to $\fv$ (see \cite{curvedMink} or Theorem \ref{tetracorresp}). These data are perfect analog of LQG spin-network data on $\G$: spins $j_\fl$ on links and coherent intertwiners $||\vec{j},\vec{\xi}\rangle_\fv$ at nodes. The coherent intertwiners 1-to-1 correspond to convex flat tetrahedra whose face areas are proportional to $j_\fl$ \cite{LS,shape,CF}. The boundary data of $A$ is a deformation of the spin-network data due to the cut-off $\frac{k-1}{2}$ of $j_\fl$ and $\rho_\fv$ for constant curvature tetrahedra versus $||\vec{j},\vec{\xi}\rangle_\fv$ for flat tetrahedra. When $k\to\infty$ while fixing $j_\fl$ (different from the semiclassical limit $j,k\to\infty$ fixing $j/k$), the cut-off is removed and the constant curvature $\L$ given by \eqref{kandLambda} reduces to be flat, then the boundary data of $A$ reduces to the spin-network data. 

We expect that $A$ defines transition amplitudes of boundary states that are the eigenstates of area operators at links and coherent with respect to quantum tetrahedra at nodes, similar to spin-network states with coherent intertwiners. The coherent states at nodes are expected to quantize the phase space $\widetilde{\cm}_{\vec{j}}$: the moduli space of SU(2) flat connections on 4-holed sphere with fixed conjugacy classes. The quantization of $(\widetilde{\cm}_{\vec{j}},\frac{k}{2\pi}\O)$ is known to give the Hilbert space of quantum group $\Su_{\fq}$ intertwiners with $\fq=e^{\pi i/(k+2)}$ (see e.g. \cite{Gawedzki:1999bq,Engle2011}). By these arguments, we conjecture that the boundary Hilbert space of $A$ are spanned by $\fq$-deformed spin-network states $|\G,j_\fl,i_\fv\rangle$ where $j_\fl, i_\fv$ are unitary irreps and intertwiners of $\Su_{\fq}$ respectively. The proof of this conjecture is a research undergoing. It Involves the coherent intertwiner of $\Su_{\fq}$ and showing the relation to the curved tetrahedron labelled by the SU(2) flat connection. Some earlier studies of the quantum group coherent intertwiner is given in \cite{Ding:2011hp}. Constructing geometrical operators for the boundary Hilbert space is also a research in progress (see \cite{Han:2016dnt} for the first step).

\subsection{Ambiguities}\label{Ambiguities}

The construction of the spinfoam amplitude with cosmological constant depends on several choices, which may relate to ambiguities of the model. In the following we classify and discuss these choices: 

(1) The spinfoam amplitude depends on choices of coherent states in Section \ref{Coherent state}. This dependence is a result from the proposal of imposing the simplicity constraint on coherent state labels. In this work we choose the coherent states \eqref{coheretntpsiz} and \eqref{xixym}. But a different set of coherent states may be chosen, as far as their are peaked semiclassically at points in the phase space. 

(2) There are freedom of choosing edge and face amplitudes $A_e,A_f$ in \eqref{fullamplitudeA}. See e.g. \cite{Kaminski:2009cc,face} for some existing discussion about preferred choices of $A_e,A_f$ in the absence of $\L$. The freedom of $A_e$ contains the freedom of the integration measure for $\vec{\xi}$. Moreover the freedom of $A_e$ has an overlap with the freedom of coherent states discussed in (1). Namely if we make a change of coherent state $\Psi_{\rho_{\vec{j},\vec{\xi}}}\mapsto\Psi'_{\rho_{\vec{j},\vec{\xi}}}=\int \rmd \xi_e K(\vec{\xi}_e,\vec{\xi}'_e)\Psi_{\rho_{\vec{j},\vec{\xi}'}}$ with certain function $K$ of $\vec{\xi}_e,\vec{\xi}'_e$ of the tetrahedron $e$, the spinfoam amplitude constructed with the new state $\Psi'_{\rho_{\vec{j},\vec{\xi}}}$ can be written in the same form as \eqref{fullamplitudeA} with $A_v$ of the old state $\Psi_{\rho_{\vec{j},\vec{\xi}}}$, while $A_e$ transforms as $A_e(\vec{\xi}_e,\vec{\xi}'_e)\mapsto \int \rmd\zeta_{e}\rmd\zeta_{e}'K(\vec{\zeta}_e,\vec{\xi}_e)A_e(\vec{\zeta}_e,\vec{\zeta}'_e)\overline{K(\vec{\zeta}'_e,\vec{\xi}'_e)}$.

(3) The vertex amplitude depends on the positive angle structure $(\vec{\a},\vec{\b})\in \Fp_{\rm new}$, since $\cz_{S^3\setminus\G_5}$ depends on $(\vec{\a},\vec{\b})$. More precisely $\cz_{S^3\setminus\G_5}$ only depends on $\vec{\a}$ but is independent of specific $\vec{\b}$ as far as $(\vec{\a},\vec{\b})\in \Fp_{\rm new}$, by the discussion below \eqref{STUFpoct}. The dependence on angles $\vec{\a}=(\{\a_{ab}\}_{a<b},\{\a_a\}_{a=1}^5)$ in $\cz_{S^3\setminus\G_5}$ may be analogous to the framing anomaly of CS theory with compact group \cite{Witten1991,Witten1989a}. For the consistency ``4d area = 3d area'' at the quantum level, it is preferred to restrict all $\a_{ab}$ in $A_v$ to vanish and still be inside $\Fp_{\rm new}$, whereas there still exists some freedom of $\{\a_a\}_{a=1}^5$.

The spinfoam amplitude depends on $\{\b_a\}_{a=1}^5$ because they enter the vertex amplitude $A_v$ via the prefactor $e^{-\sqrt{2}\b_a\re(z_a)}$ of the coherent state $\psi_{z_a}$ in \eqref{coheretntpsiz}. But this prefactor can be absorbed in $A_e$ (or definition of integration measure of $\vec{\xi}$). Thus this dependence on $\{\b_a\}_{a=1}^5$ is a part of the freedom of (1) and (2). In more detail, by the freedom of coherent states, we choose $\psi^0_{z_a}$ instead of $\psi_{z_a}$ in the definition of $\cz_{S^3\setminus\G_5}(\iota)$. Then \eqref{upperboundcz} for the bound of $|\cz_{S^3\setminus\G_5}(\iota)|$ is modified by
\be
&&\lt|\int_{\R^{5}}\prod_{a=1}^5\rmd \mu_a'\ \cz_{S^3\setminus\G_5}(\vec{\mu}'+i\vec{\a}\mid \vec{m}')\prod_{a=1}^5\psi^0_{z_a}(\vec{\mu}'_a)\rt|\nonumber\\
&\leq& Ck^{5/4}\prod_ae^{\beta_a  \left(\frac{\pi  \beta_a }{k}+\sqrt{2} \Re(z_a)\right)}.\label{upperboundcznew}
\ee
The bound diverges if $\mathrm{Re}(z_a)$ approaches to $\infty$ or $-\infty$ depending on $\sgn(\b_a)$. This can happen even after imposing the simplicity constraint since $x_{E},x_{E'}$ can approach infinity when a pair of $\xi_i'$ becomes collinear in \eqref{FGintxi}, particularly when the constant curvature tetrahedron approach to degenerate. Then we need to require in addition the following behavior to $A_e$ as $\mathrm{Re}(z)$ approaching to $\infty$ or $-\infty$ correspondingly
\be
|A_e(\vec{j},\vec{\xi},\vec{\xi}')|\leq C' e^{-\sqrt{2}\b_e\mathrm{Re}(z_e(\vec{j},\vec{\xi}))}e^{-\sqrt{2}\b'_e\mathrm{Re}(z_e(\vec{j},\vec{\xi}'))}\nonumber
\ee
where the exponential decay factors should cancel the exponential grows in \eqref{upperboundcznew} of 2 vertex amplitudes sharing the tetrahedron $e$. The freedom of $\b_e$ becomes part of the freedom of $A_e$. The integrand of $\int[\rmd\xi]$ in \eqref{fullamplitudeA} still has bounded absolute value, then $A$ is finite. 

(4) The amplitude $A$ generally depends on the choice of the simplicial complex $\ck$, similar to spinfoam models in the absence of cosmological constant. 

\section{Semiclassical analysis}\label{Semiclassical analysis}

In this section, we examine the semiclassical behavior of the vertex amplitude $A_v$ and show that the semiclassical limit of $A_v$ reproduces the 4d Regge action with $\L$. 

The semiclassical limit of quantum gravity is $\ell_P\to 0$ while keeping geometrical quantities e.g. areas, shapes, curvature, etc, fixed. $A_v$ is the LQG transition amplitude associated to a 4-simplex whose boundary is made by 5 tetrahedra labelled by $a,b=1,\cdots,5$. $A_v$ depends on $k$, $\g$, $j_{ab}$, and $\xi_{ab}$. By the result of \cite{curvedMink} (to be reviewed in Section \ref{Critical points of vertex amplitude and constant curvature 4-simplex}), $\xi_{ab}$'s parametrize geometrical shapes of 5 boundary constant curvature tetrahedra as boundary data of $A_v$, while $j_{ab}/k$ (up to $\eps_{ab}/k$) is proportional to $|\L|\fa_{ab}$. Here $\fa_{ab}$ is the area of the face $f_{ab}$ shared by tetrahedra $a$ and $b$. The cosmological constant $\L$ equals to the constant curvature of tetrahedra. Therefore the semiclassical limit in our context is $\ell_P\to 0$ while keeping $\xi_{ab}$'s, $\fa_{ab}$'s, and $\L$ fixed. The Barbero-Immirzi parameter $\g$ is also fixed. The relation between $k$ and $\L$ in \eqref{kandLambda} indicates that $k\to \infty$ in the semiclassical limit. These motivate the following definition:

\begin{definition}

The semiclassical limit of $A_v$ is the asymptotic behavior of $A_v$ when we scale uniformly all $j_{ab}\to\infty$ and $k\to \infty$ (so $\sig=ik\g\to i\infty$) while keeping $j_{ab}/k$ fixed.

\end{definition}

This limit generalizes the semiclassical limit of the Turaev-Viro model in 3d gravity, and is studied in \cite{HHKR} for 4d spinfoam vertex amplitude. 

The semiclassical limit of spinfoam amplitude is the same as the semiclassical limit of CS theory. Indeed, the flat connection position variables $\scrq_I$ depending on $j_{ab}$ only through the ratio $j_{ab}/k$ (see \eqref{e2Labjab}). The above semiclassical limit send $k\to\infty$ but leaves $\scrq_I$ finite. The limit effectively removes the dependence of $\a_{ab},\epsilon_{ab}$ in $e^{2L_{ab}}$. The limit $k\to\infty$ keeping $\scrq_I$ finite is the same as the semiclassical limit of CS partition function. Therefore, it is useful to firstly study the semiclassical limit of the CS partition function $\cz_{S^3\setminus\G_5}$ in Section \ref{Semiclassical analysis of Chern-Simons partition function}, then the result can be applied straightforwardly to the semiclassical limit of $A_v$ in Section \ref{Critical points of vertex amplitude and constant curvature 4-simplex} and \ref{Asymptotics of vertex amplitude}.

\subsection{Semiclassical analysis of Chern-Simons partition function}\label{Semiclassical analysis of Chern-Simons partition function}

Recall the construction of $\cz_{S^3\setminus\G_5}$ in Section \ref{S3-G5 partition function}. Eqs.\eqref{Utrans}, \eqref{Ttrans}, \eqref{Strans}, and \eqref{affinetrans} lead to
\begin{widetext}
\be
&&\cz_{S^3\setminus\G_5}\left(\vec{\mu}\mid \vec{m}\right)
=\frac{4i}{k^{15}} \sum_{\vec{n} \in(\mathbb{Z} / k \mathbb{Z})^{15}} \int_{\mathcal{C}} \mathrm{d}^{15} \nu\, e^{S_0}  Z_{\times}\left(-\mathbf{B}^{T} \vec{\nu} \mid-\mathbf{B}^{T} \vec{n}\right)\label{sumintZ}\\
&&\quad S_0=\frac{ \pi i}{k}\lt[-2\lt(\vec{\mu}-\frac{i Q}{2} \vec{t}\rt) \cdot \vec{\nu}+2\vec{m} \cdot \vec{n}-\vec{\nu} \cdot \mathbf{A B}^{T} \cdot \vec{\nu}+(k+1)\vec{n} \cdot \mathbf{A B}^{T} \cdot \vec{n}\right]\\
&&\quad Z_{\times}(\vec{\mu} \mid \vec{m})=\prod_{a=1}^{5} \Psi_{\Delta}\left(\mu_{X_a} \mid m_{X_a}\right) \Psi_{\Delta}\left(\mu_{Y_a} \mid m_{Y_a}\right) \Psi_{\Delta}\left(\mu_{Z_a} \mid m_{Z_a}\right) \Psi_{\Delta}\left(\mu_{W_a} \mid m_{W_a}\right)\\
&&\quad \mu_{W_a}=i Q-\mu_{X_a}-\mu_{Y_a}-\mu_{Z_a},\quad m_{W_a}=-m_{X_a}-m_{Y_a}-m_{Z_a}
\ee
and $\Psi_{\Delta}$ is given by \eqref{qdilog}. 

We use \eqref{muImI} to change variables from $\mu_I,m_I$ to $\scrq_I'=\scrq_I-i\pi t_I$ and $\widetilde{\scrq}_I'=\widetilde{\scrq}_I-i\pi t_I$. 
It is intuitive to make similar change of variables from $\nu_I,n_I$ to $\scrp_I$,$\widetilde{\scrp}_I$ for studying the semiclassical limit
\be
\nu_I = \frac{b k \left(\widetilde{\scrp}_I+\scrp_I\right)}{2 \pi  \left(b^2+1\right)},&\quad & n_I= \frac{i k \left(\scrp_I-b^2 \widetilde{\scrp}_I\right)}{2 \pi  \left(b^2+1\right)}.\label{nuInI}
\ee
Semiclassically $\vec{\scrp}$ here is identical to the classical momenta conjugate to $\vec{\scrq}$ (recall \eqref{DecompSympl} and the discussion there). By the change of variables,
\be
S_0&=&-\frac{1}{2}\vec t\cdot(\vec{\scrp}+\vec{\widetilde{\scrp}})-\frac{i k}{4\pi (1+b^2)}\lt[\vec{\scrp}\cdot\lt(\mathbf{AB}^T\cdot \vec{\scrp}+2\vec{\scrq}'\rt)+b^2\vec{\widetilde \scrp}\cdot\lt(\mathbf{AB}^T\cdot \vec{\widetilde \scrp}+2\vec{\widetilde \scrq}{}'\rt) \rt]\nonumber\\
&&-\frac{i k^2}{4\pi (1+b^2)^2}\left(\vec{\scrp}-b^2 \vec{\widetilde{\scrp}}\right)\cdot\mathbf{AB}^T\cdot\left(\vec{\scrp}-b^2 \vec{\widetilde{\scrp}}\right).
\ee
\end{widetext}
We treat the sum $\sum_{n_I\in\mathbb{Z}/k\mathbb{Z}}$ by the Poisson resummation 
\be
&&\sum_{\vec{n}\in(\mathbb{Z}/k\mathbb{Z})^{15}}\!\!\!\! f(\vec{n})=\sum_{n_I=0}^{k-1}f(\vec{n})=\sum_{\vec{p}\in\mathbb{Z}^{15}}\int\limits_{-\delta}^{k-\delta} \rmd^{15} n\,f(\vec{n})e^{2\pi i \vec{p}\cdot\vec{n}}\nonumber\\
&&\qquad\quad=\lt(\frac{k}{2\pi}\rt)^{15}\sum_{\vec{p}\in\mathbb{Z}^{15}}\int\limits_{-\delta'}^{2\pi-\delta'} \rmd^{15} \mathcal{J}\,f'(\vec{\mathcal{J}})e^{i k \vec{p}\cdot\vec{\cj_{I}}}\label{poissonresumm}\\
&&\cj_{I}=\frac{2\pi n_I}{k}=\frac{i \left(\scrp_I-b^2 \widetilde{\scrp}_I\right)}{ b^2+1},\quad f'(\vec{\mathcal{J}})=f(\vec{n}).\label{cjInIk}
\ee
Here $f(\vec{n})=\chi(\vec{n})g(\vec{n})$ where $g(\vec{n})$ is the summand in \eqref{sumintZ} extended from $\vec{n}\in(\mathbb{Z}/k\mathbb{Z})^{15} $ to $\vec{n}\in\R^{15} $. $\chi(\vec{n})$ is a compact support function satisfying $\chi(\vec{n})=1$ for $\vec{n}\in\mathbb{Z}^{15}$. $\chi(\vec{n})$ vanishes outside $[-\delta,k-\delta]^{15}\setminus \cu$ (with arbitraily small $\delta>0$) where $\cu$ is an open neighborhood of singularities of $g(\vec{n})$ and $\cu\cap \mathbb{Z}^{15}=\emptyset$ \footnote{When extend $\Psi_\Delta (\mu|m)$ to $m\in\R$, poles of $\Psi_\Delta (\mu|m)$ are given by e.g. $\mu_{\rm pole}=ib u+ib^{-1} v$ with $v=-j$ and $u=-j-m+k\mathbb{Z}$ ($j\in\mathbb{N}_0$) when $\mathrm{Im}(b)>0$. Poles with $u\geq 1$ cancels with zeros when $m\in\mathbb{Z}/k\mathbb{Z}$, but this cancellation does not apply for non-integer $m$. At poles $\mathrm{Im}(\mu_{\rm pole})=\mathrm{Re}(b)(u+v)=\mathrm{Re}(b)(-2j-m+k\mathbb{Z})$. There exists $m$'s such that $\mathrm{Im}(\mu_{\rm pole})=\alpha$, i.e. the pole lies on the integration contour $\cc$ and may cause the integral to diverge. Therefore open neighborhoods of these $m$'s should be removed.}. The result does not depend on details of $\chi$ at $\vec{n}\not\in \mathbb{Z}^{15}$ because $\sum_{p_I\in\mathbb{Z}}e^{2\pi i p_In_I}=\sum_{n_I'\in\mathbb{Z}}\delta(n_I-n_I')$. By changing integration variables
\be
\rmd \nu_I\rmd\cj_I=\frac{k}{2\pi iQ}\rmd \scrp_I\rmd \widetilde{\scrp}_{I}.
\ee




The following large-$k$ asymptotic formula of the quantum dilogarithm is useful \cite{Faddeev:1993rs,Dimofte2011}
\be
\Psi_{\Delta}&=&e^{-\frac{ik}{2\pi(1+b^2)}\mathrm{Li}_2(e^{-Z})-\frac{ik}{2\pi(1+b^{-2})}\mathrm{Li}_2(e^{-\widetilde{Z}})}\nonumber\\
&&\times \lt[1+O(1/k)\rt].
\ee
The large-$k$ asymptotic behavior of $Z_{\times}$ is given by
\begin{widetext}
\be
Z_{\times}(\vec{\mu} \mid \vec{m})&=&e^{S_1+\widetilde{S}_1}\lt[1+O(1/k)\rt],\\
S_1&=&-\frac{ik}{2\pi(1+b^2)}\sum_{a=1}^{5}\lt[\mathrm{Li}_2(e^{-X_a})+\mathrm{Li}_2(e^{-Y_a})+\mathrm{Li}_2(e^{-Z_a})+\mathrm{Li}_2(e^{-W_a})\rt],\\
\widetilde{S}_1&=&-\frac{ik}{2\pi(1+b^{-2})}\sum_{a=1}^{5}\lt[\mathrm{Li}_2(e^{-\widetilde{X}_a})+\mathrm{Li}_2(e^{-\widetilde{Y}_a})+\mathrm{Li}_2(e^{-\widetilde{Z}_a})+\mathrm{Li}_2(e^{-\widetilde{W}_a})\rt].
\ee
\end{widetext}
Here $(X_a,Y_a,Z_a)_{a=1}^5\equiv -\mathbf{B}^T\vec{\scrp}$ and $(\widetilde{X}_a,\widetilde{Y}_a,\widetilde{Z}_a)_{a=1}^5\equiv -\mathbf{B}^T\vec{\widetilde \scrp}$. $W_a,\widetilde{W}_a$ are given by
\be
X_a+Y_a+Z_a+W_a&=&2\pi i+\frac{2\pi i}{k}(1+b^2),\nonumber\\
\widetilde{X}_a+\widetilde{Y}_a+\widetilde{Z}_a+\widetilde{W}_a&=&2\pi i+\frac{2\pi i}{k}(1+b^{-2})\label{XYZWconstraint1}
\ee
coinciding with the classical octahedron constraint \eqref{CXYZW} up to $O(1/k)$.

Therefore we rewrite $\cz_{S^3\setminus\G_5}$ for large $k$ by
\be
\cz_{S^3\setminus\G_5}&=&\cn_0\!\sum_{\vec{p}\in\mathbb{Z}^{15}}\int\limits_{\cc_\scrp}\rmd^{15}\!\scrp\rmd^{15}\!\widetilde{\scrp}\, e^{S_{\vec{p}}}\chi \lt[1+O(1/k)\rt],\label{Zintgral}\\
S_{\vec{p}}&=&S_0(\scrp,\widetilde{\scrp},\scrq,\widetilde{\scrq})+S_1(-\mathbf{B}^T \scrp)+\widetilde{S}_1(-\mathbf{B}^T \widetilde{\scrp})\nonumber\\
&&-\frac{k}{ b^2+1} \vec{p}\cdot\lt(\vec{\scrp}-b^2\vec{\widetilde \scrp}\rt).
\ee
where $\cn_0=-\frac{4k^{15} }{(2\pi)^{30}Q^{15}}$
The integration domain $\cc_\scrp$ is the 30 (real) dimensional submanifold of $(\vec{\scrp},\vec{\widetilde \scrp})\in \C^{30}$ satisfying $\vec{\nu}\in\cc$ and $\vec{\cj}\in [-\delta',2\pi-\delta']^{15}$.

The large-$k$ asymptotics of $\cz_{S^3\setminus\G_5}$ can be analyzed by the stationary phase approximation. The dominant contributions of integrals in \eqref{Zintgral} come from critical points that are solutions of the critical equations $\partial_{\scrp_I} S_{\vec{p}}=\partial_{\widetilde{\scrp}_I} S_{\vec{p}}=0$ (see Appendix \ref{Derivatives of Sp} for details). 

We make the linear transformation from $\vec{\scrq}',\vec{\scrp}$ to $\vec{\Phi} \equiv\left(X_{a}, Y_{a}, Z_{a}\right)_{a=1}^{5}$ and $ \vec{\Pi} \equiv\left(P_{X_{a}}, P_{Y_{a}}, P_{Z_{a}}\right)_{a=1}^{5}$, and similar to tilded variables  
\be
\vec{\scrq}'-2\pi i(\vec{n}+\vec{p})&=&\mathbf{A}\cdot \vec{\Phi} +\mathbf{B}\cdot \vec{\Pi} 
,\label{eom1}\\
\vec{\widetilde{\scrq}}{}'+2\pi i(\vec{n}+\vec{p})&=&\mathbf{A } \cdot \vec{\widetilde \Phi}+\mathbf{B}\cdot \vec{\widetilde \Pi}
,\label{eom2}\\
\vec{\scrp}=-(\mathbf{B}^{T})^{-1}\vec{\Phi},&\quad& \vec{\widetilde \scrp}=-(\mathbf{B}^{T})^{-1}\vec{\widetilde \Phi}.\label{eom3}
\ee
In terms of $\vec{\Phi},\vec{\Pi}$, the critical equations reduces to 
\be
&P_{X_a}=X_a^{\prime \prime}-W_a^{\prime \prime}, \quad P_{Y_a}=Y_a^{\prime \prime}-W_a^{\prime \prime},&\label{eom4}\\
&P_{Z_a}=Z_a^{\prime \prime}-W_a^{\prime \prime},\quad \widetilde{P}_{X_a}=\widetilde{X}_a^{\prime \prime}-\widetilde{W}_a^{\prime \prime},& \label{eom5}\\
&\widetilde{P}_{Y_a}=\widetilde{Y}_a^{\prime \prime}-\widetilde{W}_a^{\prime \prime}, \quad \widetilde{P}_{Z_a}=\widetilde{Z}_a^{\prime \prime}-\widetilde{W}_a^{\prime \prime},&\label{eom6}
\ee
where
\be
&&X''_a=\log\lt(1-e^{-X_a}\rt),\quad Y''_a=\log\lt(1-e^{-Y_a}\rt),\nonumber\\
&&Z''_a=\log\lt(1-e^{-Z_a}\rt),\quad W_a''=\log\lt(1-e^{-W_a}\rt),\label{XYZWlag1}\\
&&\widetilde{X}''_a=\log\lt(1-e^{-\widetilde{X}_a}\rt),\quad \widetilde{Y}''_a=\log\lt(1-e^{-\widetilde{Y}_a}\rt),\nonumber\\
&&\widetilde{Z}''_a=\log\lt(1-e^{-\widetilde{Z}_a}\rt),\quad \widetilde{W}_a''=\log\lt(1-e^{-\widetilde{W}_a}\rt), \label{XYZWlag2}
\ee 
Eqs.\eqref{XYZWlag1} and \eqref{XYZWlag2} reproduce e.g $z^{-1}+z^{\prime \prime}-1=0$ with $z=e^Z$ and $z''=e^{Z''}$, i.e. the Lagrangian submanifold $\cl_\Delta\subset \calp_{\partial\Delta}$ of framed flat $\PSlc$ connections on the ideal tetrahedron $\Delta$. $W_a,\widetilde{W}_a$ are given by \eqref{XYZWconstraint1}. The above logarithms are defined with the canonical lifts same as in \eqref{ZZZipi}. Moreover $X_a,Y_a,Z_a,P_{X_a},P_{Y_a},P_{Z_a}$ satisfying Eqs.\eqref{eom4} - \eqref{eom6} parametrizes the moduli space of framed flat $\PSlc$ connections on the ideal octahedron $\mathrm{oct}(a)$ made by gluing 4 ideal tetrahedra. Therefore any solution of critical equations gives 5 flat connections respectively on 5 ideal octahedra and vice versa. As a result, all possible critical points are in $\cl_{S^3\setminus \G_5}$, since the set of 5 flat connections on 5 ideal octahedra respectively is equivalent to a flat connection on $S^3\setminus\G_5$ (see the discussion below \eqref{Odarboux3fold}). Given a $\PSlc$ flat connections on $S^3\setminus\G_5$, $\vec{\scrq}',\vec{\scrp}$ at the critical point are determined by \eqref{eom1} - \eqref{eom3}, the same as the \eqref{ABCDt} up to $2\pi i(\vec{n}+\vec{p})$. 


We set $n_I\in\mathbb{Z}$ in \eqref{eom1} and \eqref{eom2} as an approximation up to $O(1/k)$, because for large $k$ any $\cj_I\in \R$ in \eqref{cjInIk} can be approximated by $n_I\in \mathbb{Z}$ up to $O(1/k)$ \footnote{e.g. When $k=10000$, $\cj_I/2\pi=0.5624587\cdots$ can be approximated by $n_I=5625$, the error bound is $|\cj_I/2\pi-n_I/k|<1/k$. }. Semiclassically critical equations are insensitive to $O(1/k)$. Then \eqref{eom1} - \eqref{eom3} are the same as \eqref{ABCDt} (only up to gauge shifts $m_I\to m_I+k\mathbb{Z}$ of $m_I\in\mathbb{Z}/k\mathbb{Z}$).

Fixing the range of $m_I$ (e.g. fixing $m_I=0,\cdots,k-1$) in $\mathcal{Z}_{S^{3}\setminus\Gamma_{5}}(\vec{\mu} \mid \vec{m})$ fixes the lifts of $\scrq_I,\widetilde{\scrq}_I$ from $e^{\scrq_I},e^{\widetilde{\scrq}_I}$, then uniquely fixes $\vec{p}=\vec{p}_0\in\mathbb{Z}$, given the lifts of logarithms in \eqref{XYZWlag1} and \eqref{XYZWlag2}, since different ${p}_I\in\mathbb{Z}$ change $\scrq_I,\widetilde{\scrq}_I$ by $\mp 2\pi i p_I$ ($n_I$ is determined by $\scrp_I$). Therefore only one term with $\vec{p}=\vec{p}_0$ in \eqref{Zintgral} has critical point and contributes to the leading order, whereas other terms with $\vec{p}\neq\vec{p}_0$ have no critical point thus are suppressed faster than $O(k^{-N})$ for all $N>0$.


Given $\vec{\mu},\vec{m}$ or $\vec{\scrq},\vec{\widetilde \scrq}$ such that there exists a $\PSlc$ flat connection on $S^3\setminus\G_5$ satisfying \eqref{eom1} and \eqref{eom2}, $\cz_{S^3\setminus\G_5}(\vec{\mu}\mid\vec{m})$ has a critical point thus is not suppressed fast, or in physics terms, $\cz_{S^3\setminus\G_5}(\vec{\mu}\mid\vec{m})$ has a semiclassical approximation. In this case, the critical point is generally non-unique, namely, there exists multiple critical points corresponding to the same $\vec{\scrq},\vec{\widetilde \scrq}$. Indeed different $\vec{\scrp} $ thus different $\vec{\Phi},\vec{\Pi}$ satisfying \eqref{eom3} - \eqref{XYZWlag2} can give the same $\vec{\scrq}$ via \eqref{eom1} (the critical equations expressed in terms of $e^{\scrq_I},e^{\scrp_I}$ are polynomial equations of degree higher than 1) and similar for tilded variables. The critical points 1-to-1 correspond to the solutions of $(\vec{\scrp},\vec{\widetilde \scrp})$ with given $\vec{\scrq},\vec{\widetilde \scrq}$,. The solutions are denoted by $({\scrp}^{(\a)}({\scrq}),{\widetilde \scrp}{}^{(\a)}({\widetilde \scrq}))$, $\a\in\ci$ where $\ci$ is a set of index labelling the solutions. $\a$ labels the branches of $\cl_{S^3\setminus \G_5}$. Given any $\a$, the coordinates $\vec{\scrq}$ provide a local parametrization of $\cl_{S^3\setminus \G_5}$. 

The asymptotic behavior of $\cz_{S^3\setminus\G_5}$ relates to the action $S_{\vec{p}=\vec{p}_0}$ evaluated at critical points
\be
S_{\vec{p}_0}^{(\a)}({\scrq},{\widetilde \scrq})=S_{\vec{p}_0}\lt({\scrq},{\widetilde \scrq},{\scrp}^{(\a)}({\scrq}),{\widetilde \scrp}^{(\a)}(\widetilde{\scrq})\rt).
\ee
The derivative of $S_{\vec{p}_0}^{(\a)}$ with respect to $\vec{\scrq},\vec{\widetilde \scrq}$ are
\be
\partial_{\vec{\scrq}}S_{\vec{p}_0}^{(\a)}&=&-\frac{ik}{2\pi (1+b^2)}\vec{\scrp}^{(\a)}({\scrq}),\\ 
\partial_{\vec{\widetilde \scrq}}S_{\vec{p}_0}^{(\a)}&=&-\frac{ik}{2\pi (1+b^{-2})}\vec{\widetilde \scrp}{}^{(\a)}(\widetilde{\scrq}).
\ee
where we have used $\partial_{{\scrp}} S_{\vec{p}_0}=\partial_{{\widetilde \scrp}} S_{\vec{p}_0}=0$ since $\{{\scrp}^{(\a)}(\scrq),{\widetilde \scrp}^{(\a)}({\widetilde{\scrq}})\}_{\a\in\ci}$ satisfy the critical equations. It implies that 
\footnote{Given $S(\vec{x})$ function on $\R^n$ and $\vec{\nabla}S(\vec{x})=\vec{f}(\vec{x})$, we choose a curve $c\subset \R^n$ parametrized by $t\in[0,1]$ ends at $x_0$. we denote by $\vec{t}$ the tangent vector of $c$. Then $\frac{\rmd}{\rmd t}S(\vec{x}(t))=\vec{t}\cdot\vec{\nabla}S(\vec{x}(t))=\vec{t}\cdot\vec{f}(\vec{x}(t))$. Therefore $S(\vec{x}_0)=\int^{x_0}_c \vec{f}(\vec{x})\cdot \rmd\vec{x} +C$.}
\begin{widetext}
\be
S_{\vec{p}_0}^{(\a)}({\scrq},{\widetilde \scrq})
=-\frac{ik}{2\pi (1+b^2)}\int^{\vec{\scrq}}\vec{\scrp}^{(\a)}(\scrq')\cdot \rmd \vec{\scrq}'-\frac{ik}{2\pi (1+b^{-2})}\int^{\vec{\widetilde \scrq}}\vec{\widetilde \scrp}{}^{(\a)}(\tilde{\scrq}{}')\cdot \rmd \vec{\widetilde \scrq}{}'+C^\a.\label{Sthth}
\ee
\end{widetext}
where $C^\a$ is an integration constant. The integrals are along certain curves embedded in $\cl_{S^3\setminus \G_5}$. The result is independent of smooth deformations of the integration contour in $\cl_{S^3\setminus \G_5}$, since $\O=0$ on the Lagrangian submanifold $\cl_{S^3\setminus \G_5}$. By this result $\mathrm{exp} (S_{\vec{p}_0}^{(\a)})$ is expressed as analog of WKB wave function. The large-$k$ asymptotics of $\cz_{S^3\setminus\G_5}$ is given by a finite sum over critical points
\be
\!\!\!\!\!\cz_{S^3\setminus\G_5}(\vec{\mu}\mid\vec{m})&=&\sum_{\a}\cn_0^{(\a)}e^{S_{\vec{p}_0}^{(\a)}({\scrq},{\widetilde \scrq})}\lt[1+O(1/k)\rt],\label{czalpha}\\
\cn_0^{(\a)}&=&\frac{\cn_0}{\sqrt{\det( -H_\a/2\pi)}}
\ee
where $H_\a$ is the Hessian matrix $\partial^2S_{\vec p_0}$ evaluated at the critical point. $H_\a$ is generically nondegenerate as supported by a large number of numerical experiments.

\subsection{Critical points of vertex amplitude and constant curvature 4-simplex}\label{Critical points of vertex amplitude and constant curvature 4-simplex}

Let's recall $\mathcal{Z}_{S^{3} \backslash \Gamma_{5}}(\iota)$ 
and coherent states $\psi_{z_{a}},\xi_{\left(x_{a}, y_{a}\right)}$ defined in \eqref{coheretntpsiz} and \eqref{xixym}. Restricting $\iota =\iota_{\vec{j},\vec{\xi}}$ to satisfy the simplicity constraint, $A_v=\mathcal{Z}_{S^{3} \backslash \Gamma_{5}}(\iota_{\vec{j},\vec{\xi}})$ is the vertex amplitude with cosmological constant. 

The simplicity constraint restrict $\mathrm{Re}(\mu_{ab})=0$ (the semiclassical behavior is insensitive to $\a_{ab}$), thus 
\be
&&
e^{2 L_{a b}}=\exp \left[\frac{2 \pi i}{k}\left(b \alpha_{a b}+2j_{a b}+\frac{\epsilon_{ab}}{2}\right)\right]\simeq e^{\frac{4 \pi i}{k}j_{ab}},\nonumber\\
&&
e^{2 \widetilde{L}_{a b}}= \exp \left[\frac{2 \pi i}{k}\left(b^{-1} \alpha_{a b}-2j_{a b}-\frac{\epsilon_{ab}}{2}\right)\right]\simeq e^{-\frac{4 \pi i}{k}j_{ab}}\nonumber
\ee
Here $\simeq$ stands for the semiclassical approximation.  
\begin{widetext}
We make the change of variable \eqref{muImI} in $\psi_{z_{a}}$ (recall $\mathscr{Q}_{I}^{\prime}=\mathscr{Q}_{I}-i \pi t_{I}, \ \widetilde{\mathscr{Q}}_{I}^{\prime}=\widetilde{\mathscr{Q}}_{I}-i \pi t_{I}$)
\be
\psi_{z_a}=\left(\frac{2}{k}\right)^{1 / 4} e^{S_{z_a}},\quad 
S_{z_a}\simeq\frac{b k\left(\widetilde{\mathscr{Q}}'_{a}+\mathscr{Q}'_{a}\right)}{2 \pi\left(b^{2}+1\right)}\left[\sqrt{2} z_a-\frac{b \left(\widetilde{\mathscr{Q}}_{a}'+\mathscr{Q}_{a}'\right)}{2 \left(b^{2}+1\right)}\right]-\frac{k(\bar{z}_a+z_a)^{2}}{8 \pi}.
\ee
where we neglect the term $-\sqrt{2}\b_a \re(z_a)$ since it is subleading as $k\to\infty$. $\xi_{(x_a, y_a)}$ is simplified by $k\to\infty$ and restricting $m_a=0,\cdots,k-1$ and $x_a,y_a\in(0,2\pi)$. After neglecting exponentially small contributions,  
\be
\xi_{(x_a, y_a)} &\simeq& \left(\frac{2}{k}\right)^{\frac{1}{4}} e^{\frac{i k x_a y_a}{4 \pi}} e^{-\frac{k}{4 \pi}\left(\frac{2 \pi m_a}{k}-x_a\right)^{2}} e^{-{i } y_a m_a}=\left(\frac{2}{k}\right)^{\frac{1}{4}} e^{S_{(x_a,y_a)}},\\
S_{(x_a,y_a)}&=&\frac{i k x_a y_a}{4 \pi}-\frac{k}{4 \pi}\left[\frac{i \left(\mathscr{Q}'_{a}-b^{2} \widetilde{\mathscr{Q}}'_{a}\right)}{b^{2}+1}-x_a\right]^{2}+ \frac{k\left(\mathscr{Q}_{a}'-b^{2} \widetilde{\mathscr{Q}}'_{a}\right)}{2 \pi\left(b^{2}+1\right)}y_a.
\ee
The vertex amplitude $A_v$ is expressed as below
\be
A_v&=&\cn\sum_{\vec{m} \in(\mathbb{Z} / k \mathbb{Z})^{5}}\sum_{\vec{n} \in(\mathbb{Z} / k \mathbb{Z})^{15}} \int_{\mathbb{R}^{5}\times\mathcal{C}}\,\mathrm{d}^{5} \mu   \mathrm{d}^{15} \nu\, e^{\ci(\scrp, \widetilde{\scrp}, \mathscr{Q}, \widetilde{\mathscr{Q}})},\qquad\qquad
\cn=\frac{4 i}{k^{15}} \left(\frac{2}{k}\right)^{5/2},\\
\ci&=&S_{0}(\scrp, \widetilde{\scrp}, \mathscr{Q}, \widetilde{\mathscr{Q}})+S_{1}\left(-\mathbf{B}^{T} \scrp\right)+\widetilde{S}_{1}\left(-\mathbf{B}^{T} \widetilde{\scrp}\right)+\sum_{a=1}^5\lt[S_{z_a}(\mathscr{Q}_a, \widetilde{\mathscr{Q}}_a)+S_{(x_a,y_a)}(\mathscr{Q}_a, \widetilde{\mathscr{Q}}_a)\rt].\nonumber
\ee
For finite $z_a$, the integrand is a Schwartz function of both $\vec{\mu}$ and $\vec{\nu}$ along the integration cycle ($\psi_{z_a}$ is a Gaussian function, and see the discussion below \eqref{Strans}), so interchanging $\vec{\mu}$-integral with $\vec{\nu}$-integral does not affect the result.
We apply the Poisson resummation similar to \eqref{poissonresumm},
\be
A_v&=&\cn'\sum_{(\vec{p},\vec{s}) \in\mathbb{Z}^{20} } \int_{\cc_\scrq\times\cc_\scrp}\mathrm{d}^{5} \scrq\, \mathrm{d}^{5} \widetilde{\scrq}\, \mathrm{d}^{15} \scrp\, \mathrm{d}^{15} \widetilde{\scrp}\, e^{\ci_{\vec{p},\vec{s}}(\scrp, \widetilde{\scrp}, \mathscr{Q}, \widetilde{\mathscr{Q}})},\qquad \cn'=\frac{i\left({k}/{2}\right)^{45/2}}{8192 \pi ^{40}  Q^{20}}\\
\ci_{\vec{p},\vec{s}}&=&\ci(\scrp, \widetilde{\scrp}, \mathscr{Q}, \widetilde{\mathscr{Q}})-\frac{k}{b^{2}+1} \vec{p} \cdot\left(\vec{\scrp}-b^{2} \vec{\widetilde{\scrp}}\right)-\frac{k}{b^{2}+1}\sum_{a=1}^5 {s}_a\left({\scrq}_a-b^{2} \widetilde{\scrq}_a\right),
\ee
\end{widetext}
where $\cc_{\scrq}$ is 10-dimensional real manifold satisfying $\mu_a\in\R$ and $m_a\in [0,k)$ (here $\mu_a,m_a$ are understood as continuous variables relating $\scrq_a,\widetilde{\scrq}_a$ by \eqref{muImI}).

We again apply the stationary phase analysis to the integral as $k\to\infty$. The critical equations $\partial_\scrp\ci_{\vec{p},\vec{s}}=\partial_{\widetilde{\scrp}}\ci_{\vec{p},\vec{s}}=0$ give the same results as \eqref{eom1} - \eqref{XYZWlag2} whose solutions are flat connections on $S^3\setminus\G_5$. Other set of critical equations $\partial_\scrq\ci_{\vec{p},\vec{s}}=\partial_{\widetilde{\scrq}}\ci_{\vec{p},\vec{s}}=0$ imply
\be
&&\frac{2\pi}{k}\mathrm{Re}(\mu_a)=\sqrt{2}\mathrm{Re}(z_a),\quad \frac{2\pi}{k}\mathrm{Re}(\nu_a)=\sqrt{2}\mathrm{Im}(z_a),\nonumber\\
&&\frac{2\pi}{k}m_a=x_a,\quad \frac{2\pi}{k}n_a=y_a,\quad s_a=0.\label{munumnzxy111}
\ee
See Appendix \ref{Derivatives of Sp} for derivations. At the critical point, the 4-holed sphere data $\scrq_a,\widetilde{\scrq}_a,\scrp_a,\widetilde{\scrp}_a
$ are determined by the coherent state label $z_a,x_a,y_a$. The determined 4-holed sphere data together with $2L_{ab},2\widetilde{L}_{ab}$ determined by $j_{ab}$ provide the boundary condition to the flat connection solving \eqref{eom1} - \eqref{XYZWlag2}.

The simplicity constraint requires that $z_{a}, x_{a}, y_{a}$ are determined by the data $\vec{j}, \vec{\xi}$ via \eqref{FGintxi}. Then \eqref{munumnzxy111} determines the 4-holed sphere FG coordinates $\cx_a,\cy_a$. Due to the 1-to-1 correspondence between values of FG coordinates $\{x_{E}\}_E$ and framed $\PSlc$ flat connections on $\cs_a$ \cite{FG03}, the resulting $\cx_a,\cy_a$ together with $e^{\scrq_{ab}}=e^{2L_{ab}}$ (belonging to U(1) as $k\to\infty$) determine uniquely a $\PSu\simeq \mathrm{SO}(3)$ flat connection on $\cs_a$. 
We denote by $\mathcal{M}_{flat}\left(\mathcal{S}_{a}, \mathrm{PSU}(2)\right)$ the moduli space of $\PSu$ flat connections on the 4-holed sphere $\cs_a$. Flat connections in this moduli space have following geometrical interpretations as constant curvature tetrahdra. 

\begin{theorem} \label{tetracorresp}

There is a bijection between flat connections in $\mathcal{M}_{\text {flat}}\left(\mathcal{S}_{a}, \mathrm{PSU}(2)\right)$ and convex constant curvature tetrahedron geometries in $3 d$, excepting degenerate geometries. Non-degenerate tetrahedral geometries are dense in $\mathcal{M}_{\text {flat}}\left(\mathcal{S}_{a}, \mathrm{PSU}(2)\right)$.

\end{theorem}

The proof of this theorem is given in \cite{curvedMink}. 
Both positive and negative constant curvature tetrahedra are included in $\mathcal{M}_{flat}\left(\mathcal{S}_{a}, \operatorname{PSU}(2)\right)$. 

Given the boundary condition leading to $\PSu$ flat connections on $\{\cs_a\}_{a=1}^5$, if there exists a $\PSlc$ flat connections on $S^3\setminus\G_5$ satisfying the boundary condition, it is a critical point of $A_v=\cz_{S^3\setminus\G_5}(\iota_{\vec{j},\vec{\xi}})$ and has the geometrical interpretation as a constant curvature 4-simplex. 

\begin{theorem} \label{tetrareconstr}

	There is a bijection between $\PSlc$ flat connections on $S^3\setminus\G_5$ satisfying the boundary condition, and nondegenerate, convex, oriented, geometrical 4-simplex with constant curvature in Lorentzian signature. 

\end{theorem}

The proof of this theorem is given in \cite{HHKR}. Note that not every flat connection on $\times_{a=1}^5\cs_a$ can extend to a flat connection $S^3\setminus\G_5$. It is shown in \cite{HHKR} that there is a subset of $\PSu$ flat connections on $\times_{a=1}^5\cs_a$ that can serve as the boundary of $\PSlc$ flat connections on $S^3\setminus\G_5$, and these boundary $\PSu$ flat connections correspond to 5 constant curvature tetrahedra that can be glued\footnote{Namely, they have the same constant curvature $\L$, and satisfy triangle-shapes matching and orientation matching when they are glued. } to form the close boundary of a nondegenerate 4-simplex with the same constant curvature $\L$. $A_v$ with these boundary data has critical points. However, any boundary $\PSu$ flat connection corresponding to 5 tetrahedra that cannot be glued to form 4-simplex boundary cannot extend to a $\PSlc$ flat connection on $S^3\setminus\G_5$, then results in that $A_v$ has no critical point thus is suppressed faster than $O(k^{-N})$ for all $N>0$. 

We do not discuss the possible flat connections corresponding to degenerate 4-simplex or tetrahedron. We also do not consider the boundary condition with $z_a\to\infty$ which leads to critical points located at the infinity of the integration cycle \footnote{Critical points at infinity give $z$, $z'$ or $z''\to\infty$ of certain $\Delta\subset S^3\setminus\G_5$. They either correspond to degenerate 4-simplex or correspond to special 4-simplices which become close to degenerate if $|\L\fa|\ll 1$, i.e. scales of 4-simplices are small (see \cite{hanSUSY} and Appendix E therein).}. 

In this geometrical correspondence between flat connection and 4-simplex geometry, the holonomy's squared eigenvalue $e^{2L_{ab}}$ relates to the area $\fa_{ab}$ of the 4-simplex boundary triangle $f_{ab}$ shared by the pair of tetrahedra $a,b$ (corresponding to $\cs_a,\cs_b$), i.e. semiclassically
\be
e^{2L_{ab}}\simeq e^{i\frac{|\L|}{3}\fa_{ab}},\quad \fa_{ab}\in[0,6\pi/|\L|]. \label{Landfa}
\ee
The framing flag $s_{\ell_{ab}}$ evaluated at $\fp_a\in\cs_a$, $s_{\ell_{ab}}(\fp_a)=\xi_{ab}$, relates to the unit normal $\vec{n}_{ab}$ (located at a vertex of the curved tetrahedron) of the face $f_{ab}$ viewed in the frame of tetrahedra $a$ by $\vec{n}_{ab}=\xi_{ab}^\dagger\vec{\bm{\sig}}\xi_{ab}$. Note that $\xi_{ab}$ is not alway the same as $\xi_i$ in \eqref{Hi}, see the discussion in the paragraph above \eqref{Hiprim}. Given the tetrahedra $a$, if we denote by $\vec{\fn}_i$ the geometrical outward pointing face-normal of the tetrahedron, we have $\vec{n}_{ab}=\sgn(\L)\vec{\fn}_i$ if $\xi_{ab}=\xi_i=(\xi_i^1,\xi_i^2)^T$, and $\vec{n}_{ab}=-\sgn(\L)\vec{\fn}_i$ if $\xi_{ab}=(-\bar{\xi}^2_i,\bar{\xi}^1_i)^T$ \cite{3dblockHHKR}.

\begin{figure}[h]
\begin{center}
\includegraphics[width=7cm]{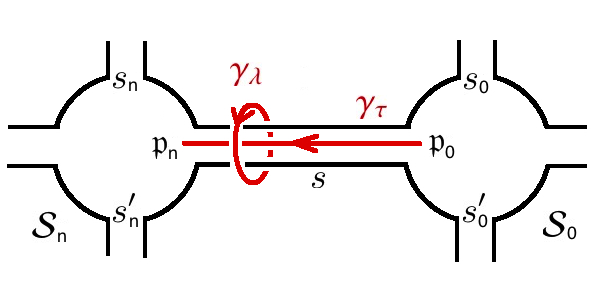}
\caption{The contour $\g_\t$ used to define the complex FN twist $\t_\ell$, and the meridian cycle $\g_\l$ used to define the complex FN length $\l_\ell$.}
\label{xandtau}
\end{center}
\end{figure}

In order to obtain the geometrical interpretation of the conjugate $\ct_{ab}$, we review the definition of the complex FN twist variable: Let's consider the annulus cusps $\ell$ connecting a pair of 4-holed spheres $\cs_0,\cs_n$. Let $s$ be the framing flag for $\ell$, and $s_{0,n},s_{0,n}'$ be the framing flags for a pair of other cusps connecting $\cs_{0,n}$. Then the complex FN twist is defined by (see e.g. \cite{DGV})
\be
\t_\ell=-\frac{\lag s_{0}\wedge s_{0}'\rag}{\lag s_{0}\wedge s\rag\lag s_{0}'\wedge s \rag}\frac{\lag s_{n}\wedge s\rag\lag s_{n}'\wedge s\rag}{\lag s_{n}\wedge s_{n}'\rag}.
\ee 
where $\lag s\wedge s'\rag$ are evaluated at a common point after parallel transportation. Without loss of generality, we evaluate the first ratio with factors ${\lag s_{0}\wedge s_{0}'\rag},\lag s_{0}\wedge s\rag,\lag s_{0}'\wedge s \rag$ at a point $\fp_0 \in\cs_0$, and evaluate the second ratio with factors ${\lag s_{n}\wedge s\rag,\lag s_{n}'\wedge s\rag},{\lag s_{n}\wedge s_{n}'\rag}$ at a point $\fp_n\in\cs_n$. The evaluation involves both $s(\fp_0)$ and $s(\fp_n)$ at two ends of $\ell$, while the parallel transportation between $s(\fp_0)$ and $s(\fp_n)$ depends on a choice of contour $\g_\t$ connecting $\fp_0,\fp_n$ (FIG.\ref{xandtau}). Different $\g_\t$ may transform $s(\fp_n)\to\l_\ell s(\fp_n)$ but keep $s(\fp_0)$ invariant. Moreover by definition, $\t_\ell$ also depend on the choice of two other auxiliary cusps for each of $\cs_0,\cs_n$. The choices of $\g_\t$ and the auxiliary cusps are part of the definition for $\t_\ell$. The choices in defining $\t_\ell$ doesn't affect our later result. The Atiyah-Bott symplectic form implies $\log(\tau_\ell)$ is the conjugate variable of the FN length variable $L_\ell=\log(\l_\ell)$ associated to the same annulus $\ell$:
\be
\{L_\ell,\log(\tau_{\ell'})\}_\O=\delta_{\ell,\ell'}.
\ee

Applying the above definition to $S^3\setminus\G_5$, we set $\cs_0=\cs_b$, $\fp_0\equiv \fp_b$ and $\cs_n=\cs_a$, $\fp_n\equiv\fp_a$. Framing flags associated to holes in $\cs_a$ (or $\cs_b$) evaluated at $\fp_a$ (or $\fp_b$) are $\{\xi_{ac}\}_{c\neq a}$ (or $\{\xi_{bc}\}_{c\neq b}$). In particular, $s(\fp_a)=\xi_{ab}$ and $s(\fp_{b})=\xi_{ba}$. We denote by $G_{ab}$ the flat connection holonomy along $\g_\t$ staring at $\fp_a$ and ending at $\fp_b$. $G_{ab}$ satisfies \cite{HHKR,3dblockHHKR,hanSUSY}
\be
G_{ab} \xi_{ab}=e^{-\frac{1}{2} \nu \operatorname{sgn}\left(V_{4}\right)  \Theta_{ab}+i\theta_{ab}} \xi_{ba},\quad \nu=\sgn(\L).\label{Gabxixi}
\ee
By the geometrical correspondence of the flat connection, $\Theta_{ab}$ is the hyper-dihedral (boost) angle hinged by the face $f_{ab}$ shared by the tetrahedra $a,b$ on the boundary of the 4-simplex. $\operatorname{sgn}\left(V_{4}\right)=\pm 1$ is the orientation of the 4-simplex. $\theta_{ab}\in[0,2\pi)$ is an angle relating to the phase convention of $\xi$'s. Inserting \eqref{Gabxixi} in the definition of $\t_\ell$ we obtain
\be
&&\t_{\ell_{ab}}\equiv\t_{ab}=e^{- \nu\operatorname{sgn}\left(V_{4}\right)  \Theta_{ab}+2i\theta_{ab}}\chi_{ab}(\xi),\\
&&\chi_{a b}(\xi)=\frac{\left\langle\xi_{b d} \wedge \xi_{b h}\right\rangle}{\left\langle\xi_{b d} \wedge \xi_{b a}\right\rangle\left\langle\xi_{b h} \wedge \xi_{b a}\right\rangle} \frac{\left\langle\xi_{a c} \wedge \xi_{a b}\right\rangle\left\langle\xi_{a e} \wedge \xi_{a b}\right\rangle}{\left\langle\xi_{a c} \wedge \xi_{a e}\right\rangle}\nonumber
\ee
where we have set $s_0(\fp_b)=\xi_{bd}, s_0'(\fp_b)=\xi_{bh}$ and $s_n(\fp_a)=\xi_{ac}, s_n'(\fp_a)=\xi_{ae}$. $\chi(\xi)$ is a function only depending on the boundary condition on $\{\cs_a\}_{a=1}^5$.

\begin{theorem}

	Given a $\PSlc$ flat connection $\Fa$ on $S^3\setminus\G_5$ corresponding to a nondegenerate convex constant curvature 4-simplex, there exists a unique flat connection $\widetilde{\Fa}\neq \Fa$ sharing the same boundary condition. $\Fa,\widetilde{\Fa}$ correspond to the same constant curvature 4-simplex geometry, but opposite orientations: $\mathrm{sgn}\left(V_{4}\right)|_\Fa=-\mathrm{sgn}\left(V_{4}\right)|_{\widetilde \Fa}$. 

\end{theorem}

The detailed proof is again given in \cite{HHKR}. The boundary condition corresponding to the boundary tetrahedra of nondegenerate 4-simplex gives exactly 2 critical points $\Fa,\widetilde{\Fa}$ which are called the \emph{parity pair}, as an analog of the similar siutation in the EPRL amplitude \cite{semiclassical}. That $\Fa,\widetilde{\Fa}$ correspond to the same geometry means that they endow the same edge-lengths, areas, angles, etc to the 4-simplex. Implied by this result, $e^{2 L_{ab}},e^{\cx_a},e^{\cy_a}$ have the same value at $\Fa,\widetilde{\Fa}$ since they are determined by the geometry, whereas $\t_{ab}$ are different
\be
\t_{ab}|_\Fa=e^{- \nu \Theta_{ab}+2i\theta_{ab}}\chi_{ab}(\xi),\quad \t_{ab}|_{\widetilde{\Fa}}=e^{ \nu\Theta_{ab}+2i\theta_{ab}}\chi_{ab}(\xi),\nonumber
\ee
since $\t_{ab}$ relates to the orientation. Here $\theta_{ab},\chi_{ab}(\xi)$ are the same at $\Fa,\widetilde{\Fa}$ since they are determined only by the boundary condition.

\begin{lemma}

At each annulus $\ell_{ab}$, $\t_{ab}=\t_{\ell_{ab}}$ relates to $\ct_{ab}$ by $\ct_{ab}=\frac{1}{2}\log(\t_{ab})+f(\{L_{ab}\},\{\cx_a,\cy_a\})$, where $f$ is a linear function of $\{L_{ab}\},\{\cx_a,\cy_a\}$.

\end{lemma}

\textbf{Proof:} Each $\t_{ab}$ is a product of $z^{\pm 1},z'{}^{\pm 1},z''{}^{\pm 1}$ of some ideal tetrahedra in the triangulation of $S^3\setminus\G_5$ (see Appendix A.3.3 in \cite{DGV}). When expressing in terms of octahedron phase space coordinates, Each $\log(\t_{ab})$ is a linear function of $X_a,P_{X_a},Y_a,P_{Y_a},Z_a,P_{Z_a}$ ($a=1,\cdots,5$) when we impose $C_a=2\pi i$, see \cite{hanSUSY} for explicit examples of $\log(\t_{ab})$
. By the symplectic transformation \eqref{ABCDt}, we express $\log(\t_{ab})=\sum_{c<d}(\a_{(ab),(cd)}\ct_{cd}+\b_{(ab),(cd)}L_{cd})+\sum_{c=1}^5(\rho_c\cx_c+\sig_c\cy_c)+i \pi\mathbb{Z}$. By $\{L_\ell,\log(\tau_{\ell'})\}_\O=\delta_{\ell,\ell'}$, we determine $\a_{(ab),(cd)}=2\delta_{(ab),(cd)}$ and define $f=-\frac{1}{2}[\sum_{c<d}\b_{(ab),(cd)}L_{cd}+\sum_{c=1}^5(\rho_c\cx_c+\sig_c\cy_c)+i \pi\mathbb{Z}] $.

$\Box$

As a result, $\ct_{ab}$ are given by
\be
\ct_{ab}|_\Fa&=&-\frac{1}{2} \nu\Theta_{ab}+i\theta_{ab}+\frac{1}{2}\log\chi_{ab}(\xi)\nonumber\\
&&+f(\{L_{ab}\},\{\cx_a,\cy_a\})+\pi i N^{(A)}_{ab}\\
\ct_{ab}|_{\widetilde \Fa}&=&\frac{1}{2}\nu \Theta_{ab}+i\theta_{ab}+\frac{1}{2}\log\chi_{ab}(\xi)\nonumber\\
&&+f(\{L_{ab}\},\{\cx_a,\cy_a\})+\pi i N^{(\widetilde{A})}_{ab}
\ee
where $N^{(A)}_{ab},N^{(\widetilde{A})}_{ab}\in\mathbb{Z}$ label the lifts of logarithms.

\subsection{Asymptotics of vertex amplitude}\label{Asymptotics of vertex amplitude}

The vertex amplitude $A_v$ has precisely 2 critical points $\Fa,\widetilde{\Fa}$ when the boundary condition corresponds to 5 tetrahedra that can be glued to form the close boundary of a nondegenerate constant curvature 4-simplex. By \eqref{czalpha}, the vertex amplitude has the following large-$k$ asymptotics
\be
A_v(\vec{j},\vec{\xi})&=&\lt[\sn_\a e^{S_{\vec{p}_{0}}^{(\alpha)}({\scrq}, {\widetilde{\scrq}})}+\sn_{\widetilde \a} e^{S_{\vec{p}_{0}}^{(\widetilde{\alpha})}({\scrq}, {\widetilde{\scrq}})}\rt],\label{AeSeS}\\
&&\times \lt[1+O\lt({1}/{k}\rt)\rt]\nonumber\\
\sn_{\a}&=&\frac{\cn'e^{\frac{ik}{4\pi}\sum_{a=1}^5\lt[4\mathrm{Re}(z_a)\mathrm{Im}(z_a)-x_ay_a\rt]}}{\sqrt{\det (-\mathscr{H}_{\a}/2\pi)}},
\ee
where $S_{\vec{p}_{0}}^{(\alpha)}$ is given in \eqref{Sthth}. The nondegeneracy of the Hessian matrix $\mathscr{H}_{\a}=\partial^2\ci_{\vec{p}_0,\vec{0}}$ is supported by many numerical experiments. ${\scrq}_I, {\widetilde{\scrq}}_I$ are the same at the critical points $\Fa,\widetilde{\Fa}$. $\a,\widetilde{\a}$ are branches of the Lagrangian submanifold $\cl_{S^3\setminus \G_5}$ containing $\Fa,\widetilde{\Fa}$ respectively. The asymptotics \eqref{AeSeS} of $A_v$ reduces to the same form as the one studied in \cite{HHKRshort,3dblockHHKR}. In the following we sketch the computation of \eqref{AeSeS} and refer the details to \cite{HHKRshort,3dblockHHKR}.

We rewrite \eqref{AeSeS} in $A_v\simeq e^{i\eta}(\sn_+ e^{S}+\sn_{-}e^{-S})$ where we factor out the overall phase $e^{i\eta}$, and we are interested in the phase difference $e^{2S}$ bewteen 2 exponentials in \eqref{AeSeS}. To extract the phase difference, we consider a small variation $\delta {\scrq}_I,\delta {\widetilde \scrq}_I$. The consequent variation of $\delta S$ is given by 
\be
2\delta S&=&-\frac{ik}{2\pi (1+b^2)}\lt(\vec{\scrp}^{(\a)}-\vec{\scrp}^{(\widetilde{\a})}\rt)\cdot \delta \vec{\scrq}-c.c.\nonumber\\
&=&-\frac{k\L}{6\pi (1+b^2)}\sum_{a<b}\lt(\Theta_{ab}+2\pi i N_{ab} \rt)\delta \fa_{ab}-c.c.\nonumber\\
&=&-\frac{i\L}{6\pi}\mathrm{Im}(t)\sum_{a<b}\Theta_{ab}\delta \fa_{ab}-\frac{i\L}{3}\mathrm{Re}(t)\sum_{a<b}N_{ab}\delta \fa_{ab},\nonumber
\ee
where $N_{ab}=\sgn(\L)(N^{(A)}_{ab}-N^{(\widetilde{A})}_{ab})\in\mathbb{Z}$. Only $\Theta_{ab}$ and $N^{(A)}_{ab},N^{(\widetilde{A})}_{ab}$ in $\ct_{ab}$ give nonvanishing contribution to $2\delta S$ because each of $\{L_{ab},\cx_a,\cy_a,\chi_{ab}(\xi),\theta_{ab}\}$ gives the same value at $\Fa$ and $\widetilde{\Fa}$ (see \cite{HHKRshort,3dblockHHKR} for details). By the Schl\"afli identity $\sum_{a<b}\delta\Theta_{ab} \fa_{ab}=\L|V_4|$ of constant curvature 4-simplex \cite{eva}, $\delta S$ can be integrated
\be
2S&=&-\frac{i\L k\g}{6\pi}\lt(\sum_{a<b}\fa_{ab}\Theta_{ab}-\L |V_4|\rt)\nonumber\\
&&-\frac{i\L k}{3}\sum_{a<b}N_{ab}\fa_{ab}+2C,
\ee
where $|V_4|$ is the 4-simplex volume. $2C$ is a geometry-independent integration constant. Eqs.\eqref{Landfa} and \eqref{conjclass} implies $\frac{|\L|}{3}\fa_{ab}=\frac{4\pi}{k}j_{ab}$, thus $\frac{i\L k}{3}\sum_{a<b}N_{ab}\fa_{ab}\in 2\pi i\mathbb{Z}$ is negligible in $e^{2S}$. As a result, we obtain the leading asymptotics of $A_v$ as
\be
A_v&=&e^{i\eta}\lt(\sn_{+}e^{iS_{\rm Regge}+C}+\sn_{-}e^{-iS_{\rm Regge}-C}\rt)\\
&&\times \lt[1+O\lt({1}/{k}\rt)\rt],\nonumber\\
\sn_{+,-}&=&\frac{\cn'}{\sqrt{\det (-\mathscr{H}_{\a,\widetilde{\a}}/2\pi)}}
\ee
where in the exponents
\be
S_{\rm Regge}&=&\frac{\L k\g}{12\pi}\lt(\sum_{a<b}\fa_{ab}\Theta_{ab}-\L |V_4|\rt).
\ee
is the Regge action of the constant curvature 4-simplex. The coefficient $\frac{|\L| k\g}{12\pi}$ is identified to be the inverse gravitational coupling $1/\ell_P^2$. This identification is consistent with \eqref{kandLambda}.

\section{Conclusion and outlook}

In this work, we propose an improved formulation of 4d spinfoam quantum gravity with cosmological constant $\L$. This formulation is featured with the finite spinfoam amplitudes on simplicial complexes and the correct semiclassical behavior of the vertex amplitude. 

Despite the above promising aspects, this formulation still has several open issues, which are expected to be addressed in the future research: Firstly, it is conjectured in Section \ref{Boundary data} that the boundary Hilbert space of the spinfoam amplitude $A$ is the Hilbert space of $\fq$-deformed spin-network states with $\fq$ root of unity. To prove this conjecture, we need to define and study coherent intertwiners of $\fq$-deformed spin-networks, and clarify if there is a canonical bijection between these coherent intertwiners and the boundary data of $A$. The expected coherent intertwiner should be a $\fq$-deformation of the Livine-Speciale coherent intertwiner \cite{LS}. 

We need to construct geometrical operator on the boundary Hilbert space to understand quantum geometrical interpretaion of boundary states. The construction may be based on the combinatorial quantization of SU(2) CS theory \cite{Alekseev:1994pa,Alekseev:1994au}. It is interesting to define coherent states that are coherent in both spins (areas) and intertwiners (shapes of curved tetrahedra). The coherent state may be a $\fq$-deformation of the complexifier coherent states in \cite{Thiemann:2000bw}. In addition, we need to direct sum over all graphs to defind the entire $\fq$-deformed LQG kinematical Hilbert space and check the cylindrical consistency of operators. This should generalize the work \cite{Lewandowski:2008ye} from real $\fq$ to $\fq$ root of unity.

It is discussed in Section \ref{Ambiguities} that the spinfoam amplitude $A$ has ambiguities in which the freedom of choosing coherent states is due to imposing semiclassical simplicity constraint to coherent state labels. It may be useful to develop an operator formalism or other ways to impose the simplicity constraint (such as the master constraint, Gupta-Bleuler, etc) at the quantum level, for reducing the freedom of the ampltiude. Another possible drawback of our implementation of simplicity constraint is that spins such that $\dim(\widetilde{\cm}_{\vec j})<2$ ($\widetilde{\cm}_{\vec j}$ only contains degenerate 4-gons) have to be excluded from our formalism.   
 
The present work only study the semiclassical behavior of the vertex amplitude. The semiclassical analysis should generalizes to the spinfoam amplitude with $\L$ on arbitrary simplicial complex, as well as taking into account the sum over $j$. 

$\L$ in this spinfoam model should be understood as the value of cosmological constant at ultraviolet. It would be interesting to apply the Wilson renormalization to the spinfoam model with $\L$ (see e.g. \cite{Bahr:2018gwf} for some earlier results). The spinfoam renormalization is expected to result in a flow of $\L$ from the ultraviolet to infrared, where the infrared value of $\L$ should relate to the observation. 

It should also be interesting to develop a group field theory (GFT) based on the spinfoam formulation with $\L$. The notion of group fields might be suitably generalized to include $\L$. The ``group fields'' might actually be fields on the moduli space of flat connections. The GFT is expected to reproduce spinfoam amplitudes $A$, which are finite order by order in the perturbative expansion.

\section*{Acknowledgements}

The author acknowledges Tudor Dimofte for communication on the positive angle structure, and acknowledges Zhe Sun for discussions on FG coordinates. This work receives support from the National Science Foundation through grant PHY-1912278.

\onecolumngrid

\appendix



\section{A plot for the polytope $\Fp(\mathrm{oct})$}\label{A plot for the polytope}

\begin{figure}[h]
	\begin{center}
	\includegraphics[width=7cm]{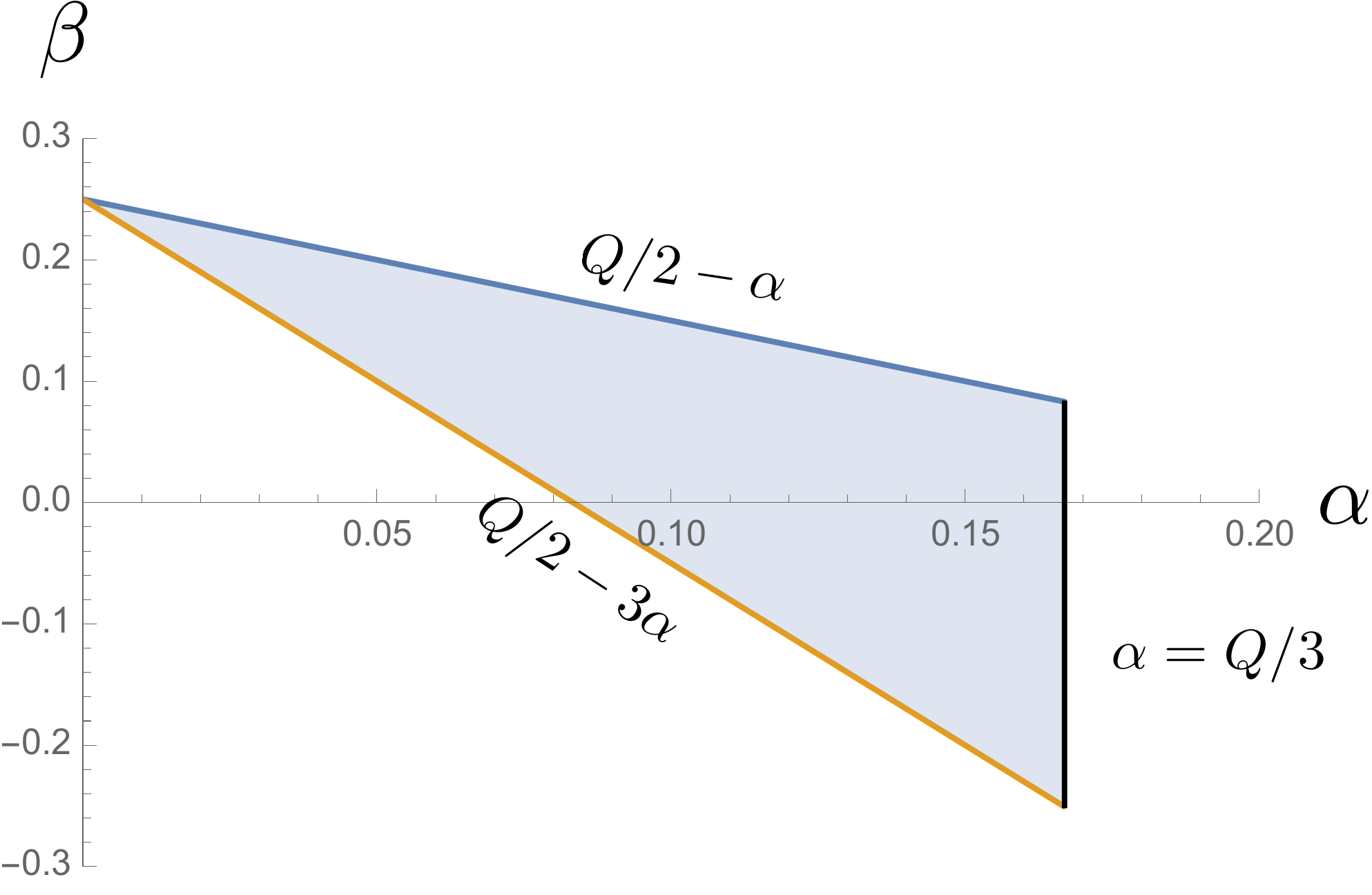}
	\caption{Setting $\a_X=\a_Y=\a_Z=\a$, $\b_X=\b_Y=\b_Z=\b$, and $Q=1/2$, $\Fp(\mathrm{oct})$ is restricted to the grey open triangle in the plot. }
	\label{polytope}
	\end{center}
\end{figure}

The open polytope $\Fp(\mathrm{oct})$ is defined by the following inequalities
\be
&&\a_{X},\a_{Y},,\a_{Z}>0,\quad \a_X+\a_Y+\a_Z<Q,\nonumber\\
&&\a_{X}+\b_{X}<\frac{Q}{2},\quad \a_{Y}+\b_{Y}<\frac{Q}{2},\quad \a_{Z}+\b_{Z}<\frac{Q}{2}, \nonumber\\
&& \a_{X}+\a_{Y}+\a_Z+\b_X>\frac{Q}{2},\quad \a_{X}+\a_{Y}+\a_Z+\b_Y>\frac{Q}{2},\nonumber\\
&&\a_{X}+\a_{Y}+\a_Z+\b_Z>\frac{Q}{2}.\nonumber
\ee 
FIG.\ref{polytope} plots the intersection between $\Fp(\mathrm{oct})$ and the plane of $\a_X=\a_Y=\a_Z$, $\b_X=\b_Y=\b_Z$

\section{Darboux coordinates of $\calp_{\partial(S^3\setminus\G_5)}$}\label{Darboux coordinates of P}


\begin{table}[h]
\begin{center}
\begin{tabular}{|c|c|c|}
\hline
$\cs_1$: & $h_2'\cap h_3':\ \ \chi^{(1)}_{23}=Z_2+Z_3$        & $h_3'\cap e_4':\ \ \chi^{(1)}_{34}=Y_3''+Z_3'+Z_4''+W_4'$\\
         & $h_2'\cap e_4':\ \ \chi^{(1)}_{24}=Z_2''+W_2'+Z_4$ & $h_3'\cap c_5':\ \ \chi^{(1)}_{35}=Z_3''+W_3'+Y_5''+Z_5'$\\
         & $h_2'\cap c_5':\ \ \chi^{(1)}_{25}=Y_2''+Z_2'+Z_5$ & $e_4'\cap c_5':\ \ \chi^{(1)}_{45}=Y''_4+Z_4'+Z_5''+W_5'$\\
\hline
$\cs_2$: & $f_1'\cap i_3':\ \ \chi^{(2)}_{13}=X_1''+Y_1'+X_3$ & $i_3'\cap f_4':\ \ \chi^{(2)}_{34}=X_3''+Y_3'+W_4''+X_4'$\\
         & $f_1'\cap f_4':\ \ \chi^{(2)}_{14}=X_1+X_4$        & $i_3'\cap b_5':\ \ \chi^{(2)}_{35}=W_3''+X_3'+X_5''+Y_5'$\\
         & $f_1'\cap b_5':\ \ \chi^{(2)}_{15}=W_1''+X_1'+X_5$ & $f_4'\cap b_5':\ \ \chi^{(2)}_{45}=X''_4+Y_4'+W_5''+X_5'$\\
\hline
$\cs_3$: & $b_1'\cap a_2':\ \ \chi^{(3)}_{12}=Z_1'+W_1''+X_2$        & $a_2'\cap d_4':\ \ \chi^{(3)}_{24}=W_2''+X_2'+Y_4'+Z_4''$\\
         & $b_1'\cap d_4':\ \ \chi^{(3)}_{14}=W'_1+X_1''+X_4'+Y_4''$ & $a_2'\cap d_5':\ \ \chi^{(3)}_{25}=X_2''+Y_2'+Z_5'+W_5''$\\
         & $b_1'\cap d_5':\ \ \chi^{(3)}_{15}=W_1+W_5'+X_5''$        & $d_4'\cap d_5':\ \ \chi^{(3)}_{45}=Y_4+W_5$\\
\hline
$\cs_4$: & $a_1'\cap c_2':\ \ \chi^{(4)}_{12}=Z_1+X_2'+Y_2''$        & $c_2'\cap j_3':\ \ \chi^{(4)}_{23}=Y_2'+Z_2''+Z_3'+W_3''$\\
         & $a_1'\cap j_3':\ \ \chi^{(4)}_{13}=Y''_1+Z_1'+W_3'+X_3''$ & $c_2'\cap j_5':\ \ \chi^{(4)}_{25}=Y_2+Y_5'+Z_5''$\\
         & $a_1'\cap j_5':\ \ \chi^{(4)}_{15}=Z_1''+W_1'+X_5'+Y_5''$ & $j_3'\cap j_5':\ \ \chi^{(4)}_{35}=W_3+Y_5$\\
\hline
$\cs_5$: & $i_1'\cap e_2':\ \ \chi^{(5)}_{12}=Y'_1+Z_1''+W_2'+X_2''$ & $e_2'\cap g_3':\ \ \chi^{(5)}_{23}=Z_2'+W_2''+Y_3'+Z_3''$\\
         & $i_1'\cap g_3':\ \ \chi^{(5)}_{13}=Y_1+X_3'+Y_3''       $ & $e_2'\cap g_4':\ \ \chi^{(5)}_{24}=W_2+Z_4'+W_4''$\\
         & $i_1'\cap g_4':\ \ \chi^{(5)}_{14}=X_1'+Y_1''+W_4'+X_4''$ & $g_3'\cap g_4':\ \ \chi^{(5)}_{34}=Y_3+W_4$\\
\hline
\end{tabular}
\end{center}
\caption{Edge coordinates $\chi^{(a)}_{mn}$ of 4-holed spheres. Recall in FIG.\ref{5oct} that the octahedra are glued through the triangles labelled by $a,b,c,d,e,f,g,h,i,j$. For example $a_2'$ labels the triangles symmetric to the triangle $a$ with respect to the equator of Oct(2). The ``primed triangles'' with the primed labels triangulate the geodesic boundary of $S^3\setminus\G_5$. Here $X_{a},Y_a,Z_a,W_a$ ($a=1,\cdots,5$) are the tetrahedron edge coordinates from the 4 tetrahedra triangulating Oct($a$).}
\label{edges}
\end{table}%

Darboux coordinates $
\mathscr{Q}_I=(2L_{ab},\cx_a),\ \mathscr{P}_I=(\ct_{ab},\cy_a)$ expressed in terms of $(X_{a}, P_{X_{a}})$,$(Y_{a}, P_{Y_{a}})$,$(Z_{a}, P_{Z_{a}})$,$(C_{a}, \Gamma_{a})$ are listed below 
\be
2L_{12}&=&-C_3-C_4-C_5+P_{Y_3}+P_{Y_4}+P_{Y_5}+X_3+X_4+X_5+Y_3+Y_4+Y_5+3 i \pi,\\
2L_{13}&=&-{C_2}-{C_5}+P_{Y_2}+P_{Y_4}-P_{Z_4}+P_{Z_5}+X_2+X_5+Y_2+Y_5+2 Z_5+i \pi,  \\
2L_{14}&=&-{C_3}+P_{Y_2}+P_{Y_5}-P_{Z_2}+P_{Z_3}-P_{Z_5}+X_3+Y_3+2 Z_3,\\
2L_{15}&=&-C_2-C_4+P_{Y_3}+P_{Z_2}-P_{Z_3}+P_{Z_4}+X_2+X_4+Y_2+Y_4+2 Z_2+2 Z_4, \\
2L_{23}&=&-P_{X_1}+P_{X_4}-P_{X_5}-P_{Y_4}+X_4-Y_4, \\
2L_{24}&=&-P_{X_3}+P_{X_5}-P_{Y_1}-P_{Y_5}-X_1+X_5-Y_1-Y_5+i \pi,  \\
2L_{25}&=&P_{X_1}+P_{X_3}-P_{X_4}-P_{Y_1}-P_{Y_3}+X_1+X_3-Y_1-Y_3,\\
2L_{34}&=&C_1-C_5+P_{X_2}+P_{X_5}-P_{Y_2}-P_{Z_1}-P_{Z_5}-X_1+X_2+X_5-Y_1-Y_2+Y_5-2 Z_1+i \pi,  \\
2L_{35}&=&-{C_1}+P_{X_1}-P_{X_2}-P_{X_4}-P_{Z_1}+P_{Z_4}+X_1-X_4+Y_1-Y_4+ 2 i \pi,  \\
2L_{45}&=&-C_3-P_{X_2}+P_{X3}+P_{Y_1}-P_{Z_1}+P_{Z_2}-P_{Z_3}-X_2+X_3-Y_2+Y_3+2 i \pi .
\ee
\be
\cx_1&=&\chi^{(1)}_{25}=P_{Y_2}-P_{Z_2}-Z_2+Z_5+i \pi,\\
\cx_2&=&\chi^{(2)}_{15}=-P_{X_1}-X_1+X_5+i \pi,\\
\cx_3&=&\chi^{(3)}_{15}=C_1-C_5+P_{X_5}-X_1+X_5-Y_1+Y_5-Z_1+Z_5+i \pi,\\
\cx_4&=&\chi^{(4)}_{15}=-C_1-P_{X_5}+P_{Y_5}+P_{Z_1}+X_1-X_5+Y_1+Z_1+2 i \pi,\\
\cx_5&=&\chi^{(5)}_{14}=-C_4-P_{X_1}+P_{X_4}+P_{Y_1}-X_1+X_4+Y_4+Z_4+2 i \pi.
\ee
\be
\ct_{12}&=&\frac{1}{2} \left(X_2-X_3-X_4+Y_1+Y_2-Y_3-Y_4+Z_2\right),\\
\ct_{13}&=&\frac{1}{2} \left(-X_2+X_3-Y_2+Y_3-Y_5+Z_1-Z_2-Z_5\right),\\
\ct_{14}&=&\frac{1}{2} \left(-Y_2-Z_2-Z_3+Z_5\right),\\
\ct_{15}&=&\frac{1}{2} \left(-X_2-Y_2-Z_2-Z_4\right),\\
\ct_{23}&=&\frac{1}{2} \left(-X_4+Y_1+Y_4-Y_5+Z_1-Z_5\right),\\
\ct_{24}&=&\frac{1}{2} \left(X_2+X_3-X_4+Y_1+Y_3-Y_4+Z_3+Z_5\right),\\
\ct_{25}&=&\frac{1}{2} \left(-X_3-X_4+Y_1+Y_3-Y_4-Z_4\right),\\
\ct_{34}&=&\frac{1}{2} \left(-X_2+X_3+Y_3-Y_5+Z_1+Z_3\right),\\
\ct_{35}&=&\frac{1}{2} \left(X_3+Y_3-Y_5+Z_1-Z_4-Z_5\right),\\
\ct_{45}&=&\frac{1}{2} \left(X_2+Z_3+Z_4+Z_5\right).
\ee
\be
\cy_1&=&\chi^{(1)}_{23} =Z_2+Z_3,\\
\cy_2&=&\chi^{(2)}_{14}=X_1+X_4,\\
\cy_3&=&\chi^{(3)}_{45}-2\pi i=-X_5+Y_4-Y_5-Z_5,\\
\cy_4&=&-\chi^{(4)}_{35}+2\pi i=X_3+Y_3-Y_5+Z_3,\\
\cy_5&=&\chi^{(5)}_{34}-2\pi i=-X_4+Y_3-Y_4-Z_4.
\ee 
We impose $C_a=2\pi i$ to all $2L_{ab}$ and $\cx_a$. We check that \eqref{octsymp} implies
\be
\{\mathscr{Q}_I,\mathscr{P}_J\}_\O=\delta_{IJ},\quad \{\mathscr{Q}_I,\mathscr{Q}_J\}_\O=\{\mathscr{P}_I,\mathscr{P}_J\}_\O=0.\quad I,J=(\ell_{ab},\cs_a).
\ee
 
\section{Symplectic transformation}\label{Symplectic transformation}

The linear symplectic transformation from $\vec{\Phi} \equiv (X_{a}, Y_{a}, Z_{a} )_{a=1}^{5}$ and $\vec{\Pi} \equiv (P_{X_{a}}, P_{Y_{a}}, P_{Z_{a}} )_{a=1}^{5}$ to $\vec{\mathscr Q},\vec{\mathscr P}$ is given by 
\be
\left(\begin{array}{c}
\vec{\mathscr Q} \\
\vec{\mathscr P}
\end{array}\right)=\left(\begin{array}{ll}
\mathbf{A} & \mathbf{B} \\
-(\mathbf{B}^{T})^{-1} & \mathbf{0}
\end{array}\right)\left(\begin{array}{l}
\vec{\Phi} \\
\vec{\Pi}
\end{array}\right)+i \pi\left(\begin{array}{l}
\vec{t} \\
\vec{0}
\end{array}\right),
\ee
Explicitly, $\mathbf{A}, \mathbf{B}, \vec{t}$ are given below
\be
\mathbf{A}=\left(
\begin{array}{ccccccccccccccc}
 0 & 0 & 0 & 0 & 0 & 0 & 1 & 1 & 0 & 1 & 1 & 0 & 1 & 1 & 0 \\
 0 & 0 & 0 & 1 & 1 & 0 & 0 & 0 & 0 & 0 & 0 & 0 & 1 & 1 & 2 \\
 0 & 0 & 0 & 0 & 0 & 0 & 1 & 1 & 2 & 0 & 0 & 0 & 0 & 0 & 0 \\
 0 & 0 & 0 & 1 & 1 & 2 & 0 & 0 & 0 & 1 & 1 & 2 & 0 & 0 & 0 \\
 0 & 0 & 0 & 0 & 0 & 0 & 0 & 0 & 0 & 1 & -1 & 0 & 0 & 0 & 0 \\
 -1 & -1 & 0 & 0 & 0 & 0 & 0 & 0 & 0 & 0 & 0 & 0 & 1 & -1 & 0 \\
 1 & -1 & 0 & 0 & 0 & 0 & 1 & -1 & 0 & 0 & 0 & 0 & 0 & 0 & 0 \\
 -1 & -1 & -2 & 1 & -1 & 0 & 0 & 0 & 0 & 0 & 0 & 0 & 1 & 1 & 0 \\
 1 & 1 & 0 & 0 & 0 & 0 & 0 & 0 & 0 & -1 & -1 & 0 & 0 & 0 & 0 \\
 0 & 0 & 0 & -1 & -1 & 0 & 1 & 1 & 0 & 0 & 0 & 0 & 0 & 0 & 0 \\
 0 & 0 & 0 & 0 & 0 & -1 & 0 & 0 & 0 & 0 & 0 & 0 & 0 & 0 & 1 \\
 -1 & 0 & 0 & 0 & 0 & 0 & 0 & 0 & 0 & 0 & 0 & 0 & 1 & 0 & 0 \\
 -1 & -1 & -1 & 0 & 0 & 0 & 0 & 0 & 0 & 0 & 0 & 0 & 1 & 1 & 1 \\
 1 & 1 & 1 & 0 & 0 & 0 & 0 & 0 & 0 & 0 & 0 & 0 & -1 & 0 & 0 \\
 -1 & 0 & 0 & 0 & 0 & 0 & 0 & 0 & 0 & 1 & 1 & 1 & 0 & 0 & 0 \\
\end{array}
\right),
\ee
\be
\mathbf{B}=\left(
\begin{array}{ccccccccccccccc}
 0 & 0 & 0 & 0 & 0 & 0 & 0 & 1 & 0 & 0 & 1 & 0 & 0 & 1 & 0 \\
 0 & 0 & 0 & 0 & 1 & 0 & 0 & 0 & 0 & 0 & 1 & -1 & 0 & 0 & 1 \\
 0 & 0 & 0 & 0 & 1 & -1 & 0 & 0 & 1 & 0 & 0 & 0 & 0 & 1 & -1 \\
 0 & 0 & 0 & 0 & 0 & 1 & 0 & 1 & -1 & 0 & 0 & 1 & 0 & 0 & 0 \\
 -1 & 0 & 0 & 0 & 0 & 0 & 0 & 0 & 0 & 1 & -1 & 0 & -1 & 0 & 0 \\
 0 & -1 & 0 & 0 & 0 & 0 & -1 & 0 & 0 & 0 & 0 & 0 & 1 & -1 & 0 \\
 1 & -1 & 0 & 0 & 0 & 0 & 1 & -1 & 0 & -1 & 0 & 0 & 0 & 0 & 0 \\
 0 & 0 & -1 & 1 & -1 & 0 & 0 & 0 & 0 & 0 & 0 & 0 & 1 & 0 & -1 \\
 1 & 0 & -1 & -1 & 0 & 0 & 0 & 0 & 0 & -1 & 0 & 1 & 0 & 0 & 0 \\
 0 & 1 & -1 & -1 & 0 & 1 & 1 & 0 & -1 & 0 & 0 & 0 & 0 & 0 & 0 \\
 0 & 0 & 0 & 0 & 1 & -1 & 0 & 0 & 0 & 0 & 0 & 0 & 0 & 0 & 0 \\
 -1 & 0 & 0 & 0 & 0 & 0 & 0 & 0 & 0 & 0 & 0 & 0 & 0 & 0 & 0 \\
 0 & 0 & 0 & 0 & 0 & 0 & 0 & 0 & 0 & 0 & 0 & 0 & 1 & 0 & 0 \\
 0 & 0 & 1 & 0 & 0 & 0 & 0 & 0 & 0 & 0 & 0 & 0 & -1 & 1 & 0 \\
 -1 & 1 & 0 & 0 & 0 & 0 & 0 & 0 & 0 & 1 & 0 & 0 & 0 & 0 & 0 \\
\end{array}
\right),
\ee
\be
\vec{t}=(-3, -3, -2, -4, 0, 1, 0, 1, 0, 0, 1, 1, 1, 0, 0)^T.
\ee

\section{Proof of Lemma \ref{trianineq}}\label{Proof of Lemma}

\begin{lemma}

$H_{i=1,\cdots,4}\in\Su$ satisfying $H_4H_3H_2H_1=1$ exist if and only if $j'_{i=1,\cdots,4}$ satisfy the triangle inequality, i.e. there exists $J$ such that
\be
&&|j'_1-j'_2|\leq J\leq \mathrm{min}\lt(j'_1+j'_2,k-j'_1-j'_2\rt),\label{trigineq1app}\\
&&|j'_3-j'_4|\leq J\leq \mathrm{min}\lt(j'_3+j'_4, k-j'_3-j'_4\rt).\label{trigineq2app}
\ee

\end{lemma}
	
\textbf{Proof:} We denote by $\frac{4\pi }{k}j'_i=r_i\in[0,2\pi)$. $H_i=\cos(r_i/2)+ i\vec{n}'_i\cdot\vec{\mathbf{\sig}}\sin(r_i/2)$ where $\vec{n}'$ is a unit vector in $\R^3$. $\vec{n}'=-\vec{n}$ in case of minus sign in \eqref{Hi} and $\vec{n}'=\vec{n}$ in case of plus sign. We denote by $H_2H_1=\cos(R/2)+i\vec{N}\cdot\vec{\mathbf{\sig}}\sin(R/2)$ with $R=\frac{4\pi }{k}J\in [0,2\pi)$ then $H_4H_3=\cos(R/2)-i\vec{N}\cdot\vec{\mathbf{\sig}}\sin(R/2)$. Taking the trace gives
\be
&&\cos\lt(\frac{R}{2}\rt)=\cos\lt(\frac{r_1}{2}\rt)\cos\lt(\frac{r_2}{2}\rt)-\vec{n}'_1\cdot\vec{n}'_2\sin\lt(\frac{r_1}{2}\rt)\sin\lt(\frac{r_2}{2}\rt),\\
&&\cos\lt(\frac{R}{2}\rt)=\cos\lt(\frac{r_3}{2}\rt)\cos\lt(\frac{r_4}{2}\rt)-\vec{n}'_3\cdot\vec{n}'_4\sin\lt(\frac{r_3}{2}\rt)\sin\lt(\frac{r_4}{2}\rt)
\ee
Since $\sin\lt(\frac{r_i}{2}\rt)\geq 0$, unit vectors $\vec{n}'_{i=1,\cdots,4}$ exists if and only if
\be
&&\cos\lt(\frac{r_1+r_2}{2}\rt)\leq \cos\lt(\frac{R}{2}\rt)\leq \cos\lt(\frac{r_1-r_2}{2}\rt),\label{Rrr1}\\
&&\cos\lt(\frac{r_3+r_4}{2}\rt)\leq \cos\lt(\frac{R}{2}\rt)\leq \cos\lt(\frac{r_3-r_4}{2}\rt),\label{Rrr2}
\ee
which is equivalent to
\be
&&|r_1-r_2|\leq R\leq \mathrm{min}\lt(r_1+r_2,4\pi-r_1-r_2\rt),\label{rrRrr1}\\
	&&|r_3-r_4|\leq R\leq \mathrm{min}\lt(r_3+r_4,4\pi-r_3-r_4\rt).\label{rrRrr2}
\ee

Conversely, \eqref{Rrr1},\eqref{Rrr2} or \eqref{rrRrr1},\eqref{rrRrr2} imply the existence of 2 spherical triangles in $S^3$ sharing a common edge. The spherical triangles form a 4-gon whose edges are geodesics in $S^3$ with length $r_i/2$ ($i=1,\cdots,4$). The diagonal of the 4-gon is a geodesic whose length is $R/2$. The 4-gon in $S^3$ implies the existence of $H_{i=1,\cdots,4}\in\Su$ satisfy $H_4H_3H_2H_1=1$ by the argument in Section \ref{Integration over PSU(2) flat conn}.


$\Box$

\section{Determining $\xi_i$'s from $\theta$ and $\phi$}\label{thetatheta}

It is useful to consider $\cos(\theta_{24})=-\frac{1}{2}[\Tr(H_1H_4^{-1})-\Tr(H_1)\Tr(H_4)]=\frac{1}{2}\Tr(H_4H_1)=\frac{1}{2}\Tr(H_2H_3)$. The following relation holds between $\phi$ and $\theta_{24}$ \cite{Nekrasov:2011bc}
\be
2\cos(\theta_{24})=\frac{\left(e^{i\phi }+e^{-i\phi}\right) \sqrt{c_{12}(A) c_{34}(A)}-2\left(m_{2} m_{3}+m_{1} m_{4}\right)+A\left(m_{1} m_{3}+m_{2} m_{4}\right)}{A^{2}-4},\label{thetaphiE}
\ee
where $m_i=\Tr(H_i)$ and 
\be
A=e^{i\theta}+e^{-i\theta},\quad c_{i j}(A)=A^{2}+m_{i}^{2}+m_{j}^{2}-A m_{i} m_{j}-4.
\ee

For SU(2) flat connections satsifying $H_4H_3H_2H_1=1$, we make a partial gauge fixing that $H_4=\mathrm{diag}(e^{i a_4},e^{-i a_4})$, $a_4\in[0,\pi)$ \footnote{We use the conjugation $\eps\,\mathrm{diag}(\l,\l^{-1})\,\eps^{-1}=\mathrm{diag}(\l^{-1},\l)$, where $\eps_{\a\b}=-\eps_{\b\a}$ and $\det(\eps)=1$, to fix $a_4\in [0,\pi)$ in $\l=e^{i a_4}$.}. Thus as a unit vector in Euclidean $\R^4$, $v_j=(v_j^0,v_j^1,v_j^2,v_j^3)$,
\be
v_4=\lt(\cos \left(a_4\right),0,0,-\sin \left(a_4\right)\right)
\ee
representing $H_4^{-1}$. For the triangle $(v_1,v_3,v_4)$, we $v_1=(1,0,0,0)$, $\lag v_1,v_3\rag=\cos(\theta_{13})$ ($\theta_{13}=\theta$), $\lag v_3,v_4\rag=\cos(a_3)$, and $\lag v_4,v_4\rag=1$ to determine $v_3$ 
\be
v_3&=&\left(\cos \left(\theta _{13}\right),0,v_3^2,v_3^3\right)\\
v_3^2&=&\sqrt{-\left(\csc ^2\left(a_4\right) \left(\cos ^2\left(a_3\right)+\cos ^2\left(\theta _{13}\right)\right)\right)+2 \cos \left(a_3\right) \cot \left(a_4\right) \csc \left(a_4\right) \cos \left(\theta _{13}\right)+1}\nonumber\\
v_3^3&=&\csc \left(a_4\right) \left(\cos \left(a_4\right) \cos \left(\theta _{13}\right)-\cos \left(a_3\right)\right),\nonumber
\ee
where we have used the remaining rotational summetry (of 1-2 plane) to fix $v_3^1=0$ and $v_3^2>0$. Then we use $\lag v_1,v_2\rag=\cos(a_1)$, $\lag v_2,v_3\rag=\cos(a_2)$, $\lag v_2,v_4\rag=\cos(\theta_{24})$, and $\lag v_2,v_2\rag=1$ to determine $v_2$
\be
v_2&=&\left(\cos \left(a_1\right),v_2^1,v_2^2,\csc \left(a_4\right) \left(\cos \left(a_1\right) \cos \left(a_4\right)-\cos \left(\theta _{24}\right)\right)\right)\\
v_2^1&=&\pm\Big(2 \cos \left(a_2\right) \csc \left(a_4\right) \left(\cot \left(a_4\right) \left(\cos \left(a_1\right) \cos \left(a_3\right)+\cos \left(\theta _{13}\right) \cos \left(\theta _{24}\right)\right)-\csc \left(a_4\right) \left(\cos \left(a_1\right) \cos \left(\theta _{13}\right)\right.\right.\nonumber\\
&&\left.\lt.+\cos \left(a_3\right) \cos \left(\theta _{24}\right)\right)\right)+\csc \left(a_4\right) \left(-2 \cos \left(a_1\right) \cot \left(a_4\right) \cos \left(\theta _{24}\right)+\csc \left(a_4\right) \left(\cos ^2\left(\theta _{13}\right)+\sin ^2\left(\theta _{13}\right) \cos ^2\left(\theta _{24}\right)\right)\rt.\nonumber\\
&&\lt.-2 \cos \left(a_3\right) \cos \left(\theta _{13}\right) \left(\cot \left(a_4\right)-\cos \left(a_1\right) \csc \left(a_4\right) \cos \left(\theta _{24}\right)\right)+\cos ^2\left(a_3\right) \csc \left(a_4\right)+\sin ^2\left(a_3\right) \cos ^2\left(a_1\right) \csc \left(a_4\right)\right)\nonumber\\
&&+\cos ^2\left(a_2\right)-1\Big)^{\frac{1}{2}}\Big(\csc ^2\left(a_4\right) \left(\cos ^2\left(a_3\right)+\cos ^2\left(\theta _{13}\right)\right)-2 \cos \left(a_3\right) \cot \left(a_4\right) \csc \left(a_4\right) \cos \left(\theta _{13}\right)-1\Big)^{-\frac{1}{2}}\nonumber\\
v_2^2&=&2 \left(\cos \left(a_1\right) \left(\cos \left(\theta _{13}\right)-\cos \left(a_3\right) \cos \left(a_4\right)\right)+\cos \left(\theta _{24}\right) \left(\cos \left(a_3\right)-\cos \left(a_4\right) \cos \left(\theta _{13}\right)\right)+\sin ^2\left(a_4\right) \left(-\cos \left(a_2\right)\right)\right)\nonumber\\
&&\times \sqrt{-\left(\csc ^2\left(a_4\right) \left(\cos ^2\left(a_3\right)+\cos ^2\left(\theta _{13}\right)\right)\right)+2 \cos \left(a_3\right) \cot \left(a_4\right) \csc \left(a_4\right) \cos \left(\theta _{13}\right)+1}\nonumber\\
&&\times \lt(-4 \cos \left(a_3\right) \cos \left(a_4\right) \cos \left(\theta _{13}\right)+\cos \left(2 a_3\right)+\cos \left(2 a_4\right)+\cos \left(2 \theta _{13}\right)+1\rt)^{-1},\nonumber
\ee
where $\pm$ of $v_2^1$ corresponds to the parity symmetry with respect to the plane of  $F_{134}$ (spanned by the $x^0,x^2,x^3$-directions in $\R^4$) where $v_1,v_3,v_4$ leave. Choosing $+$ or $-$ of $v_2^1$ is equivalent to fixing the orientation of $n_{123}\wedge n_{134}$ since $v_2^1\to -v_2^1$ transforms
\be
n_{123}\wedge n_{134}\to -n_{123}\wedge n_{134},\quad \text{where}\quad n^d_{ijk}\|n_{ijk}\|=\epsilon_{abcd}v_i^a v_j^b v_k^c.
\ee
Now all $\{H_i\}_{i=1}^4$ are fixed by 
\be
H_1&=&v_2,\quad H_4=v_4^{-1}, \quad H_3=v_4v_3^{-1},\quad H_2=v_3v_2^{-1},\\
&&\text{where}\quad v_j=v_j^0I+i\sum_{a=1}^3 v_j^a\bm{\sig}_a.
\ee
Every $H_i$ is uniquely determined by $(a_i,\theta_{13},\theta_{24})$, where $\theta_{24}$ relates to $\phi$ by \eqref{thetaphiE}, then $\xi_i$ is determined up to scaling as the eigenvector of $H_i$ for the eigenvalue whose square is $e^{2L_{ab}}$.


\section{Critical equations}\label{Derivatives of Sp}

Derivatives of $S_{\vec{p}}$ are given by
\be
-\frac{2\pi (1+b^2)}{ik}\partial_{\vec \scrp} S_{0}&=&\mathbf{AB}^T\cdot \vec{\scrp}+\vec{\scrq}+\frac{k}{(1+b^2)}\mathbf{AB}^T\cdot\left(\vec{\scrp}-b^2 \vec{\widetilde{\scrp}}\right),\\
-\frac{2\pi (1+b^{-2})}{ik }\partial_{\vec{\widetilde{\scrp}}} S_{0}&=&\mathbf{A B}^{T} \cdot \vec{\widetilde{\scrp}}+ \vec{\widetilde{\scrq}}-\frac{k}{(1+b^2)}\mathbf{AB}^T\cdot\left(\vec{\scrp}-b^2 \vec{\widetilde{\scrp}}\right),\\
-\frac{2\pi (1+b^2)}{ik}\partial_{\vec{\scrp}} S_{1}&=&-\mathbf{B}\cdot (P_{X_{a=1,\cdots,5}},P_{Y_{a=1,\cdots,5}},P_{Z_{a=1,\cdots,5}})^T,\\ 
&\text{e.g.}&\quad P_{Z_a}\equiv\log\lt(1-e^{-Z_a}\rt)-\log\lt(1-e^{X_a+Y_a+Z_a}\rt),\label{PZaequiv1}\\
-\frac{2\pi (1+b^{-2})}{ik}\partial_{\vec{\widetilde \scrp}}\widetilde{ S}_{1}&=&-\mathbf{B}\cdot (\widetilde{P}_{X_{a=1,\cdots,5}},\widetilde{P}_{Y_{a=1,\cdots,5}},\widetilde{P}_{Z_{a=1,\cdots,5}})^T,\\
&\text{e.g.}& \quad \widetilde{P}_{X_a}=\log\lt(1-e^{-\widetilde{Z}_a}\rt)-\log\lt(1-e^{\widetilde{X}_a+\widetilde{Y}_a+\widetilde{Z}_a}\rt),\label{PZaequiv2}\\
\vec{\scrp}&=&-(\mathbf{B}^T)^{-1}\cdot({X}_{a=1,\cdots,5},{Y}_{a=1,\cdots,5},{Z}_{a=1,\cdots,5})^T,\\
\vec{\widetilde \scrp}&=&-(\mathbf{B}^T)^{-1}\cdot(\widetilde{X}_{a=1,\cdots,5},\widetilde{Y}_{a=1,\cdots,5},\widetilde{Z}_{a=1,\cdots,5})^T,
\ee
where the branches of the logarithms are the same as the canonical lift in \eqref{ZZZipi}.

We define
\be
&&X''_a:=\log\lt(1-e^{-X_a}\rt),\quad Y''_a:=\log\lt(1-e^{-Y_a}\rt),\nonumber\\
&&Z''_a:=\log\lt(1-e^{-Z_a}\rt),\quad W_a'':=\log\lt(1-e^{-W_a}\rt),\label{XYZWlag1app}\\
&&\widetilde{X}''_a:=\log\lt(1-e^{-\widetilde{X}_a}\rt),\quad \widetilde{Y}''_a:=\log\lt(1-e^{-\widetilde{Y}_a}\rt),\nonumber\\
&&\widetilde{Z}''_a:=\log\lt(1-e^{-\widetilde{Z}_a}\rt),\quad \widetilde{W}_a'':=\log\lt(1-e^{-\widetilde{W}_a}\rt), \label{XYZWlag2app}
\ee 
such that e.g. $z=e^Z$ and $z''=e^{Z''}$ reproduce $z^{-1}+z^{\prime \prime}-1=0$ i.e. the Lagrangian submanifold $\cl_\Delta\subset \calp_{\partial\Delta}$ of framed flat $\PSlc$ connections on the ideal tetrahedron $\Delta$. $W_a,\widetilde{W}_a$ are given by \eqref{XYZWconstraint1}. The above logarithms are defined with the canonical lifts same as in \eqref{ZZZipi}. We define $P_{X_{a}},P_{Y_a},P_{Z_{a}}$ and $\widetilde{P}_{X_{a}},\widetilde{P}_{Y_{a}},\widetilde{P}_{Z_a}$ ($a=1,\cdots,5$) in the same way as \eqref{PXYZW}.
$X_a,Y_a,Z_a,P_{X_a},P_{Y_a},P_{Z_a}$ with Eqs.\eqref{XYZWlag1app}, \eqref{XYZWlag2app}, and \eqref{XYZWconstraint1} parametrizes the moduli space of framed flat $\PSlc$ connections on the ideal octahedron $\mathrm{oct}(a)$ made by gluing 4 ideal tetrahedra.

The critical equations $\partial_{X_I} S_{\vec{p}}=\partial_{\widetilde{X}_I} S_{\vec{p}}=0$ can be written in terms of $\vec{\Phi} \equiv\left(X_{a}, Y_{a}, Z_{a}\right)_{a=1}^{5}$ and $ \vec{\Pi} \equiv\left(P_{X_{a}}, P_{Y_{a}}, P_{Z_{a}}\right)_{a=1}^{5}$:
\be
\vec{\scrq}'&=&\mathbf{A}\cdot \vec{\Phi} +\mathbf{B}\cdot \vec{\Pi} 
+2\pi i(\vec{n}+\vec{p}),\label{eom1app}\\
\vec{\widetilde{\scrq}}'&=&\mathbf{A } \cdot \vec{\widetilde \Phi}+\mathbf{B}\cdot \vec{\widetilde \Pi}
-2\pi i(\vec{n}+\vec{p}).\label{eom2app}
\ee
where $\vec{p}\in\mathbb{Z}^{15}$. Up to $2\pi i(\vec{n}+\vec{p})$, the critical equations \eqref{eom1} and \eqref{eom2} reproduces the $\scrq$-part of \eqref{ABCDt}, whereas here $\vec{\Phi}$ and $\vec{\Pi}$ are related by \eqref{XYZWlag1app} \eqref{XYZWlag2app}, and \eqref{PXYZW}. Note that the $\scrp$-part of \eqref{ABCDt} has been reproduced by the relation between $\left(X_{a}, Y_{a}, Z_{a}\right)_{a=1}^{5}$ and $\vec{\scrp}$ (see above \eqref{XYZWconstraint1}).

For the vertex amplitude $A_v$, the critical equations $\partial_\scrq\ci_{\vec{p},\vec{s}}=\partial_{\widetilde{\scrq}}\ci_{\vec{p},\vec{s}}=0$ give
\be
\frac{2 \pi\left(1+b^{2}\right)}{k}\partial_{\mathscr{Q}_a} \ci_{\vec{p},\vec{s}}&=&-i{\scrp}_a+\sqrt{2}bz_a-\frac{b^2\lt(\scrq'_a+\widetilde{\scrq}'_a\rt)}{1+b^2}+\frac{\scrq'_a-b^2\widetilde{\scrq}'_a}{1+b^2}+y_a+ix_a-2\pi s_a=0\\
&=&-i{\scrp}_a+\sqrt{2}bz_a-\frac{2\pi b}{k}\mu_a-\frac{2\pi i}{k}m_a+y_a+ix_a-2\pi s_a=0, \\
\frac{2 \pi\left(1+b^{-2}\right)}{k}\partial_{\widetilde{\mathscr{Q}}_a} \ci_{\vec{p},\vec{s}}
&=&-i\widetilde{\scrp}_a+\sqrt{2}b^{-1}z_a-\frac{\lt(\scrq'_a+\widetilde{\scrq}'_a\rt)}{1+b^2}-\frac{\scrq'_a-b^2\widetilde{\scrq}'_a}{1+b^2}-y_a-ix_a+2\pi s_a=0\\
&=&-i\widetilde{\scrp}_a+\sqrt{2}b^{-1}z_a-\frac{2\pi b^{-1}}{k}\mu_a+\frac{2\pi i}{k}m_a-y_a-ix_a+2\pi s_a=0.
\ee
where $\mu_a$ and $m_a$ relate to $\scrq'_a$ and $\widetilde{\scrq}'_a$ by \eqref{muImI}. The above equations is solved by 
\be
\frac{2\pi}{k}\mu_a=\sqrt{2}\mathrm{Re}(z_a),\quad \frac{2\pi}{k}\nu_a=\sqrt{2}\mathrm{Im}(z_a),\quad \frac{2\pi}{k}m_a=x_a,\quad \frac{2\pi}{k}n_a=y_a-2\pi s_a,\label{munumnzxy111app}
\ee
where $\nu_a$ and $n_a$ relate to $X_a$ and $\widetilde{X}_a$ by \eqref{muImI}. Although $\mu_a,\nu_a$ have nonzero imaginary parts, $\a_a=\mathrm{Im}(\mu_a),\b_a=\mathrm{Im}(\nu_a)$ are fixed and do not scale as $k\to\infty$ (whereas $\mathrm{Re}(\mu_a),\mathrm{Re}(\nu_a)$ are not fixed and need to be determined by the critical equations), thus we can view $\mu_a,\nu_a$ to be real in \eqref{munumnzxy111app} as far as the semiclassical limit is concerned. The domain of $n_a$ has been restricted to the single period $n_a\in[-\delta,k-\delta]$ by \eqref{poissonresumm} ($\delta>0$ is arbitrarily small), so the last equation implies
\be
s_a=0.
\ee
when $y_a\in[0,2\pi)$ and $y_a$ is not infinitesimally close to $0$ or $2\pi$.

\bibliographystyle{jhep}
\bibliography{muxin}

\end{document}